\documentclass[11pt]{article}
\pdfoutput=1

\usepackage{jheppub} 

\usepackage{amsmath,amssymb,epsfig,amsfonts}
\usepackage{graphicx}
\usepackage{epsf}
\usepackage{makeidx}
\usepackage{multirow}
\usepackage[all]{xy}
\usepackage{hyperref}
\usepackage{verbatim} 
\usepackage{rotating}
\usepackage{xcolor}
\usepackage{multirow}
\usepackage{bm}
\usepackage{cleveref}
\usepackage{slashed}
\usepackage[normalem]{ulem}

\usepackage{graphicx}
\usepackage{latexsym,amsmath,amsfonts,amssymb}
\usepackage{mathrsfs}
\usepackage{bbm}
\usepackage{bm}
\usepackage{subfigure}
\usepackage{paralist}

\usepackage{tikz}
\usepackage{diagbox}
\usetikzlibrary{positioning}
\usetikzlibrary{calc}
\usetikzlibrary{decorations.pathreplacing,calligraphy}

\usepackage{xstring}
\usetikzlibrary{decorations.pathmorphing} 
\usetikzlibrary{decorations.markings} 
\usetikzlibrary{arrows} 
\usetikzlibrary{shapes} 
\usetikzlibrary{matrix} 
\usetikzlibrary{positioning} 
\usepackage[english]{babel} 
\usepackage[autostyle]{csquotes}

\tikzstyle{brane}=[draw]
\tikzset{D7/.style={circle, draw=black, inner sep=0pt, fill=white, minimum size=3mm}}
\tikzset{hasse/.style={circle, fill,inner sep=2pt}}
\tikzset{flavor/.style={regular polygon,fill=white,regular polygon sides=4,inner sep=2.5pt, draw}}
\tikzset{gauge/.style={circle, draw,inner sep=2.5pt}}
\tikzset{gaugeb/.style={circle, draw,fill=black,inner sep=2.5pt}}
\tikzset{gauger/.style={circle, draw,fill=cyan,inner sep=2.5pt}}
\tikzset{gaugeg/.style={circle, draw,fill=red,inner sep=2.5pt}}
\tikzset{bd/.style={circle, draw=black, inner sep=0pt, fill=black, minimum size=2mm}}
\tikzset{wd/.style={circle, draw=black, inner sep=0pt, fill=white, minimum size=2mm}}
\tikzset{Dynkin/.style={circle, draw=black, inner sep=0pt, fill=white, minimum size=2mm}}
\tikzstyle{ligne}=[draw, thick] 
\tikzset{doublearrow/.style={ draw=black!75, color=black!75, thick, double distance=3pt, }}

\numberwithin{equation}{section}  

\graphicspath{ {figs/} }

\newcommand{\be}{\begin{equation}}
\newcommand{\ee}{\end{equation}}
\newcommand{\ba}{\begin{aligned}}
\newcommand{\ea}{\end{aligned}}

\newcommand{\Spin}{\text{Spin}}

\def\half{{\frac{1}{2}}}

\def\unit{{1\kern-.65ex {\rm l}}}
\def\1{{1\kern-.65ex {\rm l}}}

\def\CA{{\cal A}}
\def\CB{{\cal B}}
\def\CC{{\cal C}}

\def\CF{{\cal F}}

\def\CM{{\cal M}}
\def\CN{{\cal N}}
\def\CO{{\cal O}}

\def\CR{{\cal R}}

\def\CT{{\cal T}}

\newcount\hour \newcount\minute
\hour=\time \divide \hour by 60
\minute=\time
\def\now{%
\ifnum \hour<13
  \ifnum \hour=0 \advance \hour by 12 \number\hour:\else \number\hour:\fi%
     \ifnum \minute<10 0\fi%
     \number\minute%
\ A.M.%
\else \advance \hour by -12 \number\hour:%
  \ifnum \minute<10 0\fi%
  \number\minute%
  \ P.M.%
\fi%
}

\makeatother

\def\mb{\mathbb}

\def\bp{\begin{pmatrix}}
\def\ep{\end{pmatrix}}

\newcommand{\nn}{\nonumber}

\newcommand{\bea}{\begin{equation} \begin{aligned}}
 \newcommand{\eea}{\end{aligned} \end{equation}}
\newcommand{\bit}{\begin{itemize}} 
\newcommand{\eit}{\end{itemize}} 

\newcommand{\Z}{\mathbb{Z}}
\newcommand{\C}{\mathbb{C}}
\newcommand{\R}{\mathbb{R}}
\newcommand{\Q}{\mathbb{Q}}

\renewcommand{\t}{\widetilde }

\renewcommand{\d}{\partial }

\newcommand{\m}{\mathfrak{m}}
\newcommand{\n}{\mathfrak{n}}

\DeclareMathOperator{\Tr}{Tr}

\newcommand{\ov}{\over}

\newcommand{\h}{\widehat}

\newcommand{\MG}{{\mathbf X}} 

\newcommand{\CY}{{\MG}} 

\newcommand{\FT}{{\mathcal{T}_\MG^{\rm 5d}}} 

\newcommand{\KK}{D_{S^1}}

\newcommand{\FTfour}{{\mathscr{T}_\MG^{\rm 4d}}} 
\newcommand{\FTX}[1]{{\mathcal{T}_{#1}^{\rm 5d}}} 
\newcommand{\FTXfour}[1]{{\mathscr{T}_{#1}^{\rm 4d}}} 

\newcommand{\MQ}{{\text{MQ}}} 

\newcommand{\EQfour}{{\text{EQ}^{(4)}}} 
\newcommand{\MQfour}{{\text{MQ}^{(4)}}} 
\newcommand{\EQfive}{{\text{EQ}^{(5)}}} 
\newcommand{\MQfive}{{\text{MQ}^{(5)}}} 

\newcommand{\torHtwo}{{\mathfrak{t}_2}} 

\begin{document}

\baselineskip=18pt  
\numberwithin{equation}{section}  
\allowdisplaybreaks  


%
%


\thispagestyle{empty}

\vspace*{0.8cm} 
\begin{center}
{{\Huge  Coulomb and Higgs Branches from \\ 

\smallskip Canonical Singularities: Part 0
}}

 \vspace*{1.5cm}
Cyril Closset,  Sakura Sch\"afer-Nameki, Yi-Nan Wang\\

 \vspace*{1.0cm} 
{\it Mathematical Institute, University of Oxford, \\
Andrew-Wiles Building,  Woodstock Road, Oxford, OX2 6GG, UK}\\

\vspace*{0.8cm}
\end{center}
\vspace*{.5cm}

\noindent
Five- and four-dimensional superconformal field theories with eight supercharges arise from canonical threefold singularities in M-theory and Type IIB string theory, respectively. We study their Coulomb and Higgs branches using crepant resolutions and deformations of the singularities. We propose a relation between the resulting moduli spaces, by compactifying the theories to 3d, followed by 3d $\mathcal{N}=4$ mirror symmetry and an $S$-type gauging of an abelian flavor symmetry. 
In particular, we use this correspondence to determine the Higgs branch of some 5d SCFTs and their magnetic quivers from the geometry.  As an application of the general framework, we observe that singularities that engineer Argyres-Douglas theories in Type IIB also give rise to rank-0 5d SCFTs in M-theory. We also compute the higher-form symmetries of the 4d and 5d SCFTs, including the one-form symmetries of generalized Argyres-Douglas theories of type $(G, G')$.

\newpage

\tableofcontents


\section{Introduction}

Many supersymmetric quantum field theories (SQFT) can be realized within string theory, which provides us with powerful tools to study them in a strongly coupled regime, often in terms of geometry \cite{Witten:1996qb,Katz:1996fh,Klemm:1996bj, Bershadsky:1996nh, Seiberg:1996bd, Morrison:1996xf, Intriligator:1997pq, Shapere:1999xr}. In this paper, we take yet another look at the `geometric engineering' of superconformal field theories (SCFT) in space-time dimension $d=4$ and $d=5$. Our leitmotiv will be that one can refine our understanding of these systems, in both 4d and 5d, by considering their relations to each other via toroidal compactification to 3d \cite{Hori:1997zj}. A more detailed description of our   geometric approach, together with  many examples, will appear in forthcoming papers~\cite{Closset:2021lwy, CSNWII}.

These theories are partly characterized by their moduli space of vacua, which generally includes both a Coulomb and a Higgs branch, together with mixed branches.
In the case of 5d SCFTs, their moduli spaces can be studied from their geometric realization in M-theory on a canonical singularity: the crepant ({\it i.e.} retaining the Calabi-Yau condition) resolutions model the Coulomb branch, whereas the Higgs branch corresponds to the deformations of the singularity. The former has recently been utilized in exploring the classification of 5d SCFTs in  \cite{Hayashi:2013lra, Hayashi:2014kca, DelZotto:2017pti, Jefferson:2017ahm, Closset:2018bjz, Jefferson:2018irk,Apruzzi:2018nre, Bhardwaj:2018yhy,Bhardwaj:2018vuu, Apruzzi:2019vpe, Apruzzi:2019opn, Apruzzi:2019enx, Bhardwaj:2019jtr, Apruzzi:2019kgb, Bhardwaj:2019fzv, Bhardwaj:2019xeg, Eckhard:2020jyr, Bhardwaj:2020kim}. 
On the Higgs branch side, the approach using brane-webs  \cite{Aharony:1997bh} has been particularly successful \cite{Cabrera:2018ann,Cabrera:2018jxt,Cabrera:2019izd,Bourget:2019aer,Bourget:2019rtl,Cabrera:2019dob,Grimminger:2020dmg, Bourget:2020asf, Bourget:2020gzi}. In the latter approach, one
constructs the so-called  {\it magnetic quiver} from a IIB brane web, a 3d $\CN=4$ gauge theory whose Coulomb branch is conjectured to be the identical to the 5d Higgs branch.

So far, the analysis of the 5d Higgs branch {directly} from the canonical-singularity point of view has been lagging behind. This is partly because the low-energy physics arising from M-theory on the deformed singularity is subject to M2-brane instanton corrections. In this paper and in \cite{Closset:2021lwy, CSNWII}, we explore the deformation theory of the hypersurface canonical singularities classified in \cite{yau2005classification}%
\footnote{See also \protect\cite{Davenport:2016ggc}. For more mathematical background, see \protect\cite{arnold2012singularities}. } 
and we discuss the general structure of the Higgs branch for {\it isolated hypersurface singularities.} We concurrently explore the Coulomb branch, which arises from the crepant resolutions of such singularities. We also determine the higher-form symmetries of  these 5d theories, which provide further information about their spectrum of defect operators. The exploration of this vast class of models give us a wonderful playground to test current ideas about 5d SCFTs.

In this paper, we focus on a few interesting points, which deserve particular attention.
First, we will argue that the magnetic quiver of the 5d SCFT is closely related to the 4d SCFT that arises by geometric engineering in Type IIB string theory {\it on the same singularity}. (The 4d geometric engineering in IIB has also studied in a lot detail over the years, see {\it e.g.} \cite{Shapere:1999xr, Cecotti:2010fi, Xie:2015rpa, Chen:2016bzh, Wang:2016yha, Chen:2017wkw, Xie:2017pfl}.) This gives us a complementary understanding of magnetic quivers, and of their generalizations, directly in the geometric-engineering picture. The relation between the 5d and 4d SCFT goes through dimensional reduction to 3d, thanks to 3d $\CN=4$ mirror symmetry~\cite{Intriligator:1996ex}, as first studied in \cite {Hori:1997zj}. In the IIB picture, M2-brane instantons become D3-brane instantons wrapping a 3-cycle times a circle; these correspond to monopole operators in the 3d $\CN=4$ magnetic-quiver theory, which can be more efficiently resummed \cite{Cremonesi:2013lqa, Nakajima:2015txa, Bullimore:2015lsa, Braverman:2016wma}. 

Conversely, our analysis of the geometry of canonical hypersurface singularities allows us to also make statements about 4d  SCFTs. For instance, we can study the Higgs branch of 4d SCFTs by studying the crepant resolutions of canonical singularities, and identify the additional low-energy degrees of freedom on the 4d Higgs branch (such as free vector multiplets and irreducibles SCFTs) from the resolution. Similarly, the Coulomb branch of the 4d SCFT is closely related to the 5d Higgs branch. 

There has been a lot of recent progress in studying the 4d $\CN=2$ Coulomb branch---for a limited list of references see \cite{Argyres:2015gha, Martone:2020nsy, Argyres:2020nrr, Argyres:2020wmq}---as well as the Higgs branch---see {\it e.g.} \cite{Argyres:2012fu,DelZotto:2014kka,Buican:2015ina,Song:2015wta,Maruyoshi:2016tqk,Maruyoshi:2016aim,Agarwal:2016pjo,Song:2017oew,Agarwal:2017roi,Benvenuti:2017bpg,Agarwal:2018zqi} and much of the SCFT/Vertex operator algebra literature \cite{Beem:2013sza}, as in {\it e.g.} \cite{Beem:2017ooy, Beem:2019tfp, Beem:2019snk}.
 It would be greatly desirable to better understand how much of these beautiful structures can be recovered from the geometric-engineering framework.

\medskip
\noindent
Let us now outline the logical structure of the paper, and highlight some of the results.

\subsection{Setup and summary}

Consider a canonical threefold singularity $\MG$. An important class of examples consists of the isolated hypersurface singularities (IHS). An IHS is defined by a single quasi-homogeneous polynomial $F(x)$ in $\C^4$:
\be\label{def IHS}
\MG \cong \big\{ (x_1, x_2, x_3, x_4)\in \C^4 \; | \; F(x_1, x_2, x_3, x_4) =0\big\}~,
\ee
We require this space to be singular, {\it i.e.} $\partial_{x_i}F= F=0$ at an isolated point,
{with an additional condition so that $\MG$ is canonical, as studied in M-theory {\it e.g.} in \cite{Gukov:1999ya}. }
This will provide our main class of examples in this paper.%
\footnote{Another important class of examples, which we will also consider, consists of CY$_3$ toric singularities.
} 
The canonical singularity $\MG$ in M-theory defines a 5d SCFT, which we denote by $\FT$:
\be\label{FT 5d definition}
\FT \qquad \longleftrightarrow  \qquad \text{M-theory on} \; \R^{1,4} \times \MG~.
\ee
 while the {\it same} singularity in Type IIB string theory defines a 4d $\CN=2$ SCFT, $\FTfour$:
 \be\label{FT 4d definition}
\FTfour \qquad \longleftrightarrow  \qquad \text{Type IIB on} \; \R^{1,3} \times \MG~.
\ee
These two superconformal field theories, in space-times of  different  dimensions, are of course distinct systems, but they are closely related. Consider compactifying the theory $\FT$ on a circle. We obtain a KK-theory in 4d with 4d $\CN=2$ supersymmetry, denoted by $\KK\FT$. By the M-theory/Type IIA duality, this theory is engineered by Type IIA string theory on $\R^{1,3} \times \MG$~\cite{Witten:1996qb,Nekrasov:1996cz}. If we further consider this theory on an arbitrary space-time manifold $\CM_4$, we have
\be
\KK\FT \; \text{on} \; \CM_4  \qquad \longleftrightarrow  \qquad \text{Type IIA on} \; \CM_4 \times \MG~.
\ee
In particular, we may consider $\CM_4 = \R^{1,2} \times S^1_\beta$, corresponding to a toroidal compactification of $\FT$ on a circle of radius $\beta$. This SQFT has 3d $\CN=4$ supersymmetry. Similarly, we may compactify the 4d SCFT $\FTfour$ on a circle, to obtain another 3d KK-theory with $\CN=4$ supersymmetry. These two three-dimensional systems are related by T-duality
\bea\label{T duality picture}
&D_{T^2} \FT &&\qquad \longleftrightarrow  \qquad && \text{M-theory on} \; \R^{1,2} \times T^2 \times \MG&& \cr
&  &&\qquad \longleftrightarrow  \qquad && \text{Type IIA on} \; \R^{1,2} \times S^1_\beta \times \MG&& \cr
&  &&\qquad \longleftrightarrow  \qquad && \text{Type IIB on} \; \R^{1,2} \times S^1_{1\ov \beta} \times \MG&&
\qquad \longleftrightarrow  \qquad \KK \FTfour  \,.\cr
\eea
This implies the equivalence of two distinct-looking 3d $\CN=4$ KK-theories.  In particular, the equivalence must hold in the extreme infrared (IR) limit (that is, decoupling all the KK-modes, by focussing on the low-energy limit at fixed $\beta$), which must give us a 3d $\CN=4$ {\it superconformal} field theory. This IR SCFT will often have some useful description as the IR fixed point of an intrinsically 3d $\CN=4$ QFT, for instance as a 3d $\CN=4$ gauge theory. Such 3d QFT descriptions may be derivable in the string theory engineering \eqref{T duality picture}, or it may be more subtle (for instance requiring further string dualities)---in general, it may be a `non-Lagrangian' theory, meaning that no useful UV Lagrangian in known in 3d. We will call these 3d $\CN=4$ SQFTs the {\it `electric quiverine'} (EQ) theories of the 5d and 4d SCFT, respectively, which we denote by
\be
\EQfive[\MG] \stackrel{\rm 3d}{\cong}   D_{T^2} \FT~, \qquad \qquad
\EQfour[\MG] \stackrel{\rm 3d}{\cong}   \KK \FTfour~.
\ee
Here, the equivalence relation `$ \stackrel{\rm 3d}{\cong} $' means that the two theories are equivalent as 3d $\CN=4$ QFTs below the scale set by the lightest KK-modes. In many interesting cases, the quiverine has a description as a 3d $\CN=4$ gauge-theory quiver, while in general it might not be a quiver, or not have any known Lagrangian description at all.

The T-duality \eqref{T duality picture} realizes 3d $\CN=4$ mirror symmetry \cite{Intriligator:1996ex, Hori:1997zj}, although not without an interesting subtlety. In general, the theories $\FT$ and $\FTfour$ have a $U(1)^f$ symmetry acting on their Higgs branch (which is generally a subgroup of the full flavor symmetry $G_H^{\rm 5d}$ and $G_H^{\rm 4d}$, respectively). The M-theory and IIB realizations provide background gauge fields for this $U(1)^f$ in the Coulomb-branch phase, coming from the reduction of the 3-form gauge field $C_3$ on 2-cycles in M-theory, and from the RR field $C_4$ on 3-cycles in IIB, respectively. 
Upon toroidal compactification, these background abelian gauge fields  may be effectively gauged in the 3d $\CN=4$ description, depending on whether we are looking at the far IR limit of $D_{T^2} \FT$ or $\KK \FTfour$. Recall also that in 3d one may gauge and `ungauge' any $U(1)$ current at minimal cost, and in a reversible manner---this is known as the `$S$ operation' \cite{Kapustin:1999ha, Witten:2003ya}---, because each gauged $U(1)$ gives rise to a new topological current. Let us then define the following {\it `magnetic quiverine'} (MQ) theories:
\be\label{MQ54 gauging intro}
\MQfive[\MG] \cong \EQfour[\MG]\big/U(1)^f~, \qquad \qquad  
\MQfour[\MG] \cong \EQfive[\MG]\big/U(1)^f \,.
\ee
where the quotient denotes the $S$-type gauging of the $U(1)^f$ flavor symmetry. We then 
{propose} that the electric and magnetic {quiverines} of $\FT$ and $\FTfour$, respectively, are 3d mirror of each other
\footnote{Here and in the following, we often keep the dependence on the singularity $\MG$ implicit, to avoid clutter.}
\be
\MQfive \quad   \overset{\text{3d mirror}}{\Longleftrightarrow} \quad  \EQfive~, \qquad \qquad
\MQfour \quad   \overset{\text{3d mirror}}{\Longleftrightarrow} \quad  \EQfour~.
\ee
 When it is an ordinary gauge-theory quiver, $\MQfive$ coincides with the magnetic quiver for the 5d SCFT $\FT$ as studied {\it e.g.} in \cite{Ferlito:2017xdq, Hanany:2018uhm, Cabrera:2018jxt, Bourget:2019aer, Eckhard:2020jyr, vanBeest:2020kou}. Similarly, the magnetic {quiverine} $\MQfour$ of $\FTfour$ is usually known as the `3d mirror' of  $\FTfour$ \cite{Xie:2015rpa}.

The most important aspect of 3d mirror symmetry is that is exchanges the Higgs branch (HB) and the 3d Coulomb branch (CB). Recall that, while the 5d Coulomb branch of $\FT$ is a real cone and  the 4d Coulomb branch of $\FTfour$ is a complex (special K\"ahler) cone, the CB of a 3d $\CN=4$ SCFT is itself a hyper-K\"ahler cone. On the other hand, the Higgs branch is not affected by toroidal compactification, and we then expect:
\bea
&\CM_H[\FT]\;\quad  \cong&& \;  {\rm HB}[\EQfive] \quad \cong \quad  {\rm CB}[\MQfive]~,\cr
&\CM_H[\FTfour] \quad \cong&& \;  {\rm HB}[\EQfour] \quad \cong\quad   {\rm CB}[\MQfour]~.
\eea
Therefore, the quantum Higgs branches of $\FT$ and $\FTfour$ can be studied as quantum Coulomb branches of their magnetic quiverines. This may seem to exchange one hard problem with another, but it turns out that, by now, 3d $\CN=4$ Coulomb branches are rather well understood (see e.g. \cite{Cremonesi:2013lqa, Nakajima:2015txa, Bullimore:2015lsa, Braverman:2016wma}).
This type of magnetic-quiver construction was first advocated in~\cite{Ferlito:2017xdq}, and explored in numerous papers in recent years  \cite{Cabrera:2018ann,Cabrera:2018jxt,Cabrera:2019izd,Bourget:2019aer,Bourget:2019rtl,Cabrera:2019dob,Grimminger:2020dmg, Bourget:2020asf,Bourget:2020gzi, Eckhard:2020jyr}.

\begin{table}[t]
\centering 
\begin{tabular}{|c | c | c| c|} 
\hline\hline 
Singularity $\MG$ & SCFT $\FT$   & SCFT  $\FTfour$& Dimension  \\ [0.5ex] 
\hline 
Divisor class number $\rho(\MG)$ &$\leq$ rank$(G_H^{\rm 5d})$& $\leq$ rank$(G_H^{\rm 4d})$& $f$ \\
Exceptional divisors & CB  & reduced HB    & $r$ \\
K\"ahler cone & extended CB  & HB & $r+f= \h d_H$ \\
Strictly normalizable deformations &reduced HB & CB &$\h r$\\
Normalizable deformations & HB  & extended CB & $\h r +f= d_H$\\
\hline 
\end{tabular}
\caption{5d and 4d SCFTs from a canonical singularity $\MG$, at a glance. See section~\protect\ref{sec: 4d/5d}  for details.} 
\label{table:intro summary}
\end{table}
We summarize some of the quantities of interest in Table~\ref{table:intro summary}, which we will further explain in the next section. In particular, we denote by $r$ and $\h r$ the rank of $\FT$ and $\FTfour$, respectively, and by $d_H$ and $\h d_H$ the quaternionic dimension of their respective Higgs branches. We also have $f$ the rank of the $U(1)^f$ flavor symmetry preserved on the CB, which is generally a subgroup of the flavor symmetry $G_H^{\rm 5d}$ or $G_H^{\rm 4d}$ acting on the 5d or 4d SCFT Higgs branch. We have the key relations:
\be
d_H = \h r +f~, \qquad \qquad r + f = \h d_H~,
\ee
whenever $\MG$ is an isolated hypersurface singularity. (We will also briefly discuss the case of isolated toric singularities.)

Our analysis also clarifies subtle issues in the engineering of 5d SCFTs from isolated singularities. For instance, we will discuss the rank-$N$ $E_8$ 5d theory in detail, resolving some interesting puzzle at $N\geq 2$. Another interesting piece of information that one can extract from the geometry is the set of higher-form symmetries of $\FT$ or $\FTfour$,  following the approach of \cite{Garcia-Etxebarria:2019cnb, Morrison:2020ool, Albertini:2020mdx}.
 In particular,  we will compute the one-form symmetry of many generalized Argyres-Douglas theories $\FTfour$ directly from the singularity data.~\footnote{These one-form symmetries were also studied in \protect\cite{DelZotto:2020esg}, which appeared on the arXiv at the same time as the first version of this paper.}

\subsection{Comments on rank-0 theories from singularities}

The perspective we just sketched immediately leads to the following observations. 
First,  the M-theory engineering predicts the existence of an infinite class of rank-zero theories in 5d -- which may or may not be interacting SCFTs. They correspond to terminal singularities---canonical singularities that do not admit crepant resolutions with exceptional divisors (so that $r=0$). For many of these 5d rank-zero SCFTs $\FT$, the singularity $\MG$ engineers a (generalized) Argyres-Douglas (AD) theory $\FTfour$ in IIB \cite{Argyres:1995jj, Argyres:1995xn, Minahan:1996fg, Cecotti:2010fi, Xie:2012hs}. Embedding the same terminal singularity in an elliptic Calabi-Yau threefold, one can also construct the 6d uplift of these 5d rank-zero SCFTs. Their contributions to the 6d gravitational anomaly in a compact model was studied in \cite{Arras:2016evy,Grassi:2018rva}. However, it is unclear whether they give rise to non-trivial rank-0 theories in 6d either.

 Similarly, the IIB engineering predicts the existence of infinite families of rank-zero 4d $\CN=2$ theories, which may or may not be interacting, and correspond to singularities $\MG$ whose complex-structure deformation $\h\MG$ does not give rise to any dynamical $U(1)$ vector multiplet in 4d (so that $\h r=0$). This happens, in particular, for all isolated toric singularities; this point was first noted by \cite{Chen:2017wkw} for toric orbifolds. The same singularities engineer well-studied (and conventional, higher rank) 5d SCFTs in M-theory. 

To determine the properties of these rank-0 theories in 5d and 4d, in particular to clarify whether these should be viewed as genuine interacting SCFTs, or simply as discrete gaugings of free hypermultiplets, 
remains an interesting question. 
Discrete gauging in 4d was studied {\it e.g.} in \cite{Argyres:2016yzz, Aharony:2016kai, Tachikawa:2017gyf, Apruzzi:2020pmv};  for discrete gauging in 3d and 6d, see \cite{Hanany:2018vph,Hanany:2018cgo,Hanany:2018dvd}.
 We will study these theories, mostly in 5d, by determining their magnetic quivers (or quiverines) according to \eqref{MQ54 gauging intro}, and thereby their Higgs branch. In all rank-zero cases we studied, we find that the 5d Higgs branch is of the type $\mathbb{H}^n/\Gamma$, which hints at an interpretation in terms of discrete gauging. Either way, these theories provide an interesting new toolkit for constructions of 5d SCFTs in M-theory.

Another point of view is provided by the brane-webs and associated magnetic quivers and Hasse diagrams for the 5d Higgs branches. It was observed in \cite{Bourget:2019aer, Grimminger:2020dmg, Bourget:2020asf} that some Hasse diagrams contain rank-0 slices, corresponding to {\it e.g.} $O(1)$ theories with $N$ flavors. 
Since such theories clearly appear as sub-structures in the Higgs branch of 5d SCFTs, an understanding of their geometric origin is very well motivated. 

An additional tantalizing piece of evidence for the existence of these rank-zero theories in 5d comes about by looking at higher-rank theories $\FT$. For instance, we find a large number of rank-$1$ theories $\FT$, from canonical singularities with $r=1$, that can be viewed as an `ordinary' rank-$1$ `coupled' to a rank-$0$ SCFT. These new theories differ in their Higgs branches and in their global symmetries, in particular in their higher-form symmetries, from the `underlying' rank-$1$ and rank-$0$ theories. We will give one specific example of this in this paper; many more examples, of essentially any rank, will be discussed in \cite{Closset:2020afy, CSNWII}.

Our evidence in 4d is somewhat more limited, but points towards the same conclusion, that these rank-0 4d  $\CN=2$ theories, engineered at toric singularities \cite{Chen:2017wkw}, can be free hypermultiplets or discrete gauging thereof.

\medskip
\noindent
This paper is organised as follow. In section~\ref{sec: 4d/5d}, we set the stage by reviewing some basic features of the geometric engineering of 5d and 4d SCFTs from canonical singularities, and we explore the relation between the two constructions via T-duality. In section~\ref{sec: 5d4d examples}, we explore this 5d/4d correspondence in a few examples where both sides are well understood. In sections~\ref{sec: 5d rank0}, we study the 5d theories that correspond to Argyres-Douglas in 4d, and are generally rank-0 5d theories. Finally, in section~\ref{sec: 4d rank0}, we discuss some 4d theories engineered at toric singularities.


\section{5d/4d correspondence via 3d mirror symmetry}\label{sec: 4d/5d}
In this section, we first review some aspects of the geometric engineering of the 5d SCFT $\FT$ and of the 4d SCFT  $\FTfour$ from an isolated canonical singularity $\MG$. We then discuss how they are related by 3d mirror symmetry. Finally, we shall briefly explain how to read off various higher-form symmetries from the geometry. The formalism reviewed here will be explained in more depth in \cite{Closset:2021lwy, CSNWII}.

\subsection{5d SCFTs from M-theory}
By now, there is a large amount of evidence for the proposition that the low-energy limit of M-theory on a singularity $\MG$ defines for us a five-dimensional SCFT in the transverse directions, as in \eqref{FT 5d definition}. While the 5d superconformal fixed point is necessarily a strongly-coupled system \cite{Cordova:2016xhm, Chang:2018xmx},  it is an otherwise ordinary local unitary QFT.%
\footnote{One may think of this system, at least heuristically, as the result of decoupling 5d gravity, in a compactification of M-theory on some Calabi-Yau threefold, by taking some large volume limit. Here, our point of view is to consider the canonical singularity $\MG$ irrespective of any embedding in a compact CY$_3$, which may or may not exist.}  
A standard approach, utilising its M-theory definition, is to deform and/or (crepantly) resolve the singularity $\MG$ to obtain a smooth local (that is, non-compact) Calabi-Yau (CY) threefold. Then, standard methods (chiefly, the supergravity approximation and the study of BPS branes) become available. The parameter spaces of resolutions and of deformations are identified with the extended Coulomb branch  and with the Higgs branch of the 5d SCFT,  respectively
\bea
& \text{ECB}(\FT) & \qquad  \leftrightarrow\qquad  &\text{resolutions (extended K\"ahler cone):}  \qquad  &&\t\MG  \cr
& \text{HB}(\FT) & \qquad  \leftrightarrow\qquad  &\text{deformations (complex structure parameters):}  \quad && \h \MG
\eea
The geometric approach to $\FT$ has mostly focussed on its Coulomb branch (CB), whose low-energy dynamics is well-approximated by 11d supergravity \cite{Cadavid:1995bk}. Any smooth point of the (extended) CB is described by a complete crepant resolution $\pi : \t \MG \rightarrow \MG$, which contains an exceptional locus:
\be\label{exceptional locus}
\pi^{-1}(0)= {\bigcup_{a=1}^r S_a}~,
\ee
consisting of $r$ exceptional divisors intersecting along curves, assuming $r>0$.%
\footnote{The case $r=0$ instead corresponds to  `small resolutions', {\it i.e.} resolutions whose exceptional loci consist of curves only. In that case, we have real mass terms but no dynamical CB fields in 5d.}
 The integer $r$, which is independent of the choice of crepant resolution, is called the {\it rank} of the 5d SCFT. The effective field theory on the CB consists of $r$ massless photons (and their superpartners in 5d $\CN{=}1$ vector multiplets)  with an action fully determined by the classical geometry $\t \MG$ \cite{Intriligator:1997pq}. The low-energy BPS excitations are the M2-branes (electrically charged particles) and M5-branes (magnetically charged strings) wrapping 2- and 4-cycles, respectively, within the exceptional locus \eqref{exceptional locus}. We should also note that some of the K\"ahler parameters correspond to effective curves which are dual to non-compact divisors. These correspond to non-dynamical vector multiplets, or, equivalently, to mass terms for flavor symmetries. {We denote by $f$ the rank of the flavor symmetry group from these curves, which is a subgroup of the actual flavor symmetry group, $G_H^{\rm 5d}$, of $\FT$ (see section~\ref{subsec:flavor sym} for more details)}. The resolved singularity may also contain 3-cycles, but on the other hand it is always simply connected \cite{caibar1999minimal, Caibarb3}.
We then have:
\bea\label{resolution top}
&{\rm dim} \, H_1(\t \MG,\R) = 0~, \qquad && {\rm dim} \, H_3(\t \MG,\R) = b_3~,\cr
&{\rm dim} \, H_2(\t \MG,\R) = r +f~, \qquad && {\rm dim} \, H_4(\t \MG,\R) = r ~.
\eea
 The exceptional divisors and curves can be collapsed to zero volume by varying the K\"ahler parameters, thus recovering the UV SCFT $\FT$. At that point,  the origin of the 5d CB, mutually non-local particles and strings become massless, which is a strong indicator of the existence of a gapless phase \cite{Argyres:1995jj}. The 3-cycles in the crepant resolution $\t\MG$ also provide free hypermultiplets on the five-dimensional CB. This can correspond to rather interesting physics, as we will see in some examples below, and in  \cite{CSNWII}.

The SCFT $\FT$ may also have a Higgs branch (HB), parameterized by the VEVs of scalar operators charged under the $SU(2)_R$ superconformal R-symmetry. A generic point on the HB also spontaneously breaks the flavor symmetry group, $G_H^{\rm 5d}$. The HB of any SQFT is a hyper-K\"ahler manifold. In any SCFT, it should also be a hyper-K\"ahler {\it cone}. We denote by $\CM_H[\FT]$ (or $\CM_H^{\rm 5d}$) the HB of $\FT$, and by $d_H$ its quaternionic dimension
\be
{\rm dim}\, \CM_H[\FT]=\half {\rm dim}_\C\, \CM_H[\FT] = d_H~.
\ee
Upon deforming the singularity $\MG$ to a smooth local CY threefold $\h \MG$, we obtain a number of 3-cycles,  $S^3_l$, $l=1, \cdots, \mu$, which are topologically three-spheres. The low-energy hypermultiplets consist of the complex-structure moduli:
\be
t_l = \int_{S^3_l} \Omega_3~,
\ee
very schematically, paired with the 5d axions arising from the reduction of the M-theory 3-form gauge field. Not all such hypermultiplets are dynamical, however \cite{Gukov:1999ya}. The Higgs branch dimension $d_H$ is generally smaller than $\mu$. For $\MG$ an IHS, one finds that:
\be
d_H = \h r +f~,
\ee
where $\h r$ is the dimension of the mixed Hodge structure (MHS) group $H^{1,2}(\h \MG)$ on the vanishing cohomology of $\MG$. Indeed, for this class of singularities, we have \cite{arnold2012singularities, Xie:2015rpa}:
\be\label{MHS quantities}
\mu = 2 \h r + f~, \qquad\quad  f = {\rm dim} \, H^{2,2}(\h\MG)~, \qquad\quad  \h r = {\rm dim}\, H^{1,2}(\h\MG)= {\rm dim}\, H^{2,1}(\h\MG)~.
\ee
The complex structure parameters in $H^{1,2}(\h\MG)$ corresponds to `strictly normalizable' deformations%
\footnote{By which we mean, such that $\Delta >1$, with the scaling dimension $\Delta$ defined in \protect\eqref{spec Delta} below; normalizable deformations have $\Delta \geq 1$.} of $\MG$, while the `normalizable deformations' also include the parameters from $H^{2,2}(\h\MG)$.~\footnote{It may be worth pointing out that the $H^{2,2}$ MHS group does not correspond to 4-forms (it is not a Dolbeault cohomology group). For a pedagogical review of MHS in our context, we refer to \protect\cite{CSNWII}.}  Importantly, the integer $f$ defined as in \eqref{MHS quantities} coincide with the integer $f$ appearing in \eqref{resolution top}, as we explain further below.

Unlike the situation for the CB, 11d supergravity is not enough to compute the metric on the Higgs branch of $\FT$, due to the presence of M2-brane instantons wrapping the vanishing 3-cycles \cite{Ooguri:1996me}. Instead, one must rely on other approaches, using various string-theory dualities. A particularly fruitful approach, in recent years, has been the study of $(p,q)$-brane webs engineering $\FT$ \cite{Aharony:1997bh}, and the related concept of a ``magnetic quiver'' whose 3d Coulomb branch is identical to the hyper-K\"ahler cone of the HB, $\CM_H[\FT]$,  that we would like to understand \cite{Cabrera:2018ann,Cabrera:2018jxt, Bourget:2019rtl, Bourget:2019aer, Bourget:2020gzi, Eckhard:2020jyr,Bourget:2020xdz}. One interesting outcome of the present work is that it provides a derivation from the geometry of magnetic quivers for the SCFT $\FT$ defined as in \eqref{FT 5d definition}, at least in a number of cases  \cite{CSNWII}. We will see some examples of this in section~\ref{sec: 5d4d examples} below.

\subsection{4d SCFTs from Type IIB}
The low-energy limit of Type IIB string theory on a canonical singularity $\MG$ defines a 4d $\CN=2$ SCFT $\FTfour$, \eqref{FT 4d definition}, in the transverse directions \cite{Katz:1997eq, Shapere:1999xr, Cecotti:2010fi, DelZotto:2015rca, Xie:2015rpa, Chen:2016bzh, Wang:2015mra, Wang:2016yha}.  In this context, the complex structure deformations correspond to the (extended) Coulomb branch of the SCFT, while the K\"ahler cone of $\MG$ underlies the Higgs branch of the theory: 
\bea
& \text{ECB}(\FTfour) & \qquad  \leftrightarrow\qquad  &\text{deformations (complex structure parameters):}  \quad && \h \MG \cr
& \text{HB}(\FTfour) & \qquad  \leftrightarrow\qquad &\text{resolutions (K\"ahler parameters):}  \qquad  &&\t\MG 
\eea
Here, the beautiful structure of the 4d $\CN=2$ Coulomb branch, which must be a special K\"ahler cone, is encoded in the classical geometry of the deformed singularity $\MG$. By contrast, the 4d Higgs branch metric receives quantum corrections from D1- and D3-brane instantons wrapping the exceptional locus of $\t\MG$.

For concreteness, let us focus on the case of quasi-homogeneous isolated hypersurface singularities \eqref{def IHS}, as first studied in \cite{Shapere:1999xr}. The 4d SCFT $\FTfour$ must have a superconformal $R$-symmetry $U(1)_r$ which is spontaneously broken on the Coulomb branch. This corresponds to the condition that the defining  polynomial $F(x)$ be quasi-homogeneous
\be\label{scaling sym}
F(\lambda^{q_1} x_1, \lambda^{q_2} x_2, \lambda^{q_3} x_3, \lambda^{q_4} x_4) = \lambda\, F(x_1, x_2, x_3, x_4)~,
\ee
for some scaling weights $q_i \in \Q_{>0}$. The requirement that the singularity be canonical translates to the condition $\sum_{i=1}^4 q_i >1$. The deformed singularity, corresponding to the CB, takes the form:
\be
\h \MG \cong \bigg\{\h F(x) \equiv F(x) + \sum_{l=1}^\mu t_l \, x^{\m_l}=0 \bigg\}~,
\ee
where $x^{\m_l}$ denote the monomials generating the Milnor ring 
\be
\CM(F) = \C[x_1, x_2, x_3, x_4]/(dF)\,,
\ee 
and $\mu$ is the Milnor number of the singularity, where one has the relation 
\be
\mu= \prod_{i=1}^4(q_i^{-1}-1)\,.
\ee 
At generic values of the complex-structure parameters $t_l$, the deformed singularity is smooth and has the homotopy type of a `bouquet' of $\mu$ three-spheres. Let us order the deformations $t_l$ according to their weights $Q[t]$ under~\eqref{scaling sym}. They have conformal dimensions
\be\label{spec Delta}
\Delta[t_l] = {Q[t_l] \ov \sum_{i=1}^4 q_i -1}~.
\ee
One then identifies the deformation parameters with dimensions $\Delta >1$ as the CB `$u$-parameters'---the VEVs $u_l \equiv \langle \CO_l \rangle$ of the CB operators $\CO_l$---while the parameters with $\Delta<1$ are supersymmetric deformations of the SCFT (that is, deformations of the 4d theory by the F-terms $\int d^4\theta \CO_l$). We denote by $\h r$ the number of CB operators---that is, the rank of the SCFT, whose CB is freely generated. Each  {independent} CB operator is paired with a corresponding F-term deformation, with
\be
\Delta[t_l] + \Delta[t_{\mu-l+1}]= 2~, \qquad l =1, \cdots \h r~.
\ee
 The deformations with $\Delta=1$, on the other hand, are unpaired, and correspond to complex mass terms, which are VEVs for background vector multiplets for the maximal torus of the flavor symmetry group, $G_H^{\rm 4d}$, of $\FTfour$. Let us also mention that, by using the Shapere-Tachikawa relations \cite{Shapere:2008zf}, the conformal central charges $a$ and $c$ can be easily computed from the spectrum \eqref{spec Delta} \cite{Xie:2015rpa}. See also \cite{Martone:2020nsy} for some more recent developments.

The Seiberg-Witten geometry \cite{Seiberg:1994rs} of $\FTfour$ is entirely captured by the local Calabi-Yau $\h\MG$ fibered over the extended Coulomb branch (ECB), corresponding to the `normalizable parameters' $t_l$ such that $\Delta[t_l] \geq 1$. The low-energy physics is therefore determined by an $SL(\mu, \Z)$ bundle over the ECB, with the SW periods computed classically as
\be
a_l = \int_{A_l} \Omega_3~, \qquad a_{D, l}= \int_{B_l} \Omega_3~, \quad l=1, \cdots, \h r~,  \qquad m_\alpha = \int_L \Omega_3~,\quad \alpha=1, \cdots, f~,
\ee 
in some appropriate basis of $H_3(\h\MG, \Z)$; the holomorphic 3-form $\Omega_3$ of the non-compact CY threefold $\h\MG$ generalizes the SW differential $\lambda_{\rm SW}$ \cite{Katz:1997eq}. The non-trivial structure of the SW fibration can be studied using Picard-Lefschetz theory. In particular, one can straightforwardly compute the monodromy group of the 4d ECB from the singularity data. Let us also recall that the Coulomb-branch BPS particles are realized as D3-branes wrapped over the vanishing 3-cycles \cite{Shapere:1999xr}, and that the Milnor number $\mu= 2 \h r +f$ is the dimension of the electro-magnetic charge lattice, extended with the flavor charges.

\subsection{Crepant resolution and anomaly matching on the Higgs branch of $\FTfour$}\label{subsec: U1r anomaly}

The Higgs branch of $\FTfour$ is engineered in IIB as a torus fibration over the complexified K\"ahler cone of the singularity---see {\it e.g.} \cite{Lindstrom:1999pz, deWit:2001brd, Alexandrov:2010np}. While the HB metric is strongly affected by the D-brane instantons, we can nonetheless understand the basic features of the Higgs phase by studying the topology of a generic crepant resolution, $\t \MG$. The number of hypermultiplets, $\h d_H$, at a generic point on the HB, is given by the number of 2-cycles \eqref{resolution top}, $\h d_H=r+f$---that is, the number of K\"ahler parameters. The number of massless {\it vector multiplets} at such a Higgs-branch point, on the other hand, is given by $\half b_3$, half the number of 3-cycles in the crepant resolution (they come in pairs~\cite{caibar1999minimal}). Finally, there might remain some terminal singularities $\MG_{\rm IR}$ on the exceptional locus of the resolution $\t \MG$, which cannot be resolved further. Such singularities are interpreted as `irreducible' SCFTs, which live at every point on the Higgs branch. (This general structure of the Higgs branch was discussed from the VOA perspective in \cite{Beem:2019tfp}.) To summarize, we have the following Higgs-branch low-energy effective theory:
\be\label{HB LEET}
\CT_{\rm HB}[\FTfour] \cong (\h d_H \, {\rm hypers})\oplus (\half b_3 \; {\rm vectors}) \oplus \FTXfour{\MG_{\rm IR}} \quad \leftrightarrow \quad   (\text{2-cycles}) \oplus (\text{3-cycles})\oplus \MG_{\rm IR}~,
\ee
with the indicated geometric correspondence. The IR SCFT is a tensor product of distinct SCFTs whenever there are several distinct residual terminal singularities.  We can provide a strong check of this picture by matching the 't Hooft anomaly $\Tr(U(1)_r)$, since $U(1)_r$ is preserved on the Higgs branch. For the SCFT in the UV, the anomaly is simply related to the conformal anomalies \cite{Shapere:2008zf}:
\be
\CA_r[\FTfour] = 24(c-a)= n_h- n_v~,
\ee
where $n_h=-16 a+20 c$ and $n_v= 8a-4c$ are the `effective' number of hypermultiplet and vector multiplets, respectively, at the fixed point. Upon going on the Higgs branch, we have to match with the anomaly of the IR theory \eqref{HB LEET}:
\be
 \CA_r[\CT_{\rm HB}]  = \h d_H - \half b_3 + \CA_r[\FTXfour{\MG_{\rm IR}}]~.
\ee
Thus, we must have:
\be
24(c-a) =  \h d_H - \half b_3 + 24(c^{\rm IR}- a^{\rm IR})~,
\ee
where $a^{\rm IR}$, $c^{\rm IR}$ are the central charges of the IR SCFT $\FTXfour{\MG_{\rm IR}}$. (See \cite{Shimizu:2017kzs, Chang:2019uag} for related discussions.) This anomaly matching condition provides a strong check on the geometric engineering picture, since it looks quite miraculous in terms of the geometry, relating the deformation data (which encodes $a-c$) to the crepant resolution data. We checked this relation explicitly in many examples. It would be very interesting to prove it directly from the geometry. For future reference, let us also define the quantity
\be
\Delta \CA_r \equiv 24(c-a) - \h d_H = -\half b_3 + 24(c^{\rm IR}- a^{\rm IR})~,
\ee
which vanishes if the Higgs branch consists of free hypermultiplets only.~\footnote{The converse is not true, of course, since we can have the coincidence that $\Delta \CA_r=0$ because $b_3=  48(c^{\rm IR}- a^{\rm IR})$. On the other hand, $\Delta \CA_r \neq 0$ implies that the HB theory includes additional degrees of freedom besides the hypermultiplets, and $\Delta \CA_r$ non-integer implies that there is some irreducible IR SCFT.}

\subsection{Flavor symmetries, conifold transitions and boundary five-manifold}\label{subsec:flavor sym}
For $\MG$ a hypersurface singularity, we have seen that the integer denoted by $f$ appears in two distinct ways:
\bea\label{f two defs}
&f&=&\;\; \rho(\MG)  \cr
&&=&\; \; {\rm dim} \, H^{2,2}(\h \MG)~.
\eea
On the first line, $\rho(\MG)$ is the divisor class number---that is, the number of non-compact divisors, or, equivalently,  the number of compact 2-cycles $\CC_{\alpha}$, $\alpha = 1, \cdots, f$, to which they are dual, in any crepant resolution $\t \MG$. They give rise to background vector multiplets in the Coulomb phase of the 5d SCFT $\FT$, in the M-theory construction. 
Since every conserved current in $\FT$ gives rise to an $SU(2)_R$-preserving massive deformation of the 5d fixed point \cite{Cordova:2016xhm}, we would naively think that all such mass terms should be visible as K\"ahler parameters of the resolved singularity. This is however not the case, which is related to the possible presence of $b_3\in 2\Z$ 3-cycles in the resolved geometry. We claim that the flavor rank of $\FT$ is given by:
\be
 {\rm rank}(G_H^{\rm 5d})  = f  + n_{\rm ssb}~,  \qquad n_{\rm ssb} \leq \half b_3~.
\ee
Here, the interpretation of the 3-cycles in $\t\MG$ is that some operators in $\FT$ have been given a VEV, landing us in a `partial Higgs phase,' which spontaneously breaks a subgroup $U(1)^{n_{\rm ssb}}\subset G_H^{\rm 5d}$ of the flavor symmetry of $\FT$. The number $n_{\rm ssb}$ must be computed in a case-by-case basis, as it depends on the details of the resolution (we will see examples of this in the next section)---it corresponds to the number of K\"ahler parameters that can be gained by geometric transitions on these 3-cycles, which preserve the 5d Coulomb phase. 

On the second line of \eqref{f two defs}, $f$ is defined as the number of `unpaired' three-cycles in any generic deformation of the singularity. These are the compact 3-cycles $S^3_{\alpha} \subset \h \MG$, $\alpha=1, \cdots, f$, dual to non-compact 3-cycles, which give rise to background vector multiplets in the Coulomb phase of the 4d SCFT $\FTfour$, in the IIB construction, and  $f$ is then identified as the rank of the flavor symmetry of $\FTfour$. Since there cannot be any 2- and 4-cycles in the deformation $\h\MG$ of an IHS \cite{Minor1968}, we do not have any mixed Higgs phase on the CB of $\FT$, and therefore we expect that
\be
 {\rm rank}(G_H^{\rm 4d})=f~.
\ee 

Note that, even when $b_3=0$ and $G_H^{\rm 5d}$ and $G_H^{\rm 4d}$ have the same rank, the flavor groups  $G_H^{\rm 5d}$ and $G_H^{\rm 4d}$ themselves, which act as $SU(2)_R$-preserving isometries on the quantum Higgs branches $\CM_H[\FT]$ and $\CM_H[\FTfour]$, will in general not be the same. 

Geometrically, there are two related ways to understand the `coincidence' \eqref{f two defs}. The first one is that there exists a geometric transition that shrinks the curve $\CC_{\alpha} \subset \t \MG$ to zero size, before turning on the complex structure deformation that gives rise to the 3-cycle $S^3_\alpha$, or vice versa, generalizing the well-studied conifold transition \cite{Candelas:1989js, Hori:1997zj}:
\be
\CC_\alpha \qquad \longleftrightarrow \qquad S^3_\alpha~.
\ee
Such a transition, in general, has to go through the singularity $\MG$ at the common origin the K\"ahler and complex structure parameter spaces. Physically, in either the 5d or 4d interpretation, this transition ($\CC_\alpha \rightarrow S^3_\alpha$ in M-theory, or $S^3_\alpha \rightarrow \CC_\alpha$ in IIB) corresponds to sending a mass to zero, upon which some field charged under the flavor symmetry $U(1)_\alpha$ becomes massless, and can be then given a VEV that spontaneously breaks the symmetry. 

The second, related explanation is that both the 2-cycles $\CC_\alpha$ and the 3-cycles $S^3_\alpha$ sit in the image of the embedding map for the boundary into the threefold, namely
\be
\CC_\alpha \in {\rm im}\left[ H_2(\d \t \MG, \Z) \rightarrow H_2(\t \MG, \Z)\right]~, \qquad
S^3_\alpha \in {\rm im}\left[  H_3(\d \h \MG, \Z) \rightarrow H_3(\h \MG, \Z)\right]~.
\ee
For an isolated singularity, the act of resolving or deforming the singularity does not affect the boundary at infinity, and we have
\be
\d \t \MG \cong  \d \h \MG  \cong L_5(\MG)~.
\ee
Here, $L_5(\MG)$ is the {\it link} of the singularity $\MG$.%
\footnote{If the singularity admits a Ricci-flat metric, the five-dimensional link admits a Sasaki-Einstein metric (by definition). Here, we will not assume this, as there are generally obstructions to the existence of a SE metric---see  \protect\cite{Sparks:2010sn} for a nice review. The non-existence of SE$_5$ metrics, for many singularities of physical interest for geometric engineering, raises quite interesting questions, which however go beyond the scope of this paper.} 
For quasi-homogeneous isolated hypersurface singularities, $L_5(\MG)$ is simply-connected \cite{Minor1968}, so that its homology takes the general form:
\bea\label{homology L5}
&H_0(L_5, \Z) =\Z~, \qquad
&&H_1(L_5, \Z) =0~, \qquad
&&H_2(L_5, \Z) =\Z^f\oplus \torHtwo~, \cr
& H_3(L_5, \Z) =\Z^f~, \qquad
&&H_4(L_5, \Z) =0~, \qquad
&&H_5(L_5, \Z) =\Z~,
\eea
Both $f$ and the finite abelian group $\torHtwo$ can be computed directly from the weights $q_i$ \cite{RANDELL1975347, 10.1007/BFb0070047}. (See \cite{Closset:2021lwy} for a review of the relevant combinatorial formulas.) We will discuss the physical meaning of ${\rm Tor} \, H_2(L_5,\Z)= \torHtwo$ in subsection~\ref{subsec: higher form}.

\subsection{Circle reductions and the electric quiverines}

In both the M-theory engineering of $\FT$ and the Type IIB engineering of $\FTfour$, the extended Coulomb branch geometry arises semi-classically in string theory. On the one hand, the ECB  of $\FT$ is identified with the extended K\"ahler cone of the singularity, of real dimension $r+f$, parameterizing the crepant resolutions of $\MG$ (see {\it e.g.} \cite{Closset:2018bjz} for a more detailed review). On the other hand, the ECB of $\FTfour$ is identified with the versal family of complex structure deformations, with the special-K\"ahler  conical structure arising naturally from the vanishing 3-cycles.

By contrast, the Higgs branch in either case is much harder to study, since it receives M2- or D-brane instanton corrections. From the assumed superconformal invariance, we know that it must be a hyper-K\"ahler cone, and, from the geometric engineering, we know its quaternionic dimension:
\be\label{dim rels}
{\rm dim} \, \CM_H[\FT] = d_H = \h r+f~, \qquad \qquad 
{\rm dim} \, \CM_H[\FTfour] = \h d_H = r+f~, 
\ee
in 5d and 4d, respectively. Note that, in this paper, we are always talking about the `quantum Higgs branch' of the SCFT \cite{Argyres:2012fu}, which generally does not have any Lagrangian description---in particular, the Higgs branch may not be realizable as a hyper-K\"ahler quotient. Of course, one could also consider mixed branches, which correspond to partially resolving then deforming the singularity, or vice versa.

It is clear, from the discussion so far, that the Higgs branch data and the Coulomb branch data of the 5d and 4d SCFTs are related to each other. In particular, we see from \eqref{dim rels} that the quaternionic Higgs branch dimension in  5d, $d_H$, is equal to the complex dimension of the extended Coulomb branch of $\FTfour$. In this subsection and the next, we spell out this relation in more detail. 

Let us first consider compactifying the theory $\FTfour$ on a finite-size circle. This gives rise to a three-dimensional $\CN=4$ supersymmetric field theory, whose Coulomb branch is now hyper-K\"ahler---each 4d $\CN=2$ abelian vector multiplet gives us a 3d $\CN=4$ vector multiplet, which contains three real scalars plus a (dual) photon, and the hyper-K\"ahler structure is dictated by 3d $\CN=4$ supersymmetry. Note that this happens as soon as we compactify the theory on $S^1$, at any radius \cite{Seiberg:1996nz, Gaiotto:2008cd}. The 3d $\CN=4$ CB receives quantum corrections which dramatically correct its metric---say, compared to a semi-classical description for a 4d Lagrangian theory on $\R^3 \times S^1$. In string theory, these corrections arise from D3-branes instantons wrapping the circle times any vanishing 3-cycle in $\h\MG$. Let $\KK \FTfour$ denote the KK 3d $\CN=4$ theory obtained by circle compactification. In the far IR (compared to the KK scale), we obtain a new 3d $\CN=4$ SCFT, which we denote by:
\be\label{EQ4 def}
\EQfour[\MG] \,   \stackrel{\rm 3d}{\cong} \, \KK \FTfour~.
\ee
More precisely, we should think of $\EQfour[\MG]$ (also written as $\EQfour$, with the dependence on the choice of singularity $\MG$ implicit) as a three-dimensional SQFT which coincides with $\KK\FTfour$ below the KK scale---for $\FTfour$ given by a superconformal Lagrangian, it would simply be the dimensionally-reduced Lagrangian theory; for non-Lagrangian theories (the generic case), we need to work harder. The 3d  $\CN=4$ SQFT $\EQfour$ is called the `electric quiver(ine)' of the 4d SCFT $\FTfour$. Its IR fixed point is the 3d $\CN=4$ SCFT we are interested in. It has a scale-invariant Coulomb branch, which must be a hyper-K\"ahler cone of quaternionic dimension
\be\label{dim CB EQ4}
{\rm dim} \, {\rm CB}[\EQfour[\MG]]= \h r~,
\ee
while its Higgs branch, of dimension $\h d_H$, coincides with the Higgs branch of $\FTfour$. Note that the $SU(2)_C$ $R$-symmetry acting on CB$[\EQfour]$ arises as an accidental IR symmetry, after non-trivial mixing between the $U(1)_r$ symmetry of $\FTfour$ and 3d topological symmetries that are present along the RG flow \cite{Buican:2015hsa}. 

Similarly, we may consider the 5d SCFT $\FT$ on a finite-size torus, which defines for us a KK field theory denoted by $D_{T^2}\FT$. This gives us another 3d $\CN=4$ SCFT in the deep IR, and we denote the intermediate 3d $\CN=4$ SQFT by
\be\label{EQ5 def}
\EQfive[\MG] \,   \stackrel{\rm 3d}{\cong} \, D_{T^2} \FT~.
\ee
Since $\FT$ is never given by a superconformal Lagrangian, the `electric quiver' $\EQfive$ is not something we can read off easily. (Note, in particular, that this is {\it not} the dimensional reduction of some 5d IR gauge-theory description which may arise on the ECB of the 5d SCFT.) By definition, the electric quiver of $\FT$ has a Coulomb branch of dimension
\be
{\rm dim} \, {\rm CB}[\EQfive[\MG]]= r~,
\ee
while its Higgs branch, of dimension $d_H$, coincides with the Higgs branch of $\FT$.

As anticipated in the introduction, the $\FT$ on a torus and $\FTfour$ are related by T-duality in Type II string theory, as suggested by our construction
\be
D_{T_2} \FT
\quad \leftrightarrow \quad   \text{IIA on} \; \R^3 \times S^1_{ \beta} \times \MG 
 \quad \leftrightarrow \quad   \text{IIB on} \; \R^3 \times S^1_{1\ov \beta} \times \MG 
\quad \leftrightarrow  \quad \KK \FTfour~,
\ee
and it is known that T-duality realizes 3d $\CN=4$ mirror symmetry as a perturbative string-theory symmetry \cite{Hori:1997zj} (we can keep $g_s$ very small in this whole discussion, by taking the M-theory circle to be small). Thus, we expect the two 3d $\CN=4$ effective descriptions \eqref{EQ4 def} and \eqref{EQ5 def} to be related by mirror symmetry. This, however, cannot be exactly true in our setup, since the dimension of the CB of $\EQfour$, \eqref{dim CB EQ4}, does not match the dimension of the HB of $\EQfive$ ($d_H= \h r+f$) unless $f=0$, and vice versa. The discrepancy comes from the $f$ $U(1)$ vectors in either description (4d in IIB, or 5d in M-theory), which we discussed in subsection~\ref{subsec:flavor sym}.

In the following, we give a natural {\it prescription} to explicitly relate the SQFT descriptions of the 4d and 5d SCFTs compactified to 3d. A more detailed explanation of our approach will be given in \cite{CSNWII}.

\subsection{Higgs branches, magnetic quiverines and 3d mirrors}\label{subsec: HB and 3d mirror}
Let us first {\it define} the magnetic quiverine of the 4d SCFT $\FTfour$ as the 3d $\CN=4$ mirror dual to its electric quiver:%
\footnote{Recall that we use the term `quiverine' for the 3d theories in general, which may not have a description as a standard quiver gauge theory.}
\be\label{MQ4 mirror}
\MQfour \qquad   \overset{\text{3d mirror}}{\Longleftrightarrow} \qquad  \EQfour~.
\ee
This magnetic quiverine---by the definition, the 3d mirror description of the 3d SCFT obtained by compactifying the 4d SCFT on a circle---is more often called the {\it 3d mirror} of $\FTfour$. In recent years, 3d mirror theories have been obtained for many strongly-coupled theories, in particular for many Argyres-Douglas theories \cite{Benini:2009gi, Benini:2010uu, Xie:2012hs, Dey:2020hfe}. The interesting point is that $\MQfour$ may have an explicit Lagrangian description in 3d, even if $\EQfour$ does not.

Similarly, we define the magnetic quiverine of the 5d SCFT $\FT$ as the 3d mirror
\be\label{MQ5 mirror}
\MQfive \qquad   \overset{\text{3d mirror}}{\Longleftrightarrow} \qquad  \EQfive~.
\ee
In the recent literature, the {\it magnetic quiver} ($\MQ$) of any SQFT with eight Poincar\'e supercharges, $\CT$, has been defined as a useful `auxiliary' construction (generally constructed from a brane-web diagram), from which the Higgs branch of $\CT$ can be computed as a 3d $\CN=4$ Coulomb branch
\be
\CM_H[\CT]  = {\rm CB}[\MQ]~.
\ee
The geometric-engineering approach gives us a fully geometric realization of this perspective\cite{Cabrera:2018jxt}, at least abstractly. Indeed, we really ought to define the magnetic quiver(ine)s directly from the string `compactification', as follows.

Let us first consider the 5d SCFT $\FT$ on $T^2 \cong S^1_M \times S^1_\beta$, as in \eqref{EQ5 def}. It is well appreciated that abelian flavor symmetries in 3d can be gauged and `ungauged' at minimal cost~\cite{Kapustin:1999ha, Witten:2003ya}. The theory $\EQfive$ has a $U(1)^f$ flavor symmetry,%
\footnote{Which may enhance to a larger, non-abelian flavor symmetry, in which case we are only considering the maximal torus.} 
which acts on its Higgs branch as an isometry commuting with the $SU(2)_R$ action. We may then {\it gauge} this symmetry, to obtain a new theory, denoted by:
\be\label{gauging EQ5}
\MQfour \equiv \EQfive\big/U(1)^f~.
\ee
This gauging operation takes a theory that depends on background multiplets $V_{(F)}$ for $U(1)^f$ and gives us another theory with a new topological symmetry $U(1)_T^f$ (one topological current for each gauged $U(1)$) coupled to background vector multiplets $V_{(T)}$. Schematically, we have:
\be
S_{\MQfour}[V_{(T)}]  = -\log \int [D V_{(F)}] \, \exp\left(- S_{\MQfive}[V_{(F)}]-{i\ov 2 \pi} \int A_{(T)} \wedge d A_{(F)} +\cdots \right)~,
\ee
where $A$ denotes the gauge fields in vector multiplets, and the ellipsis denotes the supersymmetric completion (including Fayet-Iliopoulos terms). In particular, this so-called `$S$ operation'  sends an SCFT to another SCFT \cite{Witten:2003ya}. The operation is reversible; if we now gauge the new topological symmetries of $\MQfour$, we obtain the original theory:%
\footnote{More precisely, we obtain the original electric quiver theory coupled to vector multiplets of opposite sign.}
\be
\EQfive \equiv \MQfour\big/U(1)_{T}^f~.
\ee
Note also that the gauging operation \eqref{gauging EQ5} acts on the Higgs branch of the theory as a hyper-K\"ahler quotient%
\footnote{This is what we called `reduced HB' in Table~\protect\ref{table:intro summary}.}
\be
{\rm HB}[\MQfour] =  \CM_H[\FT]\,\big/\!\big/\!\big/ \, U(1)^f~,
\ee
while it increases the dimension of the Coulomb branch by $f$ (roughly speaking, the ECB becomes the CB). In particular, we have:
\be
{\rm dim} \, {\rm HB}[\MQfour]= d_H- f = \h r~, \qquad {\rm dim} \, {\rm CB}[\MQfour]= r + f= \h d_H~.
\ee
This is perfectly consistent with the 3d mirror symmetry \eqref{MQ4 mirror}. 

By the same token, we may define the magnetic quiver of the $\FT$ as the gauging of the electric quiver for $\FTfour$:
\be\label{gauging EQ4}
\MQfive \equiv \EQfour \big/U(1)^f~.
\ee
Here, the $U(1)^f$ flavor symmetry acts as an isometry of the Higgs branch $\CM_H[\FT]$, and we have:
\be
{\rm dim} \, {\rm HB}[\MQfive]= r~, \qquad {\rm dim} \, {\rm CB}[\MQfive]=d_H~,
\ee
in agreement with \eqref{MQ5 mirror}. These relations are simple tools to study the Higgs branch of either $\FT$ or $\FTfour$. They also relate many previously-known results to each other. In the following sections, we will demonstrate this in a few concrete examples.

\subsection{Higher-form symmetries of SCFTs}
\label{subsec: higher form}

The higher-form symmetries \cite{Gaiotto:2014kfa} of the SCFT $\FT$ or $\FTfour$, defined from the singularity $\MG$, can be studied as in \cite{Garcia-Etxebarria:2019cnb, Albertini:2020mdx, Morrison:2020ool}. Here, we briefly discuss the case of  $\MG$ an isolated hypersurface singularity, and refer to \cite{Closset:2021lwy, CSNWII} for a more detailed discussion. 

The data of any QFT generally involves a choice of `global structure'---for instance, for gauge theories, we need to choose a gauge group $G$, as opposed to only choosing a Lie algebra ${\rm Lie}(G)$; this choice translates into distinct spectra of Wilson lines and of $(d{-}3)$-dimensional 't Hooft operators \cite{Aharony:2013hda}. In the geometric-engineering picture of the SCFT, the choice of global structure is related to a choice of consistent boundary conditions for the torsion fluxes in the non-compact threefold \cite{Freed:2006yc, Garcia-Etxebarria:2019cnb}. 

A time-honored general strategy to study the SCFTs is through the analysis of its moduli space of vacua, which corresponds to resolving and/or deforming the singularity $\MG$. For instance, one can see the `enhanced' flavor symmetry $G_H$ of the fixed point as the hyper-K\"ahler isometries of the conical Higgs branch. Similarly, one-form symmetries, or their magnetic $(d{-}3)$-form version for an SCFT$_d$, can be studied on the Coulomb branch, where they are spontaneously broken \cite{Gaiotto:2014kfa}. We can also have $(d{-}2)$-form symmetries and their `electro-magnetic' dual $0$-form symmetries, which can be spontaneously broken on the Higgs branch.

In practice, one analyses the $q$-form symmetries by looking for $q$-dimensional charged operators that can obtain VEVs, in either the Higgs phase or the Coulomb phase. These operators are realized in string-theory as branes wrapping a relative 3-cycle of $\h \MG$ which ends on a torsion 2-cycle of the boundary, $\d\h\MG$. We can then compute the higher-form symmetry \cite{Garcia-Etxebarria:2019cnb, Albertini:2020mdx, Morrison:2020ool} of the SCFT from the following relative homology group
\be\label{defect group X}
{\rm Tor} \, \left[{\rm im}\, : \; H_3(\h \MG, \d \h \MG, \Z) \rightarrow H_2(\h \MG, \Z)\right] \cong {\rm Tor} \, H_2(L_5(\MG), \Z) \equiv \torHtwo~.
\ee
A choice of global structure corresponds to choosing a consistent half-dimensional sub-lattice of  mutually-commuting fluxes. Note that, here, we always have:
\be\label{h2 ff}
 \torHtwo \cong \frak{f} \oplus \frak{f}~,
\ee
and that we {will choose the simplest sub-lattices $(\frak{f},0) \subset  \torHtwo$ or $(0,\frak{f}) \subset  \torHtwo$, for simplicity, leaving a systematic analysis of the space of global structures for future work.} We then have the following (non-exhaustive) list of possibilities, also summarized in Table~\ref{table:qforms}:
\begin{table}[t]
\centering 
\begin{tabular}{|c | c | c| c|} 
\hline\hline 
SCFT$[\MG]$ &  `electric' sym.\, /\, charged object  &`magnetic' sym.\, /\, charged object   \\ [0.5ex] 
\hline 
$\FT$ & $\Gamma_e^{(0)}$    \, /\, M2-brane & $\Gamma_m^{(3)}$    \, /\, M5-brane  \\
$\FTfour$ & $\Gamma_e^{(1)}$    \, /\, D3-brane & $\Gamma_m^{(1)}$    \, /\, D3-brane  \\
\hline 
\end{tabular}
\caption{The group $\frak{f} \subset H_2(L_5(\MG), \Z)$, interpreted as a $q$-form symmetry group $\Gamma^{(q)}$.} 
\label{table:qforms}
\end{table}

\paragraph{Three-form/zero-form symmetries of $\FT$.}  For the 5d SCFT $\FT$, the defect group \eqref{defect group X} corresponds to either M5-branes or M2-branes wrapping non-compact 3-cycles ending on a 2-cycle in $\torHtwo$. If we only allow for M5-branes in the spectrum, we have a 5d theory with a $3$-form symmetry, while if we only allow M2-branes, we have a $0$-form symmetry. Thus we have:
\be
\FT\; : \; \qquad\quad  \Gamma^{(3)}_m=\torHtwo \qquad \qquad {\rm or}\qquad \qquad
\Gamma^{(0)}_e=\torHtwo~,
\ee
for the `magnetic' or `electric' symmetry, respectively. 
These defect operators could be constructed in various ways in the field theory---similar codimension-2 defects in 5d theories were constructed in \cite{Gaiotto:2015una}. We leave their analysis for future work. 

Note that the SCFT $\FT$ arising from an hypersurface singularity does not have any one-form (or 2-form) symmetry \cite{Morrison:2020ool, Albertini:2020mdx}, since the boundary $L_5(\MG)$ is simply-connected. This is contrast with the case of toric singularities, which we will briefly discuss in section~\ref{sec: 4d rank0}.

\paragraph{One-form symmetries of $\FTfour$.} Similarly, in type-IIB string theory, we can wrap D3-branes on non-compact 3-cycles in $\h\MG$. They correspond to line operators probing the Coulomb phase. We then have the one-form symmetry:
\be
\FTfour\; : \; \qquad\quad  \Gamma^{(1)}_m= \frak{f}\qquad \qquad {\rm or}\qquad \qquad
\Gamma^{(1)}_e= \frak{f}~.
\ee
If $\FTfour$ is given by a superconformal Lagrangian, the global structure should simply correspond to the choice of gauge group $G$---in particular, we will have a one-form symmetry if a discrete subgroup of the center, $ \frak{f}\subset Z(G)$, acts trivially on the matter fields. We will see examples of this below. For non-Lagrangian theories, on the other hand, such as generalized Argyres-Douglas (AD) theories, the analysis of the one-form symmetries will necessarily be more subtle. In fact, the string-theory engineering makes very specific {\it predictions} for the one-form symmetries of some AD theories, as we will discuss in section~\ref{subsec:oneform AD} below.



\section{5d SCFTs with Lagrangian 4d SCFT counterparts}
\label{sec: 5d4d examples}

In this section, we illustrate the general discussion of the previous section with examples, where both sides of the 5d/4d correspondence have a Lagrangian description. We start by discussing the $E_8$ rank-one 5d SCFT \cite{Seiberg:1996bd}, and its rank-$N$ generalizations \cite{Intriligator:1997pq}, which already brings about interesting subtleties. 
We furthermore discuss a 5d SCFT, which has an IR description in terms of a quiver theory, as well as the SCFT associated to $G_2 + 5 \bm{F}$. Each example illustrates different aspects.

\subsection{The rank-$N$ $E_8$ theory}\label{subsec:rankNE8}
\begin{table}[t]
\centering 
\begin{tabular}{ | c ||c |c | c|| c|c| c|| c| c|c|} 
\hline\hline 
 $F$& $r$ &$f$   &$d_H$ &$\h r$& $\h d_H$  & $\Delta \CA_r$  & $b_3$ & $\frak{f}$  \\ [0.5ex] 
\hline 
$x_1^2+x_2^3+x_3^6+x_4^{6N}$ & $N$ & $6$ & $12N-1$ & $12N-7$ & $N+6$ &$-N+1$&  $2N-2$&{$\Z_N$} \\
$x_1^2+x_2^4+x_3^4+x_4^{4N}$ & $N$ & $7$ & $18N-1$ & $18N-8$ & $N+7$ &$-N+1$&  $2N-2$&{$\Z_N$} \\
$x_1^2+x_2^3+x_3^6+x_4^{6N}$ & $N$ & $8$ & $30N-1$ & $30N-9$ & $N+8$ &$-N+1$&  $2N-2$&{$\Z_N$} \\
\hline 
\end{tabular}
\caption{Isolated hypersurface singularities engineering the rank-$N$ $E_n$ theories, for $n=6,7, 8$.} 
\label{table:En theories}
\end{table}
The famous rank-one $E_n$ SCFT  \cite{Seiberg:1996bd}, for $n\leq 8$, can be engineered in M-theory on a canonical singularity, $\MG_{E_n}$, whose crepant resolution is the canonical line bundle over the del Pezzo surface $dP_{n} \cong {\rm Bl}_{n-1} \mathbb{P}^1{\times}\mathbb{P}^1$ \cite{Morrison:1996xf}. 
This resolution probes the extended Coulomb branch of the SCFT $\FTX{\MG_{E_n}}$, whose low-energy description is equivalent to the extended Coulomb branch of the 5d $\CN=1$ $SU(2)$ gauge theory with $N_f=n-1$ fundamental hypermultipets. These 5d fixed points admit an interesting rank-$N$ generalization, with a gauge-theory phase:
\be\label{rank N En gauge theory Spn}
Sp(N) + (n{-}1)\bm{F}+ \bm{AS}~.
\ee
For $N=1$, this reduces to the $Sp(1)\cong SU(2){+} (n-1) \bm{F}$ gauge theory. The Higgs branch of the UV fixed point is expected to be the moduli space of $N$ $E_n$ instantons in $\C^2$. 
For $n \geq 6$, these 5d SCFTs can be engineered in M-theory on the following isolated hypersurface singularities:
\bea\label{En sings}
&\MG_{E_6} \; : \; &&x_1^3+ x_2^3+ x_3^3+ x_4^{3N}=0~,\\
&\MG_{E_7} \; : \;&&x_1^2+ x_2^4+ x_3^4+ x_4^{4N}=0~,\\
&\MG_{E_8} \; : \;&&x_1^2+ x_2^3+ x_3^6+ x_4^{6N}=0~,
\eea
up to some interesting subtleties for $N>1$, which we will discuss below. For $N=1$, in particular, each of these singularities admits a crepant resolution obtained by blowing up the origin of $\C^4$, $x_1= x_2 = x_3 =x_4 =0$, by projective spaces with weights $(1,1,1,1)$, $(2,1,1,1)$ or $(3,2,1,1)$, respectively. The exceptional divisor is then $dP_n$ viewed as a hypersurface in (weighted) projective space.

Some of the basic properties of the singularities \eqref{En sings} are summarized in Table~\ref{table:En theories}, note that   $b_3$ and $H_2(L_5,\Z)$ were also computed in \cite{vanCoevering:2009zz}. In this section, we focus on the $E_8$ theories, for definiteness. The $E_6$ and $E_7$ families can be treated in exactly the same manner.

\subsubsection{Singularity spectrum and 4d Superconformal Quiver}\label{sec:rankN-E8:sing}
Consider the singularity $\CY_{E_8}$
\be
\label{eq-E8-rankN}
F(x)= x_1^2+x_2^3+x_3^6+x_4^{6N}=0~, \qquad (q_1, q_2, q_3, q_4) = \left(\half, {1\ov 3}, {1\ov 6}, {1\ov 6N}\right)~,
\ee
with the scaling weights $q_i$ as indicated. The Milnor ring,  $\CM(F)$, of the corresponding isolated singularity, $\MG= \{F=0\}$, is easily determined. One finds a non-trivial mixed Hodge structure on the level set of the singularity, with the dimensions \eqref{MHS quantities} given by:
\be
 f = {\rm dim} \, H^{2,2}(\h\MG) = 8~, \qquad\quad  \h r = {\rm dim}\, H^{1,2}(\h\MG)= 30N-9~,
\ee
while the Milnor number is given by $\mu=2\h r +f=60N-10$.
In particular, $H^{2,2}(\h\MG)$ is generated by the $8$ deformations of $\MG$ corresponding to the monomial in $x^\m\in \CM(F)$ with scaling dimension $Q=1-\frac{1}{6N}$, namely:
\bea
&x_3 x_4^{5N-1}~,\,\;
 &&x_3^2 x_4^{4N-1}~,\,\;
  &&x_3^3 x_4^{3N-1}~,\,\;
  &&x_3^4 x_4^{2N-1}~,\,\\
 & x_2 x_4^{4N-1}~,\,\;
 && x_2 x_3 x_4^{3N-1}~,\,\;
 && x_2 x_3^2 x_4^{2N-1}~,\,\;
 && x_2 x_3^3 x_4^{N-1}~.
\eea
In the IIB construction, they correspond to mass terms---CB operators of $\FTfour$ with scaling dimension $\Delta=1$. On general ground, as discussed in the previous section, $f$ is also the number of `unpaired' $2$-cycles in the resolution $\t \MG$ of the singularity, so that we also have $f=8$ 5d $\CN{=}1$ real mass deformations (sitting in background vector multiplets) apparent on the ECB of $\FT$. For $N>1$, there is an apparent mismatch with the number of real masses expected from the gauge-theory description \eqref{rank N En gauge theory Spn} 
\be\label{rank N E8 gauge theory Spn}
Sp(N) + 7\bm{F}+ \bm{AS}~,
\ee
from which one would expect $9$ mass deformations ($7$ from the fundamentals, $1$ from the antisymmetric, and $1$ gauge coupling). Correspondingly, the 5d global symmetry at the UV fixed point of \eqref{rank N E8 gauge theory Spn} is expected to be $E_8$ for $N=1$, and $E_8 \times SU(2)$ for $N>1$. We will explain the reason for this discrepancy (when $N>1$) in subsection~\ref{sec:rankN-E8-res} below.

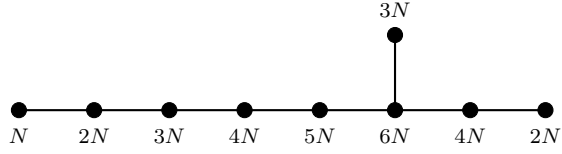
\begin{figure}
\begin{center}
$$
   \begin{tikzpicture}[x=1cm,y=1cm]
\draw[ligne, black](0,0)--(7,0);
\node[bd] at (0,0) [label=below:{{\scriptsize$N$}}] {};
\node[bd] at (1,0) [label=below:{{\scriptsize$2N$}}] {};
\node[bd] at (2,0) [label=below:{{\scriptsize$3N$}}] {};
\node[bd] at (3,0) [label=below:{{\scriptsize$4N$}}] {};
\node[bd] at (4,0) [label=below:{{\scriptsize$5N$}}] {};
\node[bd] at (5,0) [label=below:{{\scriptsize$6N$}}] {};
\draw[ligne, black](5,0)--(5,1);
\node[bd] at (5,1) [label=above:{{\scriptsize$3N$}}] {};
\node[bd] at (6,0) [label=below:{{\scriptsize$4N$}}] {};
\node[bd] at (7,0) [label=below:{{\scriptsize$2N$}}] {};
\end{tikzpicture}
$$
\caption{4d $SU$-type $\CN=2$ gauge-theory quiver for the rank-N $E_8$ theory. The same quiver graph describes  the `electric quiver' $\EQfour$, seen as a 3d $\CN=4$ theory. \label{f:rankNE8-quiver}}
\end{center}
\end{figure}

Let us first consider the 4d SCFT $\FTfour$ engineered from the singularity \eqref{eq-E8-rankN} in IIB. In this case, as first shown in \cite{Katz:1997eq}, the 4d SCFT has a Lagrangian description as a superconformal gauge theory with gauge group:
\be\label{G E8 quiver gen N}
G= \prod_{d_k=\text{Dynkin label of }\widehat{E}_8} SU(d_k N)~, 
\ee
where $d_k$ are the Dynkin labels of the affine $E_8$ algebra. This gauge group is coupled to hypermultiplets in bifundamental representations, giving us the $\h E_8$ gauge-theory quiver shown in Figure~\ref{f:rankNE8-quiver}. This result can be derived from the integer-valued singularity spectrum $\{ \Delta\}$, as in \cite{Wang:2016yha}. For instance, for $N=1$, we have the following scaling dimensions of CB operators ($\Delta \geq 1$): 
\be
\begin{array}{c||c|c|c|c|c|c}
\Delta & 1\; & 2\; & 3\; & 4\; & 5\; & 6\;\\
\hline
\# & 8 & 8 & 6 & 4 & 2 & 1 \\
\end{array}
\ee
with their multiplicities indicated on the second line. Since an $SU(K)$ gauge group contributes CB operators of dimension $\Delta=\{2, 3, \cdots, K\}$, this spectrum is compatible with:
\be
G= SU(6)\times SU(5) \times SU(4)^2 \times SU(3)^2 \times SU(2)^2~, \qquad \text{for} \; N=1.
\ee
This reasoning generalizes to $N>1$, giving us \eqref{G E8 quiver gen N}. There is then a unique way to couple all these gauge groups together with hypermultiplets to obtain an SCFT \cite{Bhardwaj:2013qia}.
As a consistency check, we can compute the conformal anomalies of $\FTfour$ directly from the singularity spectrum, to obtain:
\be
n_h=-16 a+20 c=120 N^2~, \qquad \qquad n_v= 8a-4c=120N^2-9~.
\ee
This exactly matches the number of hypermultiplets and vector multiplets, respectively, in the 4d $\CN=2$ gauge-theory quiver of Fig.~\ref{f:rankNE8-quiver}.

Since $\FTfour$ is a Lagrangian theory, its Higgs branch is computed semi-classically, as a hyper-K\"ahler quotient
\be
\CM_H^{\rm 4d} \cong \C^{240N^2} \,\big/\!\big/\!\big/ \, G~.
\ee
For $N>1$, however, there is an interesting subtlety: a subgroup $U(1)^{N-1} \subset G$ survives as free vector multiplets at low-energy.~\footnote{This is rather familiar in a different context: this  $\h E_8$  4d $\CN=2$ quiver also arises as the SCFT at low-energy on the worldvolume of $N$ D3-branes probing the (resolved) orbifold of $\C^2$ by the `$E_8$' finite subgroup of $SU(2)$, $Y= \C^2/\Gamma_{E_8}$ \protect\cite{Douglas:1996sw}. In that language, the Higgs branch of the $\CN=2$ quiver corresponds to the $8$ resolution parameters of the orbifold singularity (the so-called baryonic operators), $\pi : \t Y \rightarrow Y$, and to the $N$ positions of the D3-branes on $\t Y$ (the so-called mesonic operators). The low-energy $U(1)$ vector multiplets are the degrees of the freedom of the $N$ separated probe D3-branes (modulo the center-of-mass $U(1)$, which is decoupled).} The Higgs branch quaternionic dimension is then:
\be
\h d_H = n_h -n_v+ N-1 = 8+N~.
\ee
The $U(1)_r$ 't Hooft anomaly of course match:
\be
24(c-a) = \h d_H - (N-1)~.
\ee
According to the general discussion in section~\ref{subsec: U1r anomaly}, we should expect $2(N-1)$ 3-cycles in the {\it resolved} geometry $\t \MG$. This is indeed the case, as we will see momentarily.  The 4d $\CN=2$ theory of Fig.~\ref{f:rankNE8-quiver} also has $\Z_N$ one-form symmetry, in agreement with the prediction from the geometry.

\subsubsection{Resolutions: 5d Coulomb branch and 4d Higgs branch}
\label{sec:rankN-E8-res}

From the singular equation \eqref{eq-E8-rankN} for $\MG$, we can explicitly construct the crepant resolution $\pi : \t \MG\rightarrow \MG$. Using the notation of \cite{Lawrie:2012gg,Apruzzi:2019opn}, the resolution is obtained as a sequence of weighted blow-ups (taking the proper transform) in the ambient space:
\be
\ba
&(x_1^{(3)}~,\,x_2^{(2)}~,\,x_3^{(1)}~,\, x_4^{(1)}~;\, \delta_1)~,\cr
&(x_1^{(3)}~,\,x_2^{(2)}~,\,x_3^{(1)}~,\,\delta_i^{(1)}~;\, \delta_{i+1})~,\quad  \quad \text{for} \; i = 1, \cdots, N-1~.
\ea
\ee
The superscripts denote the weights of the blowup, and $\delta_i$ are the sections associated to the exceptional divisors of the blowup. The fully resolved singularity is completely smooth can be summarized by the equation:
\be\label{resolved singularity rankNE8}
\t \MG\; : \qquad x_1^2+x_2^3+x_3^6+x_4^{6N}\prod_{i=1}^N \delta_i^{6N-6i}=0\,.
\ee
If $N=1$, the single exceptional divisor, $S_1 \cong \{\delta_1=0\}$, is a smooth degree-6 hypersurface in the weighted projective space $\mb{P}^{3,2,1,1}$, which exactly has the topology of $dP_8$. This confirms that the 5d SCFT $\FT$ is the rank-one $E_8$ theory, with the IR gauge-theory description $SU(2){+}7\bm{F}$ \cite{Morrison:1996xf}.

For $N\geq 2$, the triple intersection numbers among the $N$ exceptional divisors $S_i$ are:
\bea\label{inter nbr rkN E8}
&S_1^3=S_2^3=\dots=S_{N-1}^3=0~, \qquad &&S_N^3=1\ ,\ \cr
&S_i^2\cdot S_{i-1}=-1, \qquad &&S_{i-1}^2\cdot S_i=1\ ,\quad (i=2,\dots,N) \,,
\eea
with all other intersection numbers vanishing. From the Riemann-Roch theorem and the adjunction formula:
\be
\label{Riemann-Roch}
S_1^2\cdot S_2+S_1\cdot S_2^2=2g(S_1\cdot S_2)-2~,
\ee
one finds that the intersection curves $S_i\cdot S_{i+1}$ $(i=1,\dots,N-1)$ all have genus $g(S_i\cdot S_{i+1})=1$. In fact, the surfaces $S_i$, $i<N$, are ruled surface over a genus-$1$ curve, $\mb{P}^1 \rightarrow S_i \rightarrow T^2$, while $S_N$ is a $dP_8$ surface. These intersection numbers are exactly the ones obtained for the gauge theory \eqref{rank N E8 gauge theory Spn} engineered in \cite{Intriligator:1997pq}.~\footnote{See equations (8.13)-(8.14) of \protect\cite{Intriligator:1997pq}, plugging in $g=0$, $g'=1$ and $n_F=7$.} The Hodge diamond of the surface $S_k$, for $k<N$, reads:
\be
h^{i,j}(S_k)=\begin{pmatrix} 1\; & 1\; & 0 \\ 1\; & 2\; & 1 \\ 0\; & 1\; & 1\end{pmatrix}~, \qquad 
{k=1, \cdots, N-1} \,.
\ee
In particular, we find a pair of 3-cycles on each of these surface. Therefore, in total, there are $2(N-1)$ 3-cycles among all the $S_k$. This agrees with the number of 3-cycles in the smooth local threefold $\t \MG$, which can be computed using the methods of \cite{Caibarb3}:
\be
b_3=H_3(\t \MG, \Z) = 2N-2~.
\ee
Note that, on the other hand, the space $\t \MG$ is simply connected.

When $N\geq 2$, interestingly, one can go through a geometric transition that flops the genus-1 ruled surfaces $S_i$ ($i<N$) into Hirzebruch surfaces, and blows up the surface $S_N$ (which becomes ${\rm Bl}_{8} \mathbb{F}_3$).
(For the case of $N=2$, this flop was explicitly constructed in~\cite{Jefferson:2018irk,Apruzzi:2019opn}.)
 After this transition, the $(N-1)$-intersection curves $S_i\cdot S_{i+1}$ all become genus-zero curves. As a consequence, all the 3-cycles disappear and there is one additional 2-cycle in the flopped geometry, $\t \MG_{\rm flopped}$. Alternatively, this geometric transition can also be understood from the $(p,q)$-web for the 5d $Sp(N)$ gauge theory, as we discuss in Appendix~\ref{app: rankN E1}.

\paragraph{Interpretation in $\FT$: the `$\bm{AS}$ hypermultiplet phase transition.'}  The resolved geometry \eqref{resolved singularity rankNE8} should correspond to a generic point on the extended Coulomb branch of the 5d SCFT $\FTX{\MG_{E_8}}$. As already mentioned, from the gauge theory perspective \eqref{rank N E8 gauge theory Spn}, one expects the ECB to be of real dimension $N+9$, instead of:
\be
d_H = r+ f= N+8~,
\ee
as dictated by the geometry. Recall that a hypermultiplet in a representation $\mathfrak{R}$ of the 5d gauge group $G_{\rm 5d}$ contributes to the prepotential as \cite{Intriligator:1997pq}:
\be
\CF_{\mathfrak{R}} =-{1\ov 12} \sum_{\rho \in \mathfrak{R}} \left|\rho(\varphi) + m_\mathfrak{R}\right|^3~,
\ee
where the sum is over all the weights $\rho$ of the representation, $\varphi$ denotes the CB VEVs of $G_{\rm 5d}$ and $m_\mathfrak{R}$ is a real `flavor' mass. For the antisymmetric of $Sp(N)$, we have:
\be
\CF_{\bm{AS}} =-{N-1\ov 12}  \left|m_{\bm{AS}} \right|^3  -{1\ov 12} \sum_{\rho \neq 0}\left|\rho(\varphi) + m_{\bm{AS}}\right|^3~.
\ee
The key point is that the $\bm{AS}$ of $Sp(N)$, of dimension $(2N+1)(N-1)$, has $N-1$ vanishing weights, $\rho=(0, \cdots, 0)$. Therefore, at a generic point on the $Sp(N)$ Coulomb branch, we have $N-1$ neutral hypermultiplets of mass $m_{\bm{AS}}$. In the limit $m_{\bm{AS}}\rightarrow 0$, these neutral hypermultiplets can be given a VEV. 

This phase transition resolves our puzzle. In M-theory, we interpret the resolved geometry \eqref{resolved singularity rankNE8} as the Coulomb branch of the SCFT in a `partial Higgs phase,' which can be described as the $Sp(N)$ gauge theory on its CB with $m_{\bm{AS}}=0$ and non-zero VEVs for the $N-1$ neutral hypermultiplets, as we just discussed. In this phase, the 5d SCFT flavor symmetry will be:
\be
G_H^{\rm 5d}= E_8~, 
\ee
for any $N$, as we will see.  The geometric transition mentioned above corresponds to turning off these VEVs and turning on the mass for the antisymmetric, $m_{\bm{AS}}\neq 0$, which corresponds to the new K\"ahler parameter for the additional curve in $S_N \subset \t \MG_{\rm flopped}$ after the flop. One could then, in principle, go to the origin of the K\"ahler cone of this flopped geometry, where we would expect an SCFT with flavor symmetry $E_8 \times SU(2)$.  For $N=2$, this flavor symmetry is manifest in the model derived from a non-isolated singularity \cite{Apruzzi:2019opn}; however, to our knowledge the geometry that makes the $SU(2)$ manifest for $N>2$ is unknown. It would be interesting to find such explicit geometries, including as isolated singularities if they exist. 

In summary, our puzzle in the counting of the mass parameters of $\FT$ arose as an order-of-limit issue. If we consider $\FT$ as {\it defined} by the canonical singularity \eqref{eq-E8-rankN}, there is no issue except that the 5d gauge-theory interpretation \eqref{rank N En gauge theory Spn} is partially lost.%
\footnote{Note that is not in contradiction with \protect\cite{Intriligator:1997pq}, where the prepotential of the $Sp(N){+}7\bm{F}{+}\bm{AS}$ gauge theory was matched to the prepotential computed from the intersection numbers \protect\eqref{inter nbr rkN E8}, because that computation was not keeping track of the mass parameters. Instead, one should redo these computations while keeping track of all the non-compact divisors, as done recently in \protect\cite{Closset:2018bjz, Saxena:2019wuy} for toric singularities, and in \protect\cite{Apruzzi:2019opn} for elliptic models.} If we insist on interpreting this SCFT as a `strong-coupling limit' (i.e. a UV completion) of the 5d  $\CN=1$ gauge theory, the correct interpretation is that one needs to first send $m_{\bm{AS}}$ to zero, go to the `partial Higgs phase' for the chargeless hypermultiplets, then take a scaling limit wherein all the remaining $N+8$ ECB parameters are sent to zero.

\subsubsection{Magnetic quivers and the 5d Higgs branch}
The 4d $\CN=2$ quiver of Fig.~\ref{f:rankNE8-quiver} can be directly reduced to 3d, since it is a Lagrangian SCFT. Thus, we see that the `electric quiver' $\MQfour$ of $\FTfour$ is just the $\widehat{E}_8$ quiver with $SU(N)$ gauge group, now viewed as a 3d $\CN=4$ gauge theory. This theory (like $\FTfour$ itself) has a flavor symmetry $U(1)^8$, which are simply the symmetries rotating the $8$ bifundamental hypermultiplets. Gauging this symmetry according to our general prescription, we obtain the magnetic quiver of  the 5d SCFT
\be\label{MQ5 E8}
   \begin{tikzpicture}[x=.7cm,y=.7cm]
\node at (-3,0.5) {$\MQfive[\MG_{E_8}]  = $};
\draw[ligne, black](0,0)--(7,0);
\node[bd] at (0,0) [label=below:{{\scriptsize$N$}}] {};
\node[bd] at (1,0) [label=below:{{\scriptsize$2N$}}] {};
\node[bd] at (2,0) [label=below:{{\scriptsize$3N$}}] {};
\node[bd] at (3,0) [label=below:{{\scriptsize$4N$}}] {};
\node[bd] at (4,0) [label=below:{{\scriptsize$5N$}}] {};
\node[bd] at (5,0) [label=below:{{\scriptsize$6N$}}] {};
\draw[ligne, black](5,0)--(5,1);
\node[bd] at (5,1) [label=above:{{\scriptsize$3N$}}] {};
\node[bd] at (6,0) [label=below:{{\scriptsize$4N$}}] {};
\node[bd] at (7,0) [label=below:{{\scriptsize$2N$}}] {};
\end{tikzpicture}
\ee
with the gauge group $\prod_{k} U(d_k N)$ modulo the diagonal $U(1)$. The CB of this $\MQfive$ can be studied by standard methods. We see, in particular, that its has the expected dimension:
\be
d_H = {\rm dim} \, \CM_H^{\rm 5d} = \h r+ f= 30 N-1~.
\ee
Moreover, this $\MQfive$ implies that $\FT$ has the enhanced flavor symmetry $E_8$;  the Cartan of this $E_8$ is the topological symmetry $U(1)^8$ of the unitary quiver. For $N>1$,  the  $\MQfive$ \eqref{MQ5 E8} is distinct from the one expected from the strong-coupling limit of the 5d $Sp(N)$ theory, for the reasons explained above. The latter quiver would be:
\be
   \begin{tikzpicture}[x=.7cm,y=.7cm]
\draw[ligne, black](-1,0)--(7,0);
\node[bd] at (-1,0) [label=below:{{\scriptsize$1$}}] {};
\node[bd] at (0,0) [label=below:{{\scriptsize$N$}}] {};
\node[bd] at (1,0) [label=below:{{\scriptsize$2N$}}] {};
\node[bd] at (2,0) [label=below:{{\scriptsize$3N$}}] {};
\node[bd] at (3,0) [label=below:{{\scriptsize$4N$}}] {};
\node[bd] at (4,0) [label=below:{{\scriptsize$5N$}}] {};
\node[bd] at (5,0) [label=below:{{\scriptsize$6N$}}] {};
\draw[ligne, black](5,0)--(5,1);
\node[bd] at (5,1) [label=above:{{\scriptsize$3N$}}] {};
\node[bd] at (6,0) [label=below:{{\scriptsize$4N$}}] {};
\node[bd] at (7,0) [label=below:{{\scriptsize$2N$}}] {};
\end{tikzpicture}
\ee
whose CB gives the (centered) moduli space of $N$ instantons---see {\it e.g.} \cite{Cremonesi:2014xha}. Note that the additional $U(1)$ node, which renders the node adjacent to it `unbalanced,' could not come from some Lagrangian matter in a 4d SCFT, since that would break four-dimensional conformal invariance.

Finally, let us highlight the fact that we have {\it derived}  the magnetic quiver \eqref{MQ5 E8} directly from the singularity $\MG_{E_8}$, without using $(p,q)$-webs. The same $\MQfive$ can be obtained from a standard web construction for the $Sp(N)$ theory in its `partial Higgs phase', as explained in Appendix~\ref{app: rankN E1}.

\subsection{Rank-two 5d SCFT with a gauge-theory phase $SU(2)_0{-}SU(2){-}5\bf{F}$}\label{subsec:expl2}
As another interesting example, consider the canonical singularity:
\be
\label{eq-E8-rank2}
F(x)= x_1^2+x_2^5+x_3^{10}+x_3 x_4^3=0~, \qquad\quad (q_1, q_2, q_3, q_4) = \left(\half, {1\ov 5}, {1\ov 10}, {3\ov 10}\right)~.
\ee
Its basic properties are summarized on the first line of Table~\ref{table:other with Lag T4d}. As we will show, this singularity engineers an SCFT $\FT$ which admits a gauge-theory phase: 
\be\label{gauge theory SU2SU25F}
SU(2)_0-SU(2)-5\bm{F}~.
\ee
The enhanced symmetry at the UV fixed point is known to be $G_H^{\rm 5d}= E_8$, through other methods~\cite{Apruzzi:2019opn, Hayashi:2018lyv, Hayashi:2019jvx}. Let us  analyse this model in the language of this paper.

\subsubsection{Singularity spectrum and 4d superconformal quiver}
\begin{table}[t]
\centering 
\begin{tabular}{ | c ||c |c | c|| c|c| c|| c| c|c|} 
\hline\hline 
 $F$& $r$ &$f$   &$d_H$ &$\h r$& $\h d_H$  & $\Delta \CA_r$  & $b_3$ & ${\frak{f}}$  \\ [0.5ex] 
\hline 
$x_1^2+x_2^5+x_3^{10}+x_3 x_4^3$ & $2$&$8$&$46$ &$38$&$10$&0 &0 &0\\
\hline 
$x_1^2+x_2^5+x_3^5+x_4^5$  & $2$ & $0$ & $32$ & $32$ & $2$ & $-6$&$12$ & {$\mb{Z}_2^{6}$}\\
\hline
\end{tabular}
\caption{The isolated singularities studied in subsections~\protect\ref{subsec:expl2} and \protect\ref{subsec:expl3}, respectively.} 
\label{table:other with Lag T4d}
\end{table}

\begin{figure}
\begin{center}
$$
   \begin{tikzpicture}[x=1cm,y=1cm]
\draw[ligne, black](0,0)--(7,0);
\node[bd] at (0,0) [label=below:{{\scriptsize$2$}}] {};
\node[bd] at (1,0) [label=below:{{\scriptsize$4$}}] {};
\node[bd] at (2,0) [label=below:{{\scriptsize$6$}}] {};
\node[bd] at (3,0) [label=below:{{\scriptsize$8$}}] {};
\node[bd] at (4,0) [label=below:{{\scriptsize$10$}}] {};
\draw[ligne, black](4,0)--(4,1);
\node[bd] at (4,1) [label=above:{{\scriptsize$5$}}] {};
\node[bd] at (5,0) [label=below:{{\scriptsize$7$}}] {};
\node[bd] at (6,0) [label=below:{{\scriptsize$4$}}] {};
\node[flavor] at (7,0) [label=below:{{\scriptsize$1$}}] {};
\end{tikzpicture}
$$
\caption{The $\CN=2$ SCFT $\FTfour$  for the singularity $x_1^2+x_2^5+x_3^{10}+x_3 x_4^3=0$, where each node is an $SU(K)$ gauge group, with $K$ as indicated. 
 \label{fig:E8 nminNO}}
\end{center}
\end{figure}
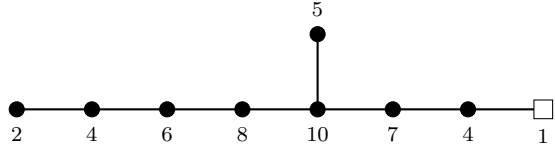

\begin{figure}
\begin{center}
$$
\begin{tikzpicture}
\node (1) [hasse] at (0,0) {};
\node (2) [hasse] at (0,-1) {};
\node (3) [hasse] at (0,-2) {};
\draw (1) edge [] node[label=left:$\mathfrak{e}_7$] {} (2);
\draw (2) edge [] node[label=left:$\mathfrak{e}_8$] {} (3);
\end{tikzpicture}
$$
\caption{Hasse diagram for $\FT$ defined by $x_1^2+x_2^5+x_3^{10}+x_3 x_4^3=0$, with magnetic quiver \protect\eqref{MQ5 NMNOE8}. \label{fig:NPModel1-Hasse}}
\end{center}
\end{figure}
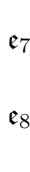

The Milnor ring of the singularity \eqref{eq-E8-rank2} has $\mu=84$ generators. Amongst these, there are exactly eight generators corresponding to 4d $\CN=2$ CB operators with scaling dimension $\Delta=1$ (that is, scaling weight $Q={9\ov 10}$):
\be
x_3^6 x_4~, \; x_3^9~, \; x_2 x_3^4 x_4~, \; x_2 x_3^7~, \; x_2^2 x_3^2 x_4~, \; x_2^2 x_3^5~, \;x_2^3 x_4~, \; x_2^3 x_3^3~.
\ee
We thus find $f=8$, which matches the rank of the flavor symmetry (including the topological symmetries) of the 5d gauge theory \eqref{gauge theory SU2SU25F}. In addition, there are $\h r= 38$ generators with $\Delta >1$, corresponding to the CB operators of $\FTfour$, whose ECB spectrum reads:
\be
\begin{array}{c||c|c|c|c|c|c|c|c|c|c}
\Delta & 1\; & 2\; & 3\; & 4\; & 5\; & 6\; & 7\; & 8\; & 9\; & 10\;\\
\hline
\# & 8 & 8 & 7 & 7 & 5 & 4 & 3 & 2 & 1 & 1\\
\end{array}
\ee
This spectrum is compatible with a 4d gauge group:
\be\label{G expl 2}
G=SU(10)\times SU(8) \times SU(7) \times SU(6) \times SU(5) \times SU(4)^2 \times SU(2)~.
\ee
We can also compute the conformal anomalies of $\FTfour$ or, equivalently, the effective number of hypermultiplets and vector multiplets:
\be
n_h= 312~, \qquad n_v= 302~.
\ee
It is then easy to see that the 4d SCFT is the $\CN=2$ superconformal  Lagrangian quiver shown in Figure~\ref{fig:E8 nminNO}, with gauge group \eqref{G expl 2}.  The 4d Higgs branch then determined semi-classically. In particular, the four-dimensional HB dimension is:
\be\label{dHhat expl2}
\h d_H = n_h - n_v= 10~.
\ee
The 4d $\CN=2$ quiver directly reduces to the electric quiver $\EQfour$, described by the same gauge theory with 3d $\CN=4$ vector multiplets. By gauging the $U(1)^8$ flavor symmetry rotating the bifundamental hypermultiplets in $\EQfour$, we obtain the magnetic quiver of the five-dimensional SCFT, $\FT$, as a unitary quiver given by
\be\label{MQ5 NMNOE8}
   \begin{tikzpicture}[x=.7cm,y=.7cm]
\node at (-3,0.5) {$\MQfive  = $};
\draw[ligne, black](0,0)--(7,0);
\node[bd] at (0,0) [label=below:{{\scriptsize$2$}}] {};
\node[bd] at (1,0) [label=below:{{\scriptsize$4$}}] {};
\node[bd] at (2,0) [label=below:{{\scriptsize$6$}}] {};
\node[bd] at (3,0) [label=below:{{\scriptsize$8$}}] {};
\node[bd] at (4,0) [label=below:{{\scriptsize$10$}}] {};
\draw[ligne, black](4,0)--(4,1);
\node[bd] at (4,1) [label=above:{{\scriptsize$5$}}] {};
\node[bd] at (5,0) [label=below:{{\scriptsize$7$}}] {};
\node[bd] at (6,0) [label=below:{{\scriptsize$4$}}] {};
\node[bd] at (7,0) [label=below:{{\scriptsize$1$}}] {};
\end{tikzpicture}\,.
\ee
From this magnetic quiver, we can see that the flavor symmetry of $\FT$ enhances to $E_8$ at the fixed point. 
This is apparent in the quiver itself, which has balanced nodes in the shape of an $E_8$ Dynkin diagram, and it can also be seen by using the quiver subtraction technique \cite{Cabrera:2018ann} to obtain the Hasse diagram of the Higgs branch $HB[\FT]$, shown in figure~\ref{fig:NPModel1-Hasse}. In fact, the magnetic quiver \eqref{MQ5 NMNOE8} exactly corresponds to the next-to-minimal nilpotent orbit of $E_8$, as studied in \cite{Hanany:2017ooe}. The 5d HB dimension (computed as the dimension of CB$[\MQfive]$) also matches the counting of generators in the Milnor ring, $d_H= \h r +f= 46$, by construction. 

\subsubsection{Resolved geometry and 5d Coulomb branch}
Given the 4d Higgs branch dimension \eqref{dHhat expl2},  $\FT$ should have rank $r= \h d_H -f=2$. This can be seen explicitly, by resolving the singularity \eqref{eq-E8-rank2}. The resolution sequence is:
\be
\ba
&(x_1^{(2)},x_2^{(1)},x_3^{(1)},x_4^{(1)};\delta_1)~,\cr
&(x_1^{(3)},x_4^{(2)},x_2^{(1)},\delta_1^{(1)};\delta_2)~,
\ea
\ee
which gives us a smooth threefold:
\be
\t \MG:\ x_1^2+x_2^5\delta_1+x_3^{10}\delta_1^6+x_3 x_4^3=0~.
\ee
The exceptional divisors, $S_1 \cong \{\delta_1=0\}$ and $S_2 \cong \{\delta_2=0\}$, are irreducible, thus the CB dimension of $\FT$ is indeed given by $r=2$. We can also compute the following triple intersection numbers:
\be
S_1^3=1~, \quad S_2^3=1~, \quad S_1^2\cdot S_2=1~, \quad S_1\cdot S_2^2=-1~.
\ee
The exceptional divisor $S_1$ is a singular surface, however:
\be
\delta_1=0\;:\; \;  x_1^2+x_3 x_4^3=0\,.
\ee
It can be viewed as ruled over a singular cubic curve $x_1^2+x_4^3=0$, with a cusp at $x_1=x_4=0$. In this case, we can perform a  flop by blowing up this singular cubic at the double point on $S_2$---see~\cite{Jefferson:2018irk, Apruzzi:2019opn} for more details and examples. After the flop, we have the following intersection numbers:
\be
S_1^{'3}=9~, \quad S_2^{'3}=0~, \quad  S_1^{'2}\cdot S'_2=-3~, \quad S'_1\cdot S_2^{'2}=1~.
\ee
Here, $S'_1$ has the topology of $\mb{P}^2$ and $S'_2$ is a Hirzebruch surface $\mb{F}_3$ blown up at eight points. After another flop (blowing up $S_1$ and blowing down $S_2$), the intersection numbers match the ones for the rank-two gauge theory  \eqref{gauge theory SU2SU25F}~\cite{Jefferson:2018irk,Apruzzi:2019opn}.

Note that there is no 3-cycle in the resolved geometry $\t \MG$, because the genus-1 curve $S_1\cdot S_2$ is a nodal curve and the 1-cycle has collapsed. Hence we have $b_3=0$. (The same result is obtained with the methods of \cite{Caibarb3}.)  This model has no higher-form symmetry, either.

\subsection{Rank-two 5d SCFT with a gauge-theory phase $G_2{+}5{\bf F}$}\label{subsec:expl3}
As our third example, consider the isolated singularity: 
\be
\label{eq:2555}
F(x)= x_1^2+x_2^5+x_3^5+x_4^5=0~, \qquad\quad (q_1, q_2, q_3, q_4) = \left(\half, {1\ov 5}, {1\ov 5}, {1\ov 5}\right)~.
\ee
whose properties are summarized on the last line of Table~\ref{table:other with Lag T4d}. It turns out that it engineers an SCFT with IR gauge theory descriptions 
\be\label{gauge theories F5 expl 3}
SU(3)_{9\ov 2}+5\bm{F}~, \quad
 Sp(2)+3\bm{F}+2\bm{AS}~,\qquad 
G_2+5\bm{F}~.
\ee
The enhanced flavor symmetry at the UV fixed point is $G_H^{\rm 5d}=Sp(6)$~\cite{Apruzzi:2019opn}. Similarly to the $SU(2)$ flavor symmetry factor for the rank-$N$ $E_8$ theory of subsection~\ref{subsec:rankNE8}, this flavor symmetry turns out to be rather subtle from the point of view of the isolated singularity. We make some preliminary comments on this issue below, and defer a complete discussion to future work.

\subsubsection{Singularity spectrum and 4d superconformal quiver}
\label{sec:G2-sing}
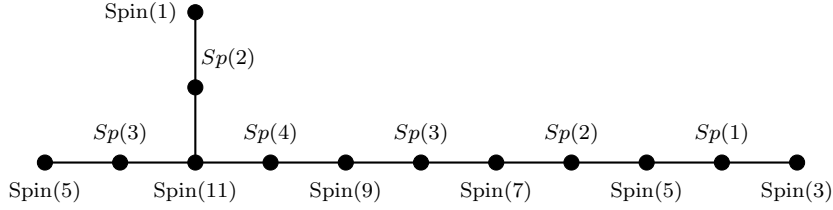
\begin{figure}
\begin{center}
  \begin{tikzpicture}[x=1cm,y=1cm]
\draw[ligne, black](0,0)--(10,0);
\node[bd] at (0,0) [label=below:{{\scriptsize$\Spin(5)$}}] {};
\node[bd] at (1,0) [label=above:{{\scriptsize$Sp(3)$}}] {};
\node[bd] at (2,0) [label=below:{{\scriptsize$\Spin(11)$}}] {};
\draw[ligne, black](2,0)--(2,2);
\node[bd] at (2,1) [label=above:{{\qquad\;\scriptsize$Sp(2)$}}] {};
\node[bd] at (2,2)[label=left:{{\scriptsize$\Spin(1)$}}] {};
\node[bd] at (3,0) [label=above:{{\scriptsize$Sp(4)$}}] {};
\node[bd] at (4,0) [label=below:{{\scriptsize$\Spin(9)$}}] {};
\node[bd] at (5,0) [label=above:{{\scriptsize$Sp(3)$}}] {};
\node[bd] at (6,0) [label=below:{{\scriptsize$\Spin(7)$}}] {};
\node[bd] at (7,0) [label=above:{{\scriptsize$Sp(2)$}}] {};
\node[bd] at (8,0) [label=below:{{\scriptsize$\Spin(5)$}}] {};
\node[bd] at (9,0) [label=above:{{\scriptsize$Sp(1)$}}] {};
\node[bd] at (10,0) [label=below:{{\scriptsize$\Spin(3)$}}] {};
\end{tikzpicture}
\caption{The 4d $\CN=2$ superconformal quiver for the singularity $x_1^2+x_2^5+x_3^5+x_4^5=0$. Here, the links are hypermultiplets in the `bifundamental' $(\bm{m}, \bm{k})$ of $Sp(m)\times \Spin(k)$.\label{fig:orthsymp quiver 1}}
\end{center}
\end{figure}

There are $\mu=64$ generators in the Milnor ring of the singularity \eqref{eq:2555}. 
The 4d ECB spectrum, corresponding to monomials with dimension $\Delta\geq 1$, is as follows:
\be\label{spec expl 3}
\begin{array}{c|c|c|c|c|c}
\Delta &2\; &4 \;&6\; & 8\;&10 \\
\hline
\# & 12&10&6&3&1\\
\end{array}
\ee
Note that there is mass term (corresponding to $\Delta=1$), in this case, and that all the scaling dimension are even. This gives us: 
\be
f=0~, \qquad d_H=\h r=32~.
\ee
The spectrum \eqref{spec expl 3} and the conformal anomalies:
\be
n_h= 232~, \qquad n_v= 236~,
\ee
can be matched by a Lagrangian 4d $\CN=2$ SCFT, which takes the form the orthosymplectic quiver shown in Figure~\ref{fig:orthsymp quiver 1}.%
\footnote{This particular quiver also appeared in \protect\cite{Bhardwaj:2013qia}. In fact, the general discussion of that paper is very useful in order to identify the possible $\CN=2$ superconformal Lagrangian (if any) associated to a given CB spectrum. Note that  the rank-$K$ gauge groups $Sp(K)$ and $SO(2K+1)$ both have CB dimensions $\Delta= \{2, 4, \cdots, 2K\}$.}
  From the resolution $\t \MG$, to be discussed below, we find $r=2$ and therefore:
 \be\label{h dH expl 3}
 \h d_H = n_h - n_v+6= 2~.
 \ee
In particular, there should be $6$ massless vector at low energy on the Higgs branch of the orthosymplectic quiver of Fig.~\ref{fig:orthsymp quiver 1}. Note that this identification is not unique, and as will be discussed in \cite{Closset:2020afy}, this particular CB spectrum has an alternative quiver realization. 

 Relatedly, this 4d $\CN=2$ quiver has a $\Z_2^6$ one-form symmetry, arising from the fact that the fundamental of $\Spin(K)$ preserves the $\Z_2$ center; this matches perfectly with the geometry of the link, since $H_2(L_5(\MG),\Z)= \Z_2^{12}$.

Since $f=0$, the orthosymplectic quiver reduced to 3d directly gives us the magnetic quiver of $\FT$, namely:
\be\label{MQ5 expl3}
   \begin{tikzpicture}[x=.7cm,y=.7cm]
\node at (-3,0.5) {$\MQfive  = $};
\draw[ligne, black](0,0)--(10,0);
\node[bd] at (0,0) [label=below:{{\scriptsize$\Spin(5)$}}] {};
\node[bd] at (1,0) [label=above:{{\scriptsize$Sp(3)$}}] {};
\node[bd] at (2,0) [label=below:{{\scriptsize$\Spin(11)$}}] {};
\draw[ligne, black](2,0)--(2,2);
\node[bd] at (2,1) [label=above:{{\qquad\;\scriptsize$Sp(2)$}}] {};
\node[bd] at (2,2.18)[label=above:{{\scriptsize$\Spin(1)$}}] {};
\node[bd] at (3,0) [label=above:{{\scriptsize$Sp(4)$}}] {};
\node[bd] at (4,0) [label=below:{{\scriptsize$\Spin(9)$}}] {};
\node[bd] at (5,0) [label=above:{{\scriptsize$Sp(3)$}}] {};
\node[bd] at (6,0) [label=below:{{\scriptsize$\Spin(7)$}}] {};
\node[bd] at (7,0) [label=above:{{\scriptsize$Sp(2)$}}] {};
\node[bd] at (8,0) [label=below:{{\scriptsize$\Spin(5)$}}] {};
\node[bd] at (9,0) [label=above:{{\scriptsize$Sp(1)$}}] {};
\node[bd] at (10,0) [label=below:{{\scriptsize$\Spin(3)$}}] {};
\end{tikzpicture}
\ee
Of course, its Coulomb branch dimension matches $d_H=\h r= 32$, by construction. 
It would be very interesting to study the Coulomb branch of \eqref{MQ5 expl3} further, for instance by computing its Hilbert series.

\subsubsection{Resolved geometry and 5d Coulomb branch}
\label{sec:G2-resolution}
Given the singular equation \eqref{eq:2555}, we apply the following resolution sequence:
\be
\ba
&(x_1^{(2)},x_2^{(1)},x_3^{(1)},x_4^{(1)};\delta_1)~,\cr
&(x_1,\delta_1;\delta_2)~,\cr
&(\delta_1,\delta_2;\delta_3)~.
\ea
\ee
The resulting space is smooth 
\be
\t \MG:\qquad \ x_1^2\delta_2+(x_2^5+x_3^5+x_4^5)\delta_1=0~.
\ee
As one can check, the set $\{\delta_1=0\}$ is empty. There are thus only two exceptional divisors $S_1\cong\{ \delta_2=0\}$ and $S_2\cong \{ \delta_3=0\}$, which are both irreducible. Therefore, the 5d SCFT $\FT$ has rank $r=2$. The triple intersection numbers are:
\be
S_1^3=-40~, \quad S_2^3=9~, \quad  S_1^2\cdot S_2=25~, \quad   S_2^2\cdot S_1=-15~.
\ee
Using \eqref{Riemann-Roch}, we can see that $S_1\cdot S_2$ is a genus-6 curve. In fact, $S_1$ is ruled over the genus-6 curve $S_1\cdot S_2$. As $S_2$ has the topology of a $\mb{P}^2$, the $S_1\cdot S_2$ is also exactly the degree-5 curve on $\mb{P}^2$ with genus $g=6$. This makes sense because the curve $S_1\cdot S_2$ has equation $\delta_2=\delta_3=x_2^5+x_3^5+x_4^5=0$, which has degree five. 
The Hodge diamond of $S_1$ takes the form:
\be
h^{i,j}(S_1)=\begin{pmatrix} 1\; & 6\; & 0 \\ 6\; & 2\; & 6 \\ 0\; & 6\; & 1\end{pmatrix}~.
\ee
Thus, there are $12$ 3-cycles on $S_1$, which also correspond to 12 3-cycles in the resolved threefold \cite{Caibarb3}. These 3-cycles give rise $6$ vector multiplets on the Higgs branch of $\FTfour$, as in \eqref{h dH expl 3}.

Note that, in the M-theory engineering, these 3-cycles must also corresponds to 6 neutral hypermultiplets obtaining a VEV on the Coulomb branch of $\FT$. Unlike for the case of the higher-rank $E_8$ theory of section~\ref{subsec:rankNE8}, it is not immediately clear whether this `partial Higgs phase' can be understood directly from the gauge-theory description \eqref{gauge theories F5 expl 3}.

To connect this resolved geometry to the 5d gauge theory, we have to go through a geometric transition: we pinch off six points on the genus-6 curve $S_1\cdot S_2$, and blow up these six double points on $S_2$. After these flops, the triple intersection numbers of the new surfaces $S_1'$, $S_2'$ will be
\be
S_1^{'3}=8~, \quad S_2^{'3}=3~, \quad S_1^{'2}\cdot S'_2=1~, \quad S_2^{'2}\cdot S'_1=-3~.
\ee
Hence $S'_1$ is a $\mb{F}_3$ and $S'_2$ is a $dP_6$. These triple intersection numbers exactly give rise to the 5d SCFT with the IR gauge-theory description \eqref{gauge theories F5 expl 3}. Before the flop, our resolved geometry has $f=0$, in agreement with the singularity spectrum. However, after the six blow-ups of double points, there are six new 2-cycles in the flopped geometry,  while all the 3-cycles have disappeared. These six new K\"ahler parameters then match the rank of the expected UV fixed point of the gauge theories  \eqref{gauge theories F5 expl 3}, which has flavor symmetry  $G_F=Sp(6)$.

Clearly, a better understanding of this apparent phase transition on the Coulomb branch of the 5d SCFT $\FT$ is desirable, both in terms of the gauge theory description and in terms of the local Calabi-Yau.

\section{Argyres-Douglas theories and their 5d SCFT counterparts}
\label{sec: 5d rank0}
In the Type IIB setup, the most studied 4d fixed points are the (generalized) Argyres-Douglas (AD) theories \cite{Argyres:1995xn} engineered at the hypersurface singularities  \cite{Shapere:1999xr, Cecotti:2010fi}
\be\label{FGG def}
F(x_1, x_2, x_3, x_4) = g_G(x_1, x_2) + g_{G'}(x_3, x_4)=0~,
\ee
where the two-variable polynomial $g_G(x, y)$ defines the ADE singularity
\bea
&g_{A_k}(x, y)= x^2 + y^{k+1}~, \qquad   
&&g_{D_k}(x,y)= x^{k+1}+ x y^2~,  &&\cr
&g_{E_6}(x,y) = x^3 +y^4~, \qquad 
&&g_{E_7}(x,y) = x^3 +x y^3~, \qquad 
&&g_{E_8}(x,y) = x^3 +y^5~.
\eea
We will denote these so-called $(G, G')$ theories \cite{Cecotti:2010fi} by
\be
\text{AD}[G, G'] \equiv  {\mathscr{T}_{\MG_{[G, G']}}^{\rm 4d}}~.
\ee
Of course, one can study many other SCFTs in a similar manner, such as, for instance, the 4d SCFTs that arise at exceptional unimodal singularities \cite{Arnold1975} which are not of the $[G,G']$ type \cite{Cecotti:2011gu}. Here, we focus on the class \eqref{FGG def} for definiteness; see \cite{Closset:2021lwy, CSNWII} for  more examples. 

We display a few examples, with their basic properties, in Table~\ref{table:AD theories}.
Note that many of these canonical singularities do not admit crepant blow ups, so that $r=0$.~\footnote{The singularities \eqref{FGG def} have $r=0$ for $(G, G')=(A_k, G')$, for any $k$ and any $G$ of type $A$, $D$ or $E$.} From our general discussion, they must therefore correspond to  `rank-zero' 5d SCFTs $\FT$. We will see that this is indeed the case, although it is not yet clear whether the rank-zero $\FT$ is a `trivial' theory---in the sense that it could consists of free 5d hypermultiplets `in disguise'.

 We will also discuss an interesting example with $r=1$, corresponding to a  singularity that engineers a non-trivial, higher-rank SCFT both in 4d and 5d.

\begin{table}[t]
\centering 
\begin{tabular}{|c | c ||c |c | c|| c|c|c|c| c|| c| c|} 
\hline\hline 
$[G,G']$& $F$& $r$ &$f$   &$d_H$ &$\h r$& $\h d_H$ & $a$ & $c$ & $\Delta \CA_r$ &$\mu$ & $b_3$  \\ [0.5ex] 
\hline 
$[A_1, A_1]$&    $ x_1^2+x_2^2+x_3^2+x_4^2$ & $0$&  $1$& $1$& $0$& $1$& ${1\ov 24}$& ${1\ov 12}$& $0$& $1$& $0$ \\
$[A_1, A_2]$&    $x_1^2+x_2^2+x_3^2+x_4^3$ & $0$&  $0$& $1$& $1$& $0$& ${43\ov 120}$& ${11\ov 30}$& ${1\ov 5}$& $2$& $0$ \\
$[A_1, A_3]$&    $x_1^2+x_2^2+x_3^2+x_4^4$ & $0$&  $1$& $2$& $1$& $1$& ${11\ov 24}$& ${1\ov 2}$& $0$& $3$& $0$ \\
$[A_1, A_4]$&    $x_1^2+x_2^2+x_3^2+x_4^5$ & $0$&  $0$& $2$& $2$& $0$& ${67\ov 84}$& ${17\ov 21}$& ${2\ov 7}$& $4$& $0$ \\
\hline
$[A_1, D_5]$&    $ x_1^2+x_2^2+x_3^4+ x_3 x_4^2$ & $0$&  $1$& $3$& $2$& $1$& ${19\ov 20}$& $1$&     ${1\ov 5}$& $5$& $0$ \\
$[A_1, D_6]$&    $ x_1^2+x_2^2+x_3^5+ x_3 x_4^2$ & $0$&  $2$& $4$& $2$& $2$& ${13\ov12}$& ${7\ov 6}$& $0$& $6$& $0$ \\
$[A_1, D_7]$&    $ x_1^2+x_2^2+x_3^6+ x_3 x_4^2$ & $0$&  $1$& $4$& $3$& $1$& ${81\ov56}$& ${3\ov 2}$& ${2\ov 7}$& $7$& $0$ \\
$[A_1, D_8]$&    $ x_1^2+x_2^2+x_3^7+ x_3 x_4^2$ & $0$&  $2$& $5$& $3$& $2$& ${19\ov 12}$& ${5\ov 3}$& $0$& $8$& $0$ \\
\hline
$[A_2, A_2]$&    $x_1^2+x_2^3+x_3^2+x_4^3$ & $0$&  $2$& $3$& $1$& $2$& ${7\ov 12}$& ${2\ov 3}$& $0$& $4$& $0$ \\
$[A_3, A_3]$&    $x_1^2+x_2^4+x_3^2+x_4^4$ & $0$&  $3$& $6$& $3$& $3$& ${15 \ov 8 }$& ${2}$& $0$& $9$& $0$ \\
$[A_4, A_4]$&    $x_1^2+x_2^5+x_3^2+x_4^5$ & $0$&  $4$& $10$& $6$& $4$& ${25\ov 6}$& ${13\ov 3}$& $0$& $16$& $0$ \\
\hline
$[A_2, D_4]$&   $ x_1^2+x_2^3+x_3^3+ x_3 x_4^2$ & $0$&  $0$& $4$& $4$& $0$& $2$& $2$& $0$& $8$& $0$ \\
\hline 
\end{tabular}
\caption{5d and 4d data for some singularities giving rise to $\FTfour= \text{AD}[G,G']$. Here, $a$ and $c$ are the conformal anomalies of $\FTfour$.  Note also that $[A_2, A_2]\cong [A_1, D_4]$.} 
\label{table:AD theories}
\end{table}

\subsection{4d fixed-points with $r=0$ and 3d Mirrors}
Let us first review some well-known fact about some of these AD theories with $r=0$, in this IIB engineering perspective \cite{Shapere:1999xr, Cecotti:2010fi, Xie:2015rpa}.

\paragraph{The theory $\text{AD}[A_1, A_{2N-1}]$.} This series includes the free hypermultiplet for $N=1$, since $\MG_{[A_1, A_1]}$ is the conifold singularity. We have:
\be
f=1~, \qquad \h r=N- 1~, \qquad  \h d_H= 1~.
\ee
The 3d mirror (that is, $\MQfour$) is known to be 3d $\CN=4$ SQED$[N_f=N]$; equivalently, it is a $U(1) \times U(1)$ quiver with $N$ bifundamentals, modulo the decoupled diagonal $U(1)$ \cite{Xie:2012hs}
\be\label{MQ4 A1A2Nm1}
  \begin{tikzpicture}[x=.7cm,y=.7cm]
\node at (-3,0) {$\MQfour \cong$};
\draw[ligne, black](-1,0)--(1,0);
\node[bd] at (-1,0) [label=above:{{\scriptsize$1$}}] {};
\node[bd] at (1,0) [label=above:{{\scriptsize$1$}}] {};
\node at (0,0.4) {\scriptsize$N$};
\end{tikzpicture}
\ee
The Higgs branch is 
\be\label{HB A2Nm1}
\CM_H^{\rm 4d} \cong \C^2/ \Z_N  \cong {\rm CB}\left[ \MQfour \right]~.
\ee
Of course, for $N=1$, we recover the elementary 3d mirror symmetry between a hypermultiplet and SQED$[N_f=1]$. The Higgs branch \eqref{HB A2Nm1} corresponds to the small resolution of the singularity:
\be
x_1^2+ x_2^2 + x_3^2+ x_4^{2N}=0~,
\ee
which leaves us with a smooth local CY threefold with a single K\"ahler parameter (for $N=1$, this is the resolved conifold).

\paragraph{The theory $\text{AD}[A_1, A_{2N}]$.} In this case, one has:
\be
f=0~, \qquad \h r=N~, \qquad  \h d_H= 0~.
\ee
Therefore the Higgs branch is trivial. The CB spectrum is $\Delta= \{{2N+2+2j\ov 2N+3}\}_{j=1}^N$.  It has been proposed that the 3d mirror consist of $N$ free hypermultiplets \cite{Benvenuti:2018bav, Dedushenko:2019mnd}:
\be
\MQfour \cong (\text{hyper})^{\otimes N}~.
\ee
From the IIB construction, we see that the HB is empty because the terminal singularity:
\be\label{terminal A2n1}
x_1^2+ x_2^2 + x_3^2+ x_4^{2N+1}=0
\ee
does not admit any (small or otherwise) crepant resolution. This is also reflected in the $\Tr(U(1)_r)$ anomaly,  $24(c-a) = {N \ov 2 N + 3}$, which is not integer.

\paragraph{The theory $\text{AD}[A_1, D_{2N+2}]$.} In this case, one finds:
\be
f=2~, \qquad \h r=N~, \qquad  \h d_H= 2~.
\ee
The 3d mirror is given by the following quiver (modulo the diagonal $U(1)$, so the gauge group is $U(1)^2$):
\be\label{A1D2Np2 MQ4}
 \begin{tikzpicture}[x=.7cm,y=.7cm]
\node at (-3,0) {$\MQfour =$};
\draw[ligne, black](-1,1-0.5)--(0,0-0.5)--(1,1-0.5);
\draw[ligne, black](-1,1-0.5)--(1,1-0.5);
\node[bd] at (0,0-0.5) [label=below:{{\scriptsize$1$}}] {};
\node[bd] at (-1,1-0.5) [label=above:{{\scriptsize$1$}}] {};
\node[bd] at (1,1-0.5) [label=above:{{\scriptsize$1$}}] {};
\node at (0,1.5-0.6) {\scriptsize$N$};
\end{tikzpicture}
\ee
The Higgs branch $\CM_H^{\rm 4d} \cong {\rm CB}\left[ \MQfour \right]$ has quaternionic dimension $2$. The flavor symmetry $G_H^{\rm 4d}$ is $SU(3)$ for $N=1$ and $SU(2) \times U(1)$ for $N>1$.

\paragraph{The theory $\text{AD}[A_1, D_{2N+1}]$.} In this case, one finds:
\be
f=1~, \qquad \h r=N~, \qquad  \h d_H= 1~.
\ee
This Higgs branch is $\C^2/\Z_2$ for any $N$. In addition, the low-energy theory on the Higgs branch also contains the irreducible SCFT $\text{AD}[A_1, A_{2N-2}]$, which itself has no Higgs branch. The 3d mirror is expected to consists of a tensor product of SQED$[N_f=2]$ (whose CB realizes the $\C^2/\Z_2$ Higgs branch) with $(N-1)$ free hypermultiplets:
\be
\MQfour \cong (\text{hyper})^{\otimes (N-1)} \otimes \, \text{SQED}[N_f=2]~,
\ee
as argued recently in \cite{Dedushenko:2019mnd}. The canonical singularity:
\be\label{AD partial resolvable}
 x_1^2+x_2^2+x_3^{2N+2}+ x_3 x_4^2=0
\ee
admits a small resolution which leaves us with the residual terminal singularity that gives the IR theory $\text{AD}[A_1, A_{2N-2}]$.  

\paragraph{The theory $\text{AD}[A_k, A_k]$.} This family of theories is particularly interesting. We have:
\be
f=k~, \qquad \h r={k(k-1)\ov 2}~, \qquad  \h d_H= k~.
\ee
The 3d mirror is the complete graph with $k+1$ nodes \cite{Nanopoulos:2010bv, Xie:2012hs}:
\bea\label{complete graph MQ}
&  \begin{tikzpicture}[x=.7cm,y=.7cm]
\node at (-7,0) {$\MQfour[\MG_{[A_1, A_1]}] =$};
\draw[ligne, black](-4,0)--(-2,0);
\node[bd] at (-4,0) [label=above:{{\scriptsize$1$}}] {};
\node[bd] at (-2,0) [label=above:{{\scriptsize$1$}}] {};
\node at (-1,0) {$~,$};

\node at (2.5,0) {$\MQfour[\MG_{[A_2, A_2]}] =$};
\draw[ligne, black](6.5,-0.2)--(7.5,1)--(5.5,1);
\draw[ligne, black](6.5,-0.2)--(5.5,1);
\node[bd] at (6.5,-0.2) [label=below:{{\scriptsize$1$}}] {};
\node[bd] at (5.5,1) [label=above:{{\scriptsize$1$}}] {};
\node[bd] at (7.5,1) [label=above:{{\scriptsize$1$}}] {};
\end{tikzpicture}
\cr
& \begin{tikzpicture}[x=.7cm,y=.7cm]
\node at (-7,0) {$\MQfour[\MG_{[A_3, A_3]}] =$};
\draw[ligne, black](-4,1)--(-2,1);
\draw[ligne, black](-4,1)--(-2,-1);
\draw[ligne, black](-4,1)--(-4,-1);
\draw[ligne, black](-4,-1)--(-2,-1);
\draw[ligne, black](-2,1)--(-2,-1);
\draw[ligne, black](-2,1)--(-4,-1);
\node[bd] at (-4,-1) [label=below:{{\scriptsize$1$}}] {};
\node[bd] at (-2,-1) [label=below:{{\scriptsize$1$}}] {};
\node[bd] at (-4,1) [label=above:{{\scriptsize$1$}}] {};
\node[bd] at (-2,1) [label=above:{{\scriptsize$1$}}] {};
\node at (-1,0) {$~,$};
\node at (1,0) {etc.,};
\end{tikzpicture}
\eea
modulo the diagonal $U(1)$. Note that $\text{AD}[A_2, A_2]\cong \text{AD}[A_1, D_4]$. From the magnetic quivers, we also learn that the flavor group $G_H^{\rm 4d}$ for $\text{AD}[A_1, D_4]$ is $SU(3)$, while the flavor symmetry for $k>2$ is $U(1)^k$. These singularities admit a smooth resolution, with the exceptional locus consisting of $k$ rational curves, therefore we have a smooth Higgs branch phase.  This family also generalizes to $\text{AD}[A_k, A_{kN+N-1}]$, whose 3d mirrors are the same complete graphs \eqref{complete graph MQ} but with all links of multiplicity $N$. 

Interestingly, we can also access the `electric quiverine' directly in certain cases, as follows. The  $\text{AD}[A_k,A_k]$ theory has $k-2$ marginal couplings, which can be tuned to reach an S-duality cusp \cite{Gaiotto:2009we}, from which we can read off some useful `quasi-Lagrangian' description. Such descriptions consist of $SU(n)$ gauge groups coupled together with strongly-coupled `matter' \cite{Cecotti:2013lda, Beratto:2020wmn}, which can be reduced to 3d `sequentially' at weak coupling.  As a simple example, consider $\text{AD}[A_3, A_3]$, which has an $S$-dual description \cite{Buican:2014hfa}:
\be
 \begin{tikzpicture}[x=.7cm,y=.7cm]
\node at (-2.8,0.5) {$\text{AD}[A_3, A_3] = $};
\node at (0,0) {$D_4$};
\node at (2,0) {$SU(2)$};
\node at (4,0) {$D_4$};
\draw[ligne, black](0.5,0)--(1.1,0);
\draw[ligne, black](2.9,0)--(3.5,0);
\draw[ligne, black](2,0.5)--(2,1.3);
\node[flavor] at (2,1.5) [label=center:{{\scriptsize$1$}}] {};
\end{tikzpicture}
\ee
Here, an $SU(2)$ subgroup of the $D_4$ theory (that is, $[A_1, D_4]$) is gauged in conformal manner; indeed, the CB spectrum of $[A_3,A_3]$ is $\Delta= \{{3\ov 2}, {3\ov 2}, 2\}$, and $D_4$ is rank-1 with $\Delta= {3\ov 2}$. Since $D_4$ itself flows to the SQED$[N_f{=}3]$ fixed point in 3d, by weakly gauging the $SU(2)$, one finds the 3d description:
\be\label{Sdual A3A3}
 \begin{tikzpicture}[x=.7cm,y=.7cm]
\node at (-4,0.5) {$\EQfour[\MG_{[A_3, A_3]}]\,  = $};
\node[bd] at (0,0) [label=above:{{\scriptsize$1$}}] {};
\node at (2,0) {$SU(2)$};
\node[bd] at (4,0) [label=above:{{\scriptsize$1$}}] {};
\node[flavor] at (-1.2,0) [label=center:{{\scriptsize$1$}}] {};
\node[flavor] at (5.2,0) [label=center:{{\scriptsize$1$}}] {};
\draw[ligne, black](-1,0)--(1.1,0);
\draw[ligne, black](2.9,0)--(5,0);
\draw[ligne, black](2,0.5)--(2,1.3);
\node[flavor] at (2,1.5) [label=center:{{\scriptsize$1$}}] {};
\end{tikzpicture}
\ee
This must be a mirror description of the bottom quiver in  \eqref{complete graph MQ}, as recently discussed in~\cite{Beratto:2020wmn}.

\subsection{5d rank-zero fixed points and their magnetic quivers}
We would like to understand better the 5d SCFTs $\FT$ arising from these same singularities. These theories have $r=0$ and $f$ mass deformations, corresponding to $f$ small-resolution K\"ahler parameters. These mass deformations trigger an RG flow, which ends on the trivial theory (with $f$ background gauge fields, from M-theory on the 2-cycles of $\t\MG$) if the small resolution $\t\MG$ is smooth. However, we might be left with a terminal singularity, such as, for instance, in \eqref{terminal A2n1}, which would signal that we flowed to a simpler `irreducible' SCFT $\FT$ without massive deformations. This is the interpretation of the partial resolution of \eqref{AD partial resolvable} in M-theory, in particular.


\paragraph{The singularity $\MG{[A_1, A_{2N-1}]}$.}
For $N=1$, the conifold, we know that $\FT$ is the free 5d hypermultiplet, while $\FTfour$ is the free 4d hypermultiplet; we have $f=1$. The electric theory for both the 5d and the 4d theory is the free hyper in 3d, while the magnetic quivers are SQED with one electron:
\be
\EQfive \cong \EQfour \cong  \text{hyper}~, \qquad \MQfive \cong \MQfour \cong \text{SQED}[N_f=1]~.
\ee
Of course, the electric and magnetic theories are related by mirror symmetry, and at the same time the gauging of the $U(1)$ flavor of the electric theory gives the correct result, as in \eqref{gauging EQ5} and \eqref{gauging EQ4}.

Next, consider the case $N=2$. The magnetic quiver $\MQfour$ \eqref{MQ4 A1A2Nm1} is self-mirror in this case, so that:
\be
\EQfour[\MG_{[A_1, A_3]}] = \text{SQED}[N_f=2]~,
\ee
with the flavor symmetry $U(1) \subset SU(2)_F$ acting on the two hypermultiplets with charge $\pm 1$. 
Using the gauging prescription \eqref{gauging EQ4}, we get the magnetic quiver of $\FT$ described by a $U(1) \times U(1)$ gauge theory coupled to two hypermultiplets with charges:
\be
\MQfive[\MG_{[A_1, A_3]}] \; : \;  \qquad 
\begin{tabular}{|c | c c|  } %
\hline
& $H_1$ & $H_2$\\
\hline
$U(1)_1$ & $1$ & $1$\\
$U(1)_2$ & $1$ & $-1$ \\
\hline 
\end{tabular}
\ee
This 3d theory obviously has a Higgs branch of dimension zero, while its Coulomb is of quaternionic dimension $2$, which by construction gives us the Higgs branch of $\FT$. To gain a better understanding of the structure of the 3d quantum CB, we may compute its Hilbert series using the monopole formula \cite{Cremonesi:2013lqa}. One finds:
\be
{\rm HS}_{\text{CB}[\MQfive]}(t) = {1+ 6 t^2 + t^4 \ov (1-t^2)^4}~.
\ee
This is consistent with $\CM_H[\FT]$ being a hyper-K\"ahler cone of dimension $d_H=2$. From the plethystic logarithm $\text{PL}[{\rm HS}]= 10 t^2 - 20 t^4+ \cdots$, we see that the 5d HB is a non-complete intersection in $\C^{10}$, and from the small-$t$ expansion ${\rm HS}= 1+ 10 t^2+ \cdots$, we find that the theory has $10$ conserved currents \cite{Gaiotto:2008ak, Gaiotto:2012uq}.  Indeed, this Higgs branch is the minimal nilpotent orbit of $Sp(2)$, and appears as a $\Z_2$ gauging of two hypermultiplets  \cite{Cremonesi:2014xha}. Thus we have the flavor group $G_H^{\rm 5d} = Sp(2)$. We provide some more details on the Hilbert series  computations in Appendix~\ref{appendix:HS}. 

The careful reader\footnote{And a thorough anonymous JHEP referee.} will have noticed that this larger symmetry group seems in contradiction with our general discussion of the global symmetry as seen from the (resolved) geometry in M-theory. In fact, such unexpected `enhanced symmetries', where even the rank of $G_H^{\rm 5d}$ is larger than expected, can often happen when $\FT$ contains a free-hypermultiplet sector (modulo discrete gauging), which is the case here. We will discuss this interesting state of affairs in more detail elsewhere.

\paragraph{The singularity $\MG{[A_1, A_{2N}]}$.} Given the 3d mirror $\MQfour$ and the fact that $f=0$, we would conclude that the rank-zero 5d SCFT $\FTX{\MG_{[A_1, A_{2N}]}}$ flows to $N$ hypers upon compactification on a torus.

\paragraph{The singularity $\MG{[A_1, D_4]}\cong \MG{[A_2,A_2]}$.} This should give an $r=0$, $d_H=3$ 5d theory. The 3d mirror of the $\MQfour$ \eqref{A1D2Np2 MQ4} with $N=1$ is simply:
\be
\EQfour[\MG_{[A_2, A_2]}] = \text{SQED}[N_f=3]~.
\ee
We now gauge the $U(1)^2$ maximal torus of the $SU(3)$ flavor symmetry to obtain the `magnetic quiver' theory:
\be
\MQfive[\MG_{[A_2, A_2]}]  \; : \; \qquad
\begin{tabular}{|c | c cc|  } %
\hline
& $H_1$ & $H_2$&$H_3$\\
\hline
$U(1)_1$ & $1$ & $1$&$1$\\
$U(1)_2$ & $1$ & $-1$&$0$ \\
$U(1)_3$ & $0$&$-1$ & $1$ \\
\hline 
\end{tabular}
\ee
with gauge group $U(1)^3$, three hypermultiplets and a zero-dimensional Higgs branch. We have $d_H=3$, with the Hilbert Series:
\be
{\rm HS}_{\text{CB}[\MQfive]}(t) = \frac{1-3 t+12 t^2-11 t^3+12 t^5-3 t+t^6}{(1-t)^6 \left(1+t+t^2\right)^3} = 1+ 9 t^2+ 30 t^3+ \cdots~.
\ee

\paragraph{The singularity $\MG{[A_k, A_k]}$.} The corresponding 5d SCFTs have:
\be
r=0~, \qquad f=k~, \qquad d_H = {k(k+1)\ov 2}~.
\ee
We studied the case $k=2$ above, and the cases $k>2$ could be studied similarly; this is left for future work. Here, let us simply mention that, for $[A_3,A_3]$, we can also directly obtain a magnetic quiver for $\FT$ by gauging the obvious $U(1)^3$ flavor symmetry in \eqref{Sdual A3A3}:
\be\label{MQ5 A3A3}
 \begin{tikzpicture}[x=.7cm,y=.7cm]
\node at (-4,0.5) {$\MQfive[\MG_{[A_3, A_3]}]\,  = $};
\node[bd] at (0,0) [label=above:{{\scriptsize$1$}}] {};
\node at (2,0) {$SU(2)$};
\node[bd] at (4,0) [label=above:{{\scriptsize$1$}}] {};
\node[bd] at (-1.2,0)[label=above:{{\scriptsize$1$}}] {};
\node[bd] at (5.2,0) [label=above:{{\scriptsize$1$}}] {};
\draw[ligne, black](-1,0)--(1.1,0);
\draw[ligne, black](2.9,0)--(5,0);
\draw[ligne, black](2,0.5)--(2,1.3);
\node[bd] at (2,1.5) [label=above:{{\scriptsize$1$}}] {};
\end{tikzpicture}
\ee
This quiver indeed has a trivial Higgs branch, and a Coulomb branch of dimension $d_H=6$.

An interesting feature of this particular family of singularities is that we also know its 5d BPS quiver---the D-brane quiver for $\KK\FT$ in IIA  \cite{Closset:2019juk}. It consists of a loop of $k+1$ nodes connected by pairs of bifundamental arrows in both directions with a quartic superpotential, generalizing the Klebanov-Witten quiver \cite{Klebanov:1998hh}  ($k=1$) to $k>1$  \cite{Fazzi:2019gvt}. It would be very interesting to study the BPS spectrum of $\KK\FT$ for $k>1$ along the lines of \cite{Closset:2019juk}.

\subsection{Further examples: higher-rank 5d theories}
Let us also mention a couple of other examples with interesting features. The first example involves a non-trivial choice of global structure, while the second example has non-zero rank, $r>0$, in 5d.

\subsubsection{The $\MG_{[A_2, D_4]}$ theory} 
This singularity has $r=0$, $f=0$ and $d_H=4$. The singularity can be equivalently written as:
\be\label{A2D4}
x_1^2+x_2^3+x_3^3+x_4^3=0~.
\ee 
The 4d SCFT has $\h r=4$ with the spectrum $\Delta= \{2, {4\ov 3}, {4 \ov 3}, {4\ov 3} \}$---thus, one marginal coupling---and $\h d_H=0$. We then propose that this theory can be described as the conformal gauging of three $\text{AD}[A_1, A_3]$ theories:
\be\label{A2D4 4d quiver}
 \begin{tikzpicture}[x=.7cm,y=.7cm]
\node at (-2.8,0.5) {$\text{AD}[A_2, D_4] = $};
\node at (0,0) {$A_3$};
\node at (2,0) {$SU(2)$};
\node at (4,0) {$A_3$};
\node at (2,1.8) {$A_3$};
\draw[ligne, black](0.5,0)--(1.1,0);
\draw[ligne, black](2.9,0)--(3.5,0);
\draw[ligne, black](2,0.5)--(2,1.3);
\node at (5.5,0.5) {or};
\node at (0+7,0) {$A_3$};
\node at (2+7,0) {$SO(3)_+$};
\node at (4+7,0) {$A_3$};
\node at (2+7,1.8) {$A_3$};
\draw[ligne, black](0.5+7,0)--(1.1+7,0);
\draw[ligne, black](2.9+7,0)--(3.5+7,0);
\draw[ligne, black](2+7,0.5)--(2+7,1.3);
\end{tikzpicture}
\ee
From the geometry, we can compute $H_2(L_5(\MG_{[A_2,D_4]}, \Z) = \Z_2^2$, so that the theory $\FTfour$ has either an electric or a magnetic  $\Z_2$ one-form symmetry; we propose that the charged line operators are precisely the Wilson or 't Hooft lines in the gauge-theory description \eqref{A2D4 4d quiver}. {(Note that we can also have $SO(3)_-$, with the discrete $\theta$-angle turned on \cite{Aharony:2013hda}, by choosing the diagonal lattice in 
\eqref{h2 ff}.~\footnote{This corresponds to lines realized by  M2/M5-brane dyonic bound states.})}

Changing the gauge coupling in \eqref{A2D4 4d quiver} corresponds to deforming the singularity \eqref{A2D4} by the monomial $x_2 x_3 x_4$. It is tempting to take a singular limit to the singularity $x_1^2 +x_2 x_3 x_4=0$, which is the toric singularity $\C^3/(\Z_2\times \Z_2)$, also known as the 5d $T_2$ model \cite{Benini:2009gi}. From the fact that  $A_3$ reduces to SQED$[N_f=2]$ in 3d, we infer:
\be\label{EQ4 A2D4}
 \begin{tikzpicture}[x=.7cm,y=.7cm]
\node at (-2.8,0.5) {$\EQfour= \MQfive=$};
\node[bd] at (0,0) [label=above:{{\scriptsize$1$}}] {};
\node at (2,0) {$SU(2)$};
\node[bd] at (4,0) [label=above:{{\scriptsize$1$}}] {};
\node[bd] at (2,1.5) [label=above:{{\scriptsize$1$}}] {};
\draw[ligne, black](0,0)--(1.1,0);
\draw[ligne, black](2.9,0)--(4,0);
\draw[ligne, black](2,0.5)--(2,1.5);
\node at (5.7,0.5) {or};
\node[bd] at (0+7,0) [label=above:{{\scriptsize$1$}}] {};
\node at (2+7,0) {$SO(3)$};
\node[bd] at (4+7,0) [label=above:{{\scriptsize$1$}}] {};
\node[bd] at (2+7,1.5) [label=above:{{\scriptsize$1$}}] {};
\draw[ligne, black](0+7,0)--(1.1+7,0);
\draw[ligne, black](2.9+7,0)--(4+7,0);
\draw[ligne, black](2+7,0.5)--(2+7,1.5);
\end{tikzpicture}
\ee
which must also be equal to $\MQfive$ since $f=0$.%
\footnote{Here, the `$SO(3)$' gauge group denotes the gauging of the overall $\Z_2$ center symmetry of the $SU(2)$ quiver on the left, which is the diagonal central $\Z_2$ between the center of $SU(2)$ and the $\Z_2$ subgroup of each $U(1)$ (thus leaving the bifundamental matter invariant).}  This magnetic quiver agrees with a direct application of the rules of \cite{vanBeest:2020kou} for computing $\MQfive$ for the toric singularity. Thus, we propose that the SCFT $\FT$ related to the $[A_2, D_4]$ Argyres-Douglas theory in IIB is precisely the 5d $T_2$. Thus, the mirror of \eqref{EQ4 A2D4}  should be given by 4 hypermultiplets. In fact, the quiver \eqref{EQ4 A2D4} is essentially `ugly' in the technical sense \cite{Gaiotto:2008ak}, which implies the presence of free fields in the IR, up to discrete gauging.  By computing the CB Hilbert series for \eqref{EQ4 A2D4} (see Appendix~\ref{app:A2D4}), we find the HS of $\CM_H^{\rm 5d}$ for either choice of the global structure:
\be
{\rm HS}_{SU(2)} = {1+28t^2+ 70t^4+28t^6+ t^8\ov (1-t^2)^8}~, \qquad \qquad
{\rm HS}_{SO(3)} = {1\ov (1-t)^8}~,
\ee
where we indicated the choice of gauge group in \eqref{EQ4 A2D4}. We see that $\FT$ has a Higgs branch $\C^8$ in the $SO(3)$ gauging case. In the $SU(2)$ gauging case, we obtain the HS of the minimal nilpotent orbit of $Sp(4)$ \cite{Hanany:2016gbz}, consistent with a $\Z_2$ gauging of four hypermultiplets. 

\subsubsection{Example of a rank-one 5d SCFT coupled to a rank-zero theory}
Our geometric setup also leads to many more examples of 5d SCFTs with $r>0$, which are, in some sense, obtained by `coupling' the rank-zero theories to higher-rank theories. While a detailed study of such a `coupling' is beyond the scope of this paper, we present one such example here. (See \cite{CSNWII} for many more examples.)

This example is a rank-1 5d SCFT coupled to the $\MG_{[A_2, D_4]}$ theory mentioned in the last section. The resulting theory is an apparently {\it new} rank-1 SCFT, not part of the $E_n$ series discovered by Seiberg \cite{Seiberg:1996bd, Morrison:1996xf}, and it also provides evidence that the 5d rank-0 SCFT $\FTX \MG_{[A_2, D_4]}$ is non-trivial as a `matter component'. This rank-one SCFT is {\it defined} as the low-energy limit of M-theory at the isolated hypersurface singularity $\MG$ given by:
\be
\label{type-I3335}
F(x)=x_1^3+x_2^3+x_3^3+x_4^5=0~.
\ee
The resolution is a weighted blow up with trivial weight $(1,1,1,1)$, which is exactly the same as for the rank-1 $E_6$ theory:
\be
(x_1,x_2,x_3,x_4;\delta_1)~.
\ee
The resolved equation:
\be
\t \MG:\ x_1^3+x_2^3+x_3^3+x_4^5\delta_1^2=0~,
\ee
has a terminal singularity of type $\MG_{[A_2, D_4]}$ at $\delta_1=x_1=x_2=x_3=0$:
\be
\delta_1^2+x_1^3+x_2^3+x_3^3=0~.
\ee
Thus the 5d SCFT $\FT$ on $\MG$ can be interpreted as a non-trivial `coupling' of a rank-1 $E_6$ theory with the $\MG_{[A_2, D_4]}$ theory. Note that, if we interpret the same resolved geometry as the Higgs branch of the 4d SCFT $\FTfour$, the interpretation is rather mundane: we have a residual 4d SCFT AD$[A_2, D_4]$ at every point on the one-dimensional Higgs branch of the larger theory, which happens to be AD$[D_4, E_8]$. On the other hand, the most straightforward interpretation in M-theory is rather more provocative: we have a rank-1 5d SCFT with an enhanced Coulomb branch (in the sense of \cite{Argyres:2016xmc})---that is, there are additional light degrees of freedom at every point on the CB. 
\begin{table}[t]
\centering 
\begin{tabular}{ | c ||c |c | c|| c|c| c|| c| c|c|} 
\hline\hline 
 $F$& $r$ &$f$   &$d_H$ &$\h r$& $\h d_H$  & $\Delta \CA_r$  & $b_3$ & $\frak{f}$  \\ [0.5ex] 
\hline 
$x_1^3+x_2^3+x_3^3+x_4^5 $ & $1$&$0$&$16$ &$16$&$1$& $-1$ & $2$ &  $\mathbb{Z}_5$ \\
\hline
\end{tabular}
\caption{The SCFT data associated to the singularity $x_1^3+x_2^3+x_3^3+x_4^5=0$.}\label{f:rank-1coupled}
\end{table}

From the deformation and resolution of $\MG$, we can compute the relevant data of $\FT$ and $\FTfour$, as shown in  Table~\ref{f:rank-1coupled}. As one can see, the flavor rank of $\FT$ is bounded by rank$(G_H^{\rm 5d})\leq f+\frac{1}{2}b_3=1$, and the HB dimension is $d_H=16$. More interestingly, there exists a non-trivial 3-form symmetry 
{$\Gamma_m^{(3)}=\Z_5$} (or a 0-form symmetry $\Z_5$). These features are distinct from any known rank-1 $E_n$ theory, including the $E_6$ theory (which has $d_H=11$). Hence, we conjecture that the $\FT$ associated to the singularity (\ref{type-I3335}) is a new rank-1 SCFT.

\subsection{One-form symmetries of the 4d SCFTs AD$[G, G']$}\label{subsec:oneform AD}
In this subsection, we further illustrate our discussion of higher-form symmetries from section \ref{subsec: higher form}, by computing the one-form symmetry $\frak{f}$ (electric or magnetic) of numerous 4d SCFTs of type AD$[G, G']$. We simply need to compute the torsion subgroup $\torHtwo= \frak{f}\oplus \frak{f} \subset H_2(L_5(\MG), \Z)$.

 For instance, for AD$[A_k, A_l]$, we find that this group is always trivial. On the other hand, for $[A_k, D_m]$, we can have non-trivial torsion:
\be
 [A_k, D_m] \; : \; \qquad
\begin{array}{|c || c ccccccccccc|  } %
\hline
\Gamma^{(1)}= \frak{f} &D_4 &D_5& D_6&D_7&D_8&D_9&D_{10}&D_{11}&D_{12}&D_{13}&D_{14}&D_{15}\\
\hline\hline
 A_1&0 & 0 & 0 & 0 & 0 & 0 & 0 & 0 & 0 & 0 & 0 & 0 \\
 A_2& \Z_2 & 0 & 0 & \Z_2 & 0 & 0 & \Z_2 & 0 & 0 & \Z_2 & 0 &
   0 \\
  A_3&0 & \Z_2 & 0 & 0 & 0 & \Z_2 & 0 & 0 & 0 & \Z_2 & 0 & 0 \\
 A_4& 0 & 0 & \Z_2^2 & 0 & 0 & 0 & 0 & \Z_2^2 & 0 & 0 & 0 & 0 \\
 A_5& 0 & 0 & 0 & \Z_2^2 & 0 & 0 & 0 & 0 & 0 & \Z_2^2 & 0 & 0 \\
 A_6& 0 & 0 & 0 & 0 & \Z_2^3 & 0 & 0 & 0 & 0 & 0 & 0 & \Z_2^3 \\
 A_7& 0 & 0 & 0 & 0 & 0 & \Z_2^3 & 0 & 0 & 0 & 0 & 0 & 0 \\
 A_8& \Z_2 & 0 & 0 & \Z_2 & 0 & 0 & \Z_2^4 & 0 & 0 & \Z_2 & 0
   & 0 \\
\hline
\end{array}
\ee
This includes the case $[A_2, D_4]$ studied above, which has $\frak{f}= \Z_2$. Similarly, for $[D_k, D_m]$, we find:
\be
 [D_k, D_m] \; : \; \qquad
 \begin{array}{|c || c ccccccccccc|  } %
\hline
\Gamma^{(1)}= \frak{f} &D_4 &D_5& D_6&D_7&D_8&D_9&D_{10}&D_{11}&D_{12}&D_{13}&D_{14}&D_{15}\\
\hline\hline
D_4& 0 & 0 & 0 & \Z_2 & 0 & 0 & 0 & 0 & 0 & \Z_2 & 0 & 0 \\
D_5& 0 & 0 & 0 & \Z_2 & 0 & \Z_2^2 & 0 & \Z_2 & 0 & 0 & 0 &
   \Z_2 \\
D_6& 0 & 0 & 0 & 0 & 0 & 0 & 0 & \Z_2^2 & 0 & 0 & 0 & 0 \\
D_7& \Z_2 & \Z_2 & 0 & 0 & 0 & \Z_2 & \Z_2 & 0 & 0 &
   \Z_2^3 & 0 & 0 \\
D_8& 0 & 0 & 0 & 0 & 0 & 0 & 0 & 0 & 0 & 0 & 0 & \Z_2^3 \\
 D_9&0 & \Z_2^2 & 0 & \Z_2 & 0 & 0 & 0 & \Z_2 & 0 & \Z_2^2 &
   0 & \Z_2 \\
D_{10}& 0 & 0 & 0 & \Z_2 & 0 & 0 & 0 & 0 & 0 & \Z_2 & 0 & 0 \\
\hline
\end{array}
\ee
As a last set of examples, consider the series:
\be
 [E_k, A_l] \; : \; \qquad
 \begin{array}{|c || c cccccccccccccc|  } %
\hline
\Gamma^{(1)}= \frak{f} &A_1&A_2&A_3&A_4 &A_5& A_6&A_7&A_8&A_9&A_{10}&A_{11}&A_{12}&A_{13}&A_{14}&A_{15}\\
\hline\hline
E_6& 0 & 0 & \Z_3 & 0 & \Z_2 & 0 & \Z_3 & 0 & 0 & 0 & 0 & 0 & 0 &
   0 & \Z_3 \\
E_7& 0 & 0 & 0 & 0 & \Z_3 & 0 & 0 & \Z_2^3 & 0 & 0 & \Z_3 & 0 & 0
   & 0 & 0 \\
E_8& 0 & 0 & 0 & 0 & \Z_5 & 0 & 0 & 0 & \Z_3^2 & 0 & \Z_5 & 0 & 0
   & \Z_2^4 & 0 \\
\hline
\end{array}
\ee
These one-forms symmetries deserve further study. As a preliminary comment, we note that they always seem related to the existence of S-duality frames in which we have a weakly coupled gauge group. For instance, we have:
\bea\nn
&[A_3, E_6]\; : \;  && \frak{f} = \Z_3~, \; && \Delta =\left\{\frac{5}{4},\frac{5}{4},\frac{3}{2},\frac{3}{2},\frac{3}{2},2,\frac{9}{4}
   ,\frac{9}{4},3\right\}~,\cr
     &{[A_5, E_7]}\; : \;&& {\frak{f} = \Z_3~}, \; &&\Delta =\left\{\frac{5}{4},\frac{5}{4},\frac{3}{2},\frac{3}{2},\frac{3}{2},\frac{7}{4},2
   ,2,\frac{9}{4},\frac{9}{4},\frac{5}{2},\frac{11}{4},3,3,\frac{7}{2},\frac{15}
   {4},\frac{9}{2}\right\}~,\\
  &[A_5, E_8]\; : \;  && {\frak{f} = \Z_5~},  \; && \Delta = \left\{\frac{7}{6},\frac{4}{3},\frac{4}{3},\frac{3}{2},\frac{3}{2},\frac{5}{3}, \frac{5}{3},2,\frac{13}{6},\frac{7}{3},\frac{7}{3},\frac{5}{2},\frac{5}{2},3,
   \frac{19}{6},\frac{10}{3},\frac{10}{3},4,\frac{25}{6},5\right\}~,
\eea
These Coulomb-branch spectra are compatible with the existence of `partially weakly-coupled description' with gauge group $SU(3)$, $SU(3){\times}SU(3)$ and $SU(5)$, respectively. This would explain the one-form symmetries if we could understand how strongly-coupled systems are coupled to these gauge groups, and why they preserve the center symmetry.
 

\section{Rank-zero 4d SCFTs from Isolated Toric Singularities}\label{sec: 4d rank0}

In the previous section, we discussed some isolated singularities that had $r=0$, and therefore correspond to rank-zero 5d SCFTs $\FT$. Conversely, a rank-zero 4d SCFT $\FTfour$ would arise from an isolated singularity such that $\h r=0$, as computed from the geometry. If we restrict ourselves to isolated hypersurface singularities, there is a unique canonical singularity that fits the bill
\be
x_1^2+ x_2^2+ x_3^2+ x_4^2=0~,
\ee
the conifold singularity, which engineers the free hypermultiplet in Type IIB string theory. To obtain potentially more interesting rank-zero 4d SCFTs, we need to consider more general canonical singularities. In that case, we would also need to properly define what we mean by the quantities $f$ and $\h r$ geometrically,  for the deformed singularity. A natural conjecture is that there always exists a mixed Hodge structure on the deformation space of any $\MG$, in some appropriate sense, such that:
\be
 f = {\rm dim} \, H^{2,2}(\h\MG)~, \qquad\quad  \h r = {\rm dim}\, H^{1,2}(\h\MG)= {\rm dim}\, H^{2,1}(\h\MG)~,
\ee
generalizing the hypersurface case. This is bound to be somewhat more subtle because, in general, the space of deformations can have several distinct branches, which can intersect in a non-trivial manner. In any case,  $\h r=0$ would correspond to a threefold whose deformation parameters all have a scaling dimension $\Delta =1$, once we identify the correct $U(1)_r$ scaling action on the coordinate ring of the singularity. In the following, we make some preliminary comments on the case, when $\MG$ is a toric singularity. 

\subsection{Toric singularities and deformations}
An isolated toric Calabi-Yau threefold singularity  $\MG$ corresponds to a strictly convex toric diagram.%
\footnote{For recent work in the 5d context and a summary of toric geometry see {\it e.g.}  \protect\cite{Closset:2018bjz, Eckhard:2020jyr}.}  The basic data of any of its crepant resolutions, $\t\MG$, is read off from the toric diagram as:
\be
f= n_E - 3~, \qquad\qquad r= n_I~,
\ee
where $n_E$ and $n_I$ denote the number of external and internal points, respectively, in the toric diagram of $\MG$. Here, $f$ again denotes the number of compact 2-cycles in $\t \MG$ dual to non-compact divisors, and it therefore counts the number of flavor background vector multiplet on the Coulomb branch of $\FT$. On the other hand, $\h d_H= r+f$, the dimension of the K\"ahler cone, must still be the dimension of the Higgs branch of $\FTfour$.

The 5d SCFT $\FT$ from an isolated toric singularity has a 1-form (or a magnetic 2-form) symmetry  \cite{Albertini:2020mdx, Morrison:2020ool}, which arise from the geometry of the boundary five-manifold $L_5(\MG)$, which does not need to be simply-connected (unlike in the hypersurface case). This then implies that the putative 4d SCFT from IIB on $\MG$ has a 0-form or 2-form symmetry given by the same discrete group at the one-form symmetry in 5d, namely ${\rm Tor} \, H_1(L_5(\MG), \Z)$. This torsion group is readily computed from the toric data \cite{Albertini:2020mdx, Morrison:2020ool}.

The versal space of deformations of any isolated toric singularity has dimension \cite{1994alg.geom..3004A}:
\be
\t d_H= \#(\text{versal deformations})  \leq f~.
\ee
More precisely, there are always exactly $f$ {\it first-order} deformations, but obstructions generally arise at higher order.%
\footnote{Hypersurfaces are `too simple' in that particular regard, since there is no obstruction in that case.}  There also exists an explicit algorithm to construct the deformed coordinate ring, for any deformation~\cite{1994alg.geom..3004A}.

Let us assume that $\h d_H=f$ for simplicity. Then, the versal deformation space has a single branch, which is the case most similar to the hypersurface singularities that we studied so far.  A generic versal deformation of $\MG$ then leads to a deformed singularity $\h \MG$ with $f$ 3-cycles.  
All these 3-cycles can undergo geometric transitions to the $N$ `flavor' 2-cycles in the (partially) resolved geometry. Thus, even without a detailed analysis of the $U(1)_r$ action on the singularity, we see that all these deformations must correspond to mass term in the putative 4d SCFT $\FTfour$ engineered in IIB. Therefore, the 5d/4d correspondence studied in this paper directly suggests that toric singularities geometrically engineer {\it rank-zero 4d SCFTs} in Type IIB. This was first proposed in \cite{Chen:2017wkw} for toric orbifolds. Here, we will present one interesting class of example with $\t d_H=f=1$, and present evidence that such theories $\FTfour$ can be non-trivial, in a limited sense that their Higgs branch can be a non-trivial hyper-K\"ahler cone. A more detailed analysis is left for future work.

\subsection{The $Y^{N,0}$ geometry, 5d $SU(N)_0$ and rank-zero 4d SCFTs}
Consider the toric diagram, which is a lattice polygon with four vertices at position $(0,0)$, $(1,0)$, $(1,N)$ and $(2,N)$
\be
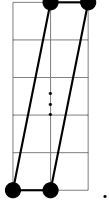
\label{toric diag SUN0}
 \begin{tikzpicture}[x=.5cm,y=.5cm]
\draw[step=.5cm,gray,very thin] (0,0) grid (2,5);
\draw[ligne] (0,0)--(1,0);
\draw[ligne] (1,5)--(2,5);
\draw[ligne] (1,0)--(2,5);
\draw[ligne] (0,0)--(1,5);
\node[bd] at (0,0) {}; 
\node[bd] at (1,0) {}; 
\node[bd] at (1,5) {}; 
\node[bd] at (2,5) {}; 
\node at (1,2.5) {$\vdots$};
\end{tikzpicture} \,.
\ee
Note that we have $f=1$ and $r= N-1$. This toric singularity is known as the cone over  $Y^{N,0}$ \cite{Gauntlett:2004hh}. It engineers a rank $N{-}1$ 5d SCFT with a single deformation, which triggers a flow to the 5d gauge theory $SU(N)_0$. In the limiting case $N=1$, we have the conifold singularity (which can also be viewed as an `$SU(1)$' theory in 5d \cite{Closset:2018bjz}), and for $N=2$ we have the complex cone over $\mathbb{F}_0$, engineering the rank-one $E_1$ SCFT. The space of versal deformations \eqref{toric diag SUN0} is one dimensional, giving us a smooth local Calabi-Yau threefold $\widetilde\MG$. For instance, the $E_1$ singularity is a non-complete intersection in $\C^9$, which admits single a consistent deformation to \cite{1994alg.geom..3004A}:
  \bea
& x_1 x_3 = (x_9{-}\varepsilon)^2~,  \, && x_2 x_4 = x_9^2~, \, && x_5 x_7= (x_9{-}\varepsilon)x_9~, \,&& x_6 x_8 =  (x_9{-}\varepsilon)x_9~, \cr
& x_5 x_8 = x_1 x_9~, \, &&  x_5 x_6 = x_2 (x_9{-}\varepsilon)~, \,  &&  x_6 x_7 = x_3 x_9~, \, &&  x_7 x_8 = x_4 (x_9{-}\varepsilon)~, \cr
& x_1 x_6 = x_5 (x_9{-} \varepsilon)~, \, &&  x_2 x_8 = x_5 x_9~, \,  &&  x_2 x_7 = x_6 x_9~, \, &&  x_3 x_5 = x_6  (x_9{-} \varepsilon)~, \cr
& x_3 x_8 = x_7  (x_9{-} \varepsilon)~, \, &&  x_4 x_6 = x_7 x_9~, \,  &&  x_1 x_7 = x_8  (x_9{-} \varepsilon)~, \, &&  x_4 x_5 = x_8 x_9~,
 \eea
with the isolated singularity at  $\varepsilon=0$.

The Higgs branch of $\FT$ can be studied using the tropical-geometric methods developed in \cite{vanBeest:2020kou}, which are easily specialized to the case of an isolated toric geometry. We summarize the relevant algorithm in Appendix~\ref{Appendix: MQ5 toric}. For the toric singularity \eqref{toric diag SUN0}, one finds the following magnetic quiver for $\FT$:
\be\label{MQ5 for SUN0}
\begin{tikzpicture}[x=.5cm,y=.5cm]
\node[] at (-3,1) {$\MQfive =\qquad $};
\draw[ligne, black](-1,1)--(1,1);
\node[bd] at (-1,1) [label=above:{{\scriptsize{$1$}}}] {};
\node[bd] at (1,1) [label=above:{{\scriptsize{$1$}}}] {};
\node at (0,1.5) {\scriptsize{$N$}};
\end{tikzpicture}
\ee
modulo the overall $U(1)$. Equivalently, $\MQfive= \text{SQED}[N_f=N]$. Its Coulomb branch gives the 5d Higgs branch, which is the $A_{N-1}$ Kleinian singularity:
\be
\CM_H^{\rm 5d} = {\rm CB}[\MQfive] = \C^2/\Z_N~.
\ee
The 3d mirror theory is the $\h a_{N-1}$ affine quiver:
\be\label{EQfive SUN0}
 \begin{tikzpicture}[x=.5cm,y=.5cm]
\node[] at (-3,1) {$\EQfive  =  $};
\draw[ligne, black](0,0)--(6,0) -- (3,2) -- (0,0); 
\node[bd] at (0,0) [label=below:{{\scriptsize$1$}}] {};
\node[bd] at (1,0) [label=below:{{\scriptsize$1$}}] {};
\node[bd] at (2,0) [label=below:{{\scriptsize$1$}}] {};
\node[bd] at (4,0) [label=below:{{\scriptsize$1$}}] {};
\node[bd] at (5,0) [label=below:{{\scriptsize$1$}}] {};
\node[bd] at (6,0) [label=below:{{\scriptsize$1$}}] {};
\node[bd] at (3,2) [label=above:{{\scriptsize$1$}}] {};
\node at (3,-0.8) {$\cdots$};
\end{tikzpicture}
\ee
This quiver has a gauge group $U(1)^N/U(1)$ and a single flavor symmetry, the `baryonic symmetry' $U(1)_B$, which assigns the same charge, $B=b_0$, to all the hypermultiplets in the loop. According to our general prescription, to obtain the magnetic quiver of the 4d SCFT $\FTfour$, we should gauge this symmetry:
\be
\MQfour = \EQfive/U(1)_B~.
\ee
There are two natural choices for the normalization of $U(1)_B$. The CB of \eqref{MQ5 for SUN0} is described as $T^+ T^-= \Phi^N$, where $\Phi$ is the vector multiplet scalar of SQED$[N_f=N]$, and $T^\pm$ are its monopole operators \cite{Borokhov:2002cg} of topological charge $B=\pm 1$. The topological symmetry is identified with the baryonic symmetry in the mirror \eqref{EQfive SUN0}, whose Higgs branch is described as  $\CB^+ \CB^- = M^N$, with $\CB^+ = H_1 \cdots H_N$ schematically, where $H_i$, $i=1, \cdots, N$ are the hypermultiplets of the circular quiver \eqref{EQfive SUN0}. Thus, we have the two possible charge assignments:
\bea\label{normalize B}
& (i) \qquad && B[T^\pm] = \pm 1 \; \;&&{\rm in} \; \MQfive \quad 
              & \leftrightarrow & \qquad B[H_i]= {1\ov N}  \; \;&&{\rm in} \; \EQfive~, \cr
& (ii) \qquad && B[T^\pm] = \pm N \; \;&&{\rm in} \; \MQfive \quad 
              & \leftrightarrow & \qquad B[H_i]= 1 \; \;&&{\rm in} \; \EQfive~. 
\eea
The magnetic quiver of the putative 4d SCFT is then given by a $U(1)^N$ gauge theory with $N$ hypermultiplets and the charges:
\be\label{MQfour SUN0}
\MQfour  \; : \; \qquad
\begin{tabular}{|c | c ccccc|  } %
\hline
& $H_1$ & $H_2$  $\cdots$ &$H_{N-2}$ && $H_{N-1}$ & $H_{N}$\\
\hline
$U(1)_1$ & $1$ &$0$&$\cdots$& $0$& $0$ &$-1$\\
$U(1)_2$ & $-1$ &$1$&$\cdots$& $0$&$0$ &$0$\\
$\vdots$ & $\vdots$&& $\ddots$& &&$\vdots$\\
$U(1)_{N-1}$ & $0$ &$0$&$\cdots$& $-1$&$1$ &$0$\\
$U(1)_B$ & $b_0$ &$b_0$&$\cdots$& $b_0$&$b_0$ &$b_0$\\
\hline 
\end{tabular}
\ee
Here, $b_0= B[H_i]$ in either normalization in \eqref{normalize B}. The mirror of \eqref{MQfour SUN0}, which is the electric quiverine of $\FTfour$, is obtained by gauging the topological symmetry of $\MQfive$. Using the results of \cite{Witten:2003ya}, it is easy to see that, depending on the normalization \eqref{normalize B}, we obtain:
\bea\label{EQ4 for SUN0}
& (i) \qquad && \EQfour = (N \, \text{hypermultiplets})~, \cr
& (ii) \qquad &&  \EQfour = (N \, \text{hypermultiplets})/\Z_N~. 
\eea
In the first case, we have $N$ free hypermultiplets, while in the second case we have a $\Z_N$ gauge theory of $N$ hypermultiplets. This discrete gauging leads to a non-trivial Higgs branch chiral ring, as can be ascertained by computing the HB Hilbert series. This is most easily done by computing the Hilbert series of the CB of $\MQfour$, as discussed in Appendix \ref{app: gauging affine aN}. Of course, in the first case, we have the Hilbert series of $\C^{2N}$, while for the $\Z_N$ gauge theory we find:
\bea\label{HS aN mq4}
& {\rm HS}_{\CM_H^{\rm 4d}}^{(N=2)} = {1 + 6 t^2 + t^4\ov (1 - t^2)^4}~, \\
& {\rm HS}_{\CM_H^{\rm 4d}}^{(N=3)}  = {1 - 3 t + 12 t^2 - 11 t^3 + 12 t^4 - 3 t^5 + t^6\ov (1 - t)^6(1 + t + t^2)^3 }~, \\
&  {\rm HS}_{\CM_H^{\rm 4d}}^{(N=4)}  ={1 + 12 t^2 + 108 t^4 + 212 t^6 + 358 t^8 + 212 t^{10} + 108 t^{12} + 
 12 t^{14} + t^{16} \ov (1 - t^2)^8 (1 + t^2)^4}~,
\eea
etc. For $N=2$, this corresponds to the minimal nilpotent orbit of $Sp(2)$, while for $N>2$ we have a Higgs branch with a $U(N)$ flavor group. From the mirror description \eqref{EQ4 for SUN0}, we know that these Higgs branches are $\Z_N$ quotients of $\C^{2N}$ preserving the hyper-K\"ahler structure---that is, quotients preserving 3d $\CN=4$ supersymmetry. In the $N=2$ case, we also know from the analysis of \cite{Shimizu:2017kzs} that the central charges $a$ and $c$ are the same as for two free hypermultiplets, yet the 4d theory has a non-trivial Higgs-branch chiral ring. 

Let us also note that the 5d SCFT $\FT$ has a $\Z_N$ 1-form symmetry or a 2-form symmetry, corresponding to having a gauge theory $SU(N)_0$ or $PSU(N)_0$ in the IR description~\cite{Morrison:2020ool}. This choice parallels the choice of global structure of $\FTfour$ apparent in \eqref{normalize B}, in which case $\FTfour$ should have a 0-form symmetry or a 2-form symmetry, respectively. Finally, it would be tempting to trivially `uplift' the 3d $\CN=4$ theory \eqref{EQ4 for SUN0} to `conclude' that the 4d $\CN=2$ SCFT $\FTfour$ consists of $N$ free hypermultiplets, or a discrete gauging thereof (which then has a two-form symmetry).  Of course, this is not guaranteed, since the flow from 4d to 3d could wash out some interesting information. It would certainly be interesting to explore these subtle issues further.

\subsubsection*{Acknowledgements}
We thank Fabio Apruzzi, Marieke van Beest, Lakshya Bhardwaj, Antoine Bourget, Stefano Cremonesi, Michele del Zotto, Julius Eckhard, Simone Giacomelli, Amihay Hanany, Zohar Komargodski, Horia Magureanu, Mario Martone, James Sparks, Yuji Tachikawa and Yifan Wang for very interesting discussions and comments on the draft. CC also gratefully acknowledges Yvette Siegert for her terminological contribution. CC is a Royal Society University Research Fellow and a Research Fellow at St John's College, Oxford. 
The work of SSN and YW is supported by the ERC
Consolidator Grant number 682608 ``Higgs bundles: Supersymmetric Gauge Theories and
Geometry (HIGGSBNDL)". SSN acknowledges support also from the Simons Foundation.



\appendix

\section{Higgs-branch Hilbert series from magnetic quivers}\label{appendix:HS}
In this Appendix, we present some relevant details on some of the Hilbert series computations mentioned in the main text. For completeness, let us first review some well-known technology. We would like to consider the Higgs branch in an SCFT with eight supercharges:
\be
\CM_H = {\rm Spec}\, \CR_H~.
\ee
Here,  $\CR_H$ denote the HB chiral ring, which is graded by the $SU(2)_R$ charge $R\in\half \Z$, and the moduli space must be a hyper-K\"ahler singularity. The Hilbert series of $\CM_H$ is the formal series over the chiral ring operators weighted by their $R$-charges:
\be
{\rm HS}_{\CM_{H}}(t) = \sum_{\rm \CO \in \CR_H}  t^{2 R[\CO]}~.
\ee
Given the Hilbert series, we have some simple `diagnostic tests' of the structure of the Higgs branch (see  \cite{Cremonesi:2017jrk} for a nice review):
\bit
\item The quaternionic dimension of $\CM_H$ corresponds to the pole at $t=1$:
\be
{\rm HS}_{\CM_{H}} \sim {1\ov (1-t)^{2 {\rm dim}(\CM_H)}}~.
\ee
\item Expanding around $t=0$, the order-$t$ term gives the number $n_{\rm free}$ of free hypermultiplets ($\C^2$ factors in $\CM_H$) while the order-$t^2$ term gives the number $n_{\rm cc}$ of conserved currents:
\be
{\rm HS}_{\CM_{H}}(t) = 1 +2 n_{\rm free} t + n_{\rm cc} t^2+ \cdots~.
\ee
Each free hypermultiplet corresponds to ${\rm HS}_{\C^2} = (1-t)^{-2}$, which factorizes from the full Hilbert series. 

\item By taking the plethystic logarithm%
\footnote{Recall the definition of the plethystic exponentional (giving us the `multi-particle states' from  the `single-particle states' \protect\cite{Benvenuti:2006qr}), and its inverse, the plethystic logarithm (PL). For a single variable $t$, we have: 
\be\nn
{\rm PE}[f(t)]=\exp\left( \sum_{p=1}^\infty {1\ov p} f(t^p)\right)~,\qquad 
{\rm PL}[g(t)]=\exp\left( \sum_{k=1}^\infty {\mu(k)\ov k} \log {g(t^k)}\right)~,
\ee
assuming $f(0)=0$ and $g(0)=1$; here, $\mu(k)$ is the M\"obius function. 
}
${\rm PL}({\rm HS}_{\CM_H}(t))$, we can find the number of generators $n_{H}$ of $\CR_H$ (with their $R$-charges), and the number of relations amongst them. The PL terminates if and only if $\CM_H$ is a complete intersection in $\C^{n_H}$. In particular, ${\rm PL}=2t$ for a free hypermultiplet.
\eit
One can refine this analysis by including fugacities for various flavor symmetries; in this paper, we only considered the `unrefined' HS, for simplicity.

For 3d $\CN=4$ theories, we can also compute the HS of the Coulomb branch, which is itself hyper-K\"ahler, with the CB operators weighted by the $SU(2)_C$ $R$-charge. Indeed, we would like to understand the 3d $\CN=4$ Coulomb branches of the `magnetic quivers' of 4d or 5d SCFTs. Assuming the magnetic quiver is given by an explicit gauge-theory UV description in 3d, with gauge group $G$, its CB Hilbert series is easily computed thanks to the monopole formula \cite{Cremonesi:2013lqa}:
\be\label{monopole form}
{\rm HS}_{\rm CB}(t) = \sum_{\m \in \Gamma/W_G} t^{2 R(\m)} P_G(t; \m)~, \qquad \qquad
P_G(t; \m)= \prod_{k=1}^{{\rm rank}(G)} {1\ov 1-t^{d_k(G_\m)}}~,
\ee
where the sum is over the GNO-quantized magnetic fluxes of $G$ (modulo the Weyl group), and $P_G(t; \m)$ denotes the Coulomb-branch index of the gauge group $G_\m$ that commutes with the flux $\m$---that is, $d_k(G_\m)$ are the dimension of the Casimir invariants of $G_\m$. 
(For instance, we have $P_{U(N)}(t; 0)= \prod_{k=1}^N (1-t^k)^{-1}$.)  Finally, $R(\m)$ is the quantum dimension of the monopole operator for this magnetic flux \cite{Gaiotto:2008ak}:
\be
 R(\m) = - \sum_{\alpha \in \Delta^+} |\alpha(\m)| +\half \sum_{\rho \in\mathfrak{R}} |\rho(\m)|~,
\ee
where $\Delta^+$ is the set of positive roots of $\mathfrak{g} = {\rm Lie}(G)$, and the second sum runs over all the weights of the (generally reducible) representation $\mathfrak{R}$ of $ \mathfrak{g}$ under which the hypermultiplets transform. Note that the monopole formulas `knows' about the global structure of $G$ through the sum over the magnetic flux lattices.~\footnote{Interestingly, the HS can also be viewed a supersymmetric partition function---under some assumptions, it is both a twisted partition function on $S^2 \times S^1$ \protect\cite{Closset:2016arn} and a limit of the 3d superconformal index \protect\cite{Razamat:2014pta}.}

Let us also note that, for abelian theories with gauge group $G= \prod_{k=1}^r U(1)_k$, the monopole formula \eqref{monopole form} simplifies to:
\be
{\rm HS}_{\rm CB}(t) = {1\ov (1-t^2)^r} \sum_{\m \in \Z^r} t^{2R(\m)}~, \qquad R(\m) = \sum_i \sum_{k=1}^r |q_i^k \m_k|~,
\ee
where the sum $\sum_i$ runs over all the hypermultiplets $H_i$, with electric charges $q_i^k$.

\subsection{The CB of $\EQfour$ for  AD$[A_2, D_4]$}\label{app:A2D4}
Consider the CB of the 3d quiver of eq.\eqref{EQ4 A2D4}, with either $SU(2)$ or $SO(3)$ as the central node. In the $SU(2)$ case, the HS takes the form:
\be\label{HS MQ5 A2D4 gen}
{\rm HS}_{\MQfive[\MG_{[A_2, D_4]}]}(t) = \sum_{\m\in \Z^3} {1\ov (1-t^2)^3} \left( {t^{2R(\m, 0)}\ov 1-t^4}+ \sum_{\n>0}{t^{2R(\m, \n)}\ov 1-t^2} \right)
\ee
with $\m= (\m_1, \m_2, \m_3)$ for the three $U(1)$ factors, and $\n \in \Z_{>0}$ the $SU(2)$ fluxes, and:
\be
 R(\m, \n) = -  |2\n| +\half \sum_{i=1}^3 \left(|\n+\m_i|+ |-\n+\m_i|\right)~,
\ee
we can write this as:
\be
{\rm HS}_{\MQfive[\MG_{[A_2, D_4]}]}^{(SU(2))}(t) ={ {\rm h}(t; 0)^3\ov 1-t^4}+ \sum_{\n=1}^\infty  { t^{-4 \n} {\rm h}(t; \n)^3 \ov 1- t^2}
\ee
corresponding to the $SU(2)$ gauging of three copies of SQED$[N_f{=}2]$, whose CB Hilbert series, with background flux $\n$ for the $SU(2)$ flavor group, can be easily computed:
\be
 {\rm h}(t; \n) \equiv {\rm HS}_{{\rm CB}[{\rm SQED}[N_f{=}2]]}(t;\n)= {t^{2|\n|} \left( 1+ t^2 + 2 |\n|(1-t^2)\right) \ov (1-t^2)^2}~.
\ee
This gives:
\be
{\rm HS}_{\MQfive[\MG_{[A_2, D_4]}]}^{(SU(2))}(t)  = {1+ 28 t^2+70 t^4+ 28 t^6+t^8\ov (1-t^2)^8}~.
\ee
Similarly, the $SO(3)$ gauging corresponds to summing over $\n \in \half \Z_{>0}$ in \eqref{HS MQ5 A2D4 gen}, and one finds:
\be
{\rm HS}_{\MQfive[\MG_{[A_2, D_4]}]}^{(SO(3))}(t)  = {1\ov  (1-t)^8}~.
\ee

\subsection{Gauging the affine $\h a_{N-1}$ quiver}\label{app: gauging affine aN}
Consider the CB Hilbert series of the $\MQfour$ \eqref{MQfour SUN0}. The monopole formula gives:
\be
{\rm HS}= {1\ov (1-t)^N}\sum_{\m_B\in \Z} \sum_{\m_1 \in \Z} \cdots \sum_{m_{N-1}\in \Z} t^{2 R(\m_B, \m)}~,
\ee
with $2 R(\m_B, \m)=\sum_{i=1}^{N-1} |m_i- m_{i+1} +b_0 \m_B|$.
For $b_0={1\ov N}$, we can perform a $SL(N, \Z)$ transformation on the electric charges, so that we have $N$ copies of SQED$[N_f=1]$, mirror to $N$ free hypers. In general, we can change basis so that:
\be\label{HS aN1}
{\rm HS}= {1\ov (1-t)}\sum_{\m_B\in \Z} \left[  {1\ov (1-t)^{N-1}}\sum_{\n_1 \in \Z} \cdots \sum_{n_N\in \Z} t^{2 R(\m_B, \n)}\right]~,
\ee
with:
\be
2 R(\m_B, \n)=  | b_0 N \m_B -\sum_{i=1}^{N-1} \n_i| + \sum_{i=1}^{N-1} |\n_i|~. 
\ee
We can first fix $\m_B$ and do the sum over $\n_i$ in \eqref{HS aN1}, which gives the HS of the CB of the $\h a_{N-1}$ quiver, corresponding to the minimal nilpotent orbit of $a_{N-1}$, in the presence of background magnetic flux $\m_B$.  
For instance:
\bea
&N=2: \; &&  {\rm HS(\m_B)} ={ t^{2 b_0 \m_B}  (1 + t^2 - 2 b_0 \m_B (1 - t^2)) \ov(-1 + t^2)^2}~, \cr
&N=3: \; &&  {\rm HS(\m_B)}=  \frac{t^{3 b_0 \m_B }
 \left(9  b_0^2  \m_B^2 (1-t^2)^2-9  b_0 \m_B  (1-t^4)+2
   (1+4 t^2+t^4)\right)}{2 (1-t^2)^4}~,
\eea
etc. Then, summing over $\m_B$, with either choice of $b_0$, we obtain the result shown in \eqref{HS aN mq4}. Note that we can view this CB (for this $\MQfour$) as a $\C^2$ fibration over the minimal nilpotent orbit of $a_{N-1}$ (the CB of the `ungauged' theory, $\EQfive$), as this computation makes rather explicit.

\section{Magnetic quivers $\MQfive$ for isolated toric singularities}
\label{Appendix: MQ5 toric}

In this appendix we summarize the construction of the magnetic quiver starting with the toric diagram \cite{vanBeest:2020kou}. Here we will limit ourselves to the strictly convex toric geometries, as these are relevant in the context of the 4d SCFTs. 

Consider then a strictly convex lattice polygons $P$ in $\mathbb{Z}^2 \subset \mathbb{R}^2$, with vertices $\bm{v}_i$, $i=1, \cdots, n$. The flavor rank is $f= n-3$ and the strict convexity requires that there are no lattice points along the edges connecting two of the vertices. The proposal in  \cite{vanBeest:2020kou} is that the magnetic quiver is obtained from a Minkowski sum decomposition of the polygon into convex polygons $P_i$
\be
P = P_1+ \cdots +  P_c \,,
\ee
where $A+B = \{a+ b; \ a\in A\,, b\in B\}$ is the Minkowski sum of two convex polygons of dimension 1 or 2 (i.e. edge segments or 2d polygons). This induces an edge coloring, by coloring each edge, that arises from 
$P_i$ into one color. 
An example is the toric polygon for the 5d $E_1$ theory, which has an IR description in 5d as 
$SU(2)_0$
\be
 \begin{tikzpicture}[x=.5cm,y=.5cm]
\draw[step=.5cm,gray,very thin] (0,-1) grid (2,1);
\node at (-3,0) {$E_1:$};
\draw[ligne, red] (0,0)--(1,1);
\draw[ligne, blue] (1,1)--(2,0);
\draw[ligne, red] (2,0)--(1,-1);
\draw[ligne, blue] (0,0)--(1,-1);
\node[bd] at (0,0) {}; 
\node[bd] at (1,1) {}; 
\node[bd] at (2,0) {}; 
\node[bd] at (1,-1) {}; 
\node  at (3,0) {$=$}; 
\draw[ligne, red] (4,-.5)--(5,0.5);
\node  at (6,0) {$+$}; 
\draw[ligne, blue](8,-.5) --(7,0.5);
\end{tikzpicture} \,.
\ee
The magnetic quiver for these isolated toric models is computed as follows:
\begin{enumerate}
\item Determine all distinct Minkowski sum decompositions of $P$, and associated edge colorings.
\item Each edge coloring can be extended by adding internal edges, such that the polygon is covered 
by polygons that have a single edge color, or by parallelograms, which are at most bi-colored, with opposite edges of the same color. 
\item Each color gets associated a node in the magnetic quiver of multiplicity $1$. 
\item The number of edges $k_{c_1, c_2}$ between nodes associated to $c_1$ and $c_2$ 
are determined by the mixed volume, i.e. 
\be
k_{c_1, c_2}= \text{Area} (G_{c_1, c_2}) = \text{area of the $c_1, c_2$ bicolored paralellogram}\,.
\ee
\end{enumerate}

For the $E_1$ there is one Minkowski sum decomposition, shown already above, 
from which the magnetic quiver follows to be two vertices and one double-line between them 
\be
\begin{tikzpicture}[x=.5cm,y=.5cm]
\node[] at (-4,1) {$\text{MQ}(E_1)= \qquad $};
\draw[ligne, black](-1,1.1)--(1,1.1);
\draw[ligne, black](-1,.9)--(1,.9);
\node[bd] at (-1,1) [blue, label=above:{{\scriptsize$1$}}] {};
\node[bd] at (1,1) [red, label=above:{{\scriptsize$1$}}] {};
\end{tikzpicture}
\ee
We will now focus on examples where the deformation theory is unobstructed, and therefore the Higgs branch dimension agrees with the flavor rank $f$. 
The $E_3$-theory (in 5d the rank 1 theory with $SU(2)+ 2 \bm{F}$ IR description) is of this type:
\be
\begin{tikzpicture}[x=.5cm,y=.5cm]
\draw[step=.5cm,gray,very thin] (0,-1) grid (2,1);
\node at (-3,0) {$E_3:$};
\draw[ligne, black] (0,0)--(1,1);
\draw[ligne, black] (1,1)--(2,1);
\draw[ligne, black] (2,1)--(2,0);
\draw[ligne, black] (2,0)--(1,-1);
\draw[ligne, black] (0,0)--(0,-1);
\draw[ligne, black] (0,-1)--(1,-1);
\node[bd] at (0,0) {}; 
\node[bd] at (1,1) {}; 
\node[bd] at (2,1) {}; 
\node[bd] at (2,0) {}; 
\node[bd] at (1,-1) {}; 
\node[bd] at (0,-1) {}; 
\node  at (3,0) {$=$}; 
\draw[ligne, blue] (4,0.5)--(5,0.5);
\node  at (6,0) {$+$}; 
\draw[ligne, red] (7,-0.5)--(7,0.5);
\node  at (8,0) {$+$}; 
\draw[ligne, cyan] (8.5,-0.5)--(9.5,0.5);
\node  at (10,0) {$=$}; 
\draw[ligne, blue] (11,.5)--(12,0.5) -- (12,-.5) --(11,0.5);
\node  at (13,0) {$+$}; 
\draw[ligne, red] (14,-0.5) -- (15,-.5) --(15,0.5)--(14,-0.5) ;
\end{tikzpicture} \,.
\ee
This implies the magnetic quiver, which has two components
\be
\begin{tikzpicture}[x=.5cm,y=.5cm]
\node[] at (-3,1) {$\text{MQ}(E_3)= \qquad $};
\draw[ligne, black](0,0)--(2,0) -- (1,1)--(0,0);
\node[bd] at (0,0) [blue, label=above:{{\scriptsize$1$}}] {};
\node[bd] at (2,0) [red, label=above:{{\scriptsize$1$}}] {};
\node[bd] at (1,1) [cyan, label=above:{{\scriptsize$1$}}] {};
\node at (3,0) {$\cup$} ;
\draw[ligne, black](4,0.1)--(6,0.1);
\draw[ligne, black](4,-0.1)--(6,-0.1);
\node[bd] at (4,0) [blue, label=above:{{\scriptsize$1$}}] {};
\node[bd] at (6,0) [red, label=above:{{\scriptsize$1$}}] {};
\end{tikzpicture}\,.
\ee

An infinite class of theories is the set of toric singularities with vertices 
\be
\{(0,0), (1,0), (1,N) , (2,N) \}\,,
\ee
i.e. 
\be
 \begin{tikzpicture}[x=.5cm,y=.5cm]
\draw[step=.5cm,gray,very thin] (0,0) grid (2,7);
\draw[ligne, red] (0,0)--(1,0);
\draw[ligne, red] (1,7)--(2,7);
\draw[ligne, blue] (1,0)--(2,7);
\draw[ligne, blue] (0,0)--(1,7);
\node[bd] at (0,0) {}; 
\node[bd] at (1,0) {}; 
\node[bd] at (1,7) {}; 
\node[bd] at (2,7) {}; 
\node at (1,4) {$\vdots$};
\end{tikzpicture} \,.
\ee
This toric diagram gives rise to the SCFT with IR description $SU(N)_0$ in M-theory. 
Here we will consider it in Type IIB. 
As the coloring indicates, there is precisely one magnetic quiver, which has two multiiplicity 1 nodes connected by $N$ edges, i.e. the $A_{N-1}$ Kleinian singularity 
\be
\begin{tikzpicture}[x=.5cm,y=.5cm]
\node[] at (-3,1) {$A_{N-1} =\qquad $};
\draw[ligne, black](-1,1)--(1,1);
\node[bd] at (-1,1) [blue, label=above:{{\scriptsize$1$}}] {};
\node[bd] at (1,1) [red, label=above:{{\scriptsize$1$}}] {};
\node at (0,1.5) {\scriptsize$N$};
\end{tikzpicture}
\ee

\section{Rank-$N$ $E_1$ magnetic quivers from the $(p,q)$-webs}\label{app: rankN E1}
In this Appendix, we discuss the brane-web construction of the magnetic quiver $\MQfive$ for the rank $N$ $E$-strings, focusing on the AS matter representation. The brane-webs for the rank $N$ $E_n$-strings were determined in \cite{Benini:2009gi, Jefferson:2017ahm} and a derivation for their entire decoupling tree will appear in \cite{vanBeest:2020civ}. 
Here, our interest is in the AS hypermultiplet phase transition, that we discussed in section \ref{subsec:rankNE8}. To illustrate this point, we simply consider the case with no fundamental flavor, {\it i.e.} the theory with an IR description:
\be
Sp(N) + 1\bm{AS} \,.
\ee
Its UV fixed point is the rank-$N$ $E_1$ theory. We present the generalized toric diagram for both the theory with the massless and massive anti-symmetric. Consider for instance the $N=2$ $E_0$-theory, which has generalized toric diagram \cite{vanBeest:2020kou} 
\be
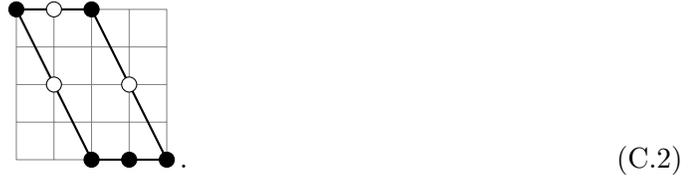
 \begin{tikzpicture}[x=.5cm,y=.5cm]
\draw[step=.5cm,gray,very thin] (4,-4) grid (0,0);
\draw[ligne] (0,0)--(2,-4);
\draw[ligne] (2,-4)--(3,-4);
\draw[ligne] (3,-4)--(4,-4);
\draw[ligne] (4,-4)--(2,0);
\draw[ligne] (0,0)--(2,0);
\node[bd] at (0,0) {}; 
\node[bd] at (2,-4) {}; 
\node[bd] at (3,-4) {}; 
\node[bd] at (4,-4) {}; 
\node[bd] at (2,0) {}; 
\node[wd] at (1,-2) {}; 
\node[wd] at (3,-2) {}; 
\node[wd] at (1,0) {}; 
\end{tikzpicture} \,.
\ee
In the dual brane-web the white dots indicate that the associated 5-branes dual to the edge segments adjacent to it end on the same 7-brane -- for a review, we refer the reader to \cite{Benini:2009gi}. 
The magnetic quiver of this theory is readily computed using the methods in \cite{vanBeest:2020kou}. For general $N$, one finds:
\be
\begin{tikzpicture}[x=.5cm,y=.5cm]
\node[] at (0,0) {$\text{MQ}^{(5)}= \qquad $};
\draw[ligne, black](2,0)--(4,0);
\draw[ligne, black](4,0.1)--(6,0.1);
\draw[ligne, black](4,-0.1)--(6,-0.1);
\node[bd] at (2,0) [label=above:{{\scriptsize$1$}}] {};
\node[bd] at (4,0) [label=above:{{\scriptsize$N$}}] {};
\node[bd] at (6,0) [ label=above:{{\scriptsize$N$}}] {};
\end{tikzpicture}\,,
\ee
This is $N$ times the affine $\mathfrak{a}_1$ Dynkin diagram, with an additional multiplicity-$1$ node. 

We can mimick the transition described in section \ref{subsec:rankNE8} at the level of the brane web. This is realized by the model where the 5-branes, which realize the $\bm{AS}$, end on one 7-brane, which effectively corresponds to decoupling of $N-1$ 5-brane segments in the transverse direction. For $N=2$, for instance, we then obtain: 
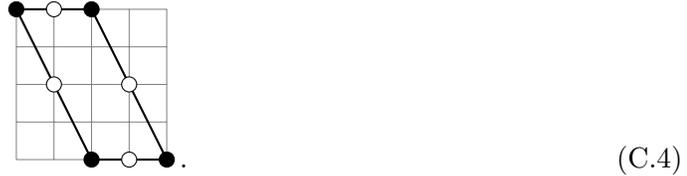
\be
 \begin{tikzpicture}[x=.5cm,y=.5cm]
\draw[step=.5cm,gray,very thin] (4,-4) grid (0,0);
\draw[ligne] (0,0)--(2,-4);
\draw[ligne] (2,-4)--(3,-4);
\draw[ligne] (3,-4)--(4,-4);
\draw[ligne] (4,-4)--(2,0);
\draw[ligne] (0,0)--(2,0);
\node[bd] at (0,0) {}; 
\node[bd] at (2,-4) {}; 
\node[wd] at (3,-4) {}; 
\node[bd] at (4,-4) {}; 
\node[bd] at (2,0) {}; 
\node[wd] at (1,-2) {}; 
\node[wd] at (3,-2) {}; 
\node[wd] at (1,0) {}; 
\end{tikzpicture} \,.
\ee
Here and for general $N$, the  $\MQfive$ is then simply $N$ times the affine $\mathfrak{a}_1$ Dynkin diagram, in agreement with our discussion in section \ref{subsec:rankNE8}: ending $N$ 5-branes on the same 7-brane, and displacing $N-1$ 5-brane segments transversely, corresponds to the `partial Higgs phase' on the CB of the 5d gauge theory.


%

\bibliography{FM}

\providecommand{\href}[2]{#2}\begingroup\raggedright\begin{thebibliography}{100}

\bibitem{Witten:1996qb}
E.~Witten, \emph{{Phase transitions in M theory and F theory}},
  \href{http://dx.doi.org/10.1016/0550-3213(96)00212-X}{\emph{Nucl. Phys.} {\bf
  B471} (1996) 195--216}, [\href{https://arxiv.org/abs/hep-th/9603150}{{\tt
  hep-th/9603150}}].

\bibitem{Katz:1996fh}
S.~H. Katz, A.~Klemm and C.~Vafa, \emph{{Geometric engineering of quantum field
  theories}},
  \href{http://dx.doi.org/10.1016/S0550-3213(97)00282-4}{\emph{Nucl. Phys. B}
  {\bf 497} (1997) 173--195}, [\href{https://arxiv.org/abs/hep-th/9609239}{{\tt
  hep-th/9609239}}].

\bibitem{Klemm:1996bj}
A.~Klemm, W.~Lerche, P.~Mayr, C.~Vafa and N.~P. Warner, \emph{{Selfdual strings
  and N=2 supersymmetric field theory}},
  \href{http://dx.doi.org/10.1016/0550-3213(96)00353-7}{\emph{Nucl. Phys. B}
  {\bf 477} (1996) 746--766}, [\href{https://arxiv.org/abs/hep-th/9604034}{{\tt
  hep-th/9604034}}].

\bibitem{Bershadsky:1996nh}
M.~Bershadsky, K.~A. Intriligator, S.~Kachru, D.~R. Morrison, V.~Sadov and
  C.~Vafa, \emph{{Geometric singularities and enhanced gauge symmetries}},
  \href{http://dx.doi.org/10.1016/S0550-3213(96)90131-5}{\emph{Nucl. Phys.}
  {\bf B481} (1996) 215--252},
  [\href{https://arxiv.org/abs/hep-th/9605200}{{\tt hep-th/9605200}}].

\bibitem{Seiberg:1996bd}
N.~Seiberg, \emph{{Five-dimensional SUSY field theories, nontrivial fixed
  points and string dynamics}},
  \href{http://dx.doi.org/10.1016/S0370-2693(96)01215-4}{\emph{Phys. Lett.}
  {\bf B388} (1996) 753--760},
  [\href{https://arxiv.org/abs/hep-th/9608111}{{\tt hep-th/9608111}}].

\bibitem{Morrison:1996xf}
D.~R. Morrison and N.~Seiberg, \emph{{Extremal transitions and five-dimensional
  supersymmetric field theories}},
  \href{http://dx.doi.org/10.1016/S0550-3213(96)00592-5}{\emph{Nucl. Phys.}
  {\bf B483} (1997) 229--247},
  [\href{https://arxiv.org/abs/hep-th/9609070}{{\tt hep-th/9609070}}].

\bibitem{Intriligator:1997pq}
K.~A. Intriligator, D.~R. Morrison and N.~Seiberg, \emph{{Five-dimensional
  supersymmetric gauge theories and degenerations of Calabi-Yau spaces}},
  \href{http://dx.doi.org/10.1016/S0550-3213(97)00279-4}{\emph{Nucl. Phys.}
  {\bf B497} (1997) 56--100}, [\href{https://arxiv.org/abs/hep-th/9702198}{{\tt
  hep-th/9702198}}].

\bibitem{Shapere:1999xr}
A.~D. Shapere and C.~Vafa, \emph{{BPS structure of Argyres-Douglas
  superconformal theories}},  \href{https://arxiv.org/abs/hep-th/9910182}{{\tt
  hep-th/9910182}}.

\bibitem{Hori:1997zj}
K.~Hori, H.~Ooguri and C.~Vafa, \emph{{NonAbelian conifold transitions and N=4
  dualities in three-dimensions}},
  \href{http://dx.doi.org/10.1016/S0550-3213(97)00529-4}{\emph{Nucl. Phys. B}
  {\bf 504} (1997) 147--174}, [\href{https://arxiv.org/abs/hep-th/9705220}{{\tt
  hep-th/9705220}}].

\bibitem{Closset:2021lwy}
C.~Closset, S.~Schafer-Nameki and Y.-N. Wang, \emph{{Coulomb and Higgs Branches
  from Canonical Singularities, Part 1: Hypersurfaces with Smooth Calabi-Yau
  Resolutions}},  \href{https://arxiv.org/abs/2111.13564}{{\tt 2111.13564}}.

\bibitem{CSNWII}
C.~Closset, S.~Schafer-Nameki and Y.-N. Wang, \emph{{Coulomb and Higgs Branches
  from Canonical Singularities: Part 2}}, .

\bibitem{Hayashi:2013lra}
H.~Hayashi, C.~Lawrie and S.~Schafer-Nameki, \emph{{Phases, Flops and F-theory:
  SU(5) Gauge Theories}},
  \href{http://dx.doi.org/10.1007/JHEP10(2013)046}{\emph{JHEP} {\bf 10} (2013)
  046}, [\href{https://arxiv.org/abs/1304.1678}{{\tt 1304.1678}}].

\bibitem{Hayashi:2014kca}
H.~Hayashi, C.~Lawrie, D.~R. Morrison and S.~Schafer-Nameki, \emph{{Box Graphs
  and Singular Fibers}},
  \href{http://dx.doi.org/10.1007/JHEP05(2014)048}{\emph{JHEP} {\bf 05} (2014)
  048}, [\href{https://arxiv.org/abs/1402.2653}{{\tt 1402.2653}}].

\bibitem{DelZotto:2017pti}
M.~Del~Zotto, J.~J. Heckman and D.~R. Morrison, \emph{{6D SCFTs and Phases of
  5D Theories}}, \href{http://dx.doi.org/10.1007/JHEP09(2017)147}{\emph{JHEP}
  {\bf 09} (2017) 147}, [\href{https://arxiv.org/abs/1703.02981}{{\tt
  1703.02981}}].

\bibitem{Jefferson:2017ahm}
P.~Jefferson, H.-C. Kim, C.~Vafa and G.~Zafrir, \emph{{Towards Classification
  of 5d SCFTs: Single Gauge Node}},
  \href{https://arxiv.org/abs/1705.05836}{{\tt 1705.05836}}.

\bibitem{Closset:2018bjz}
C.~Closset, M.~Del~Zotto and V.~Saxena, \emph{{Five-dimensional SCFTs and gauge
  theory phases: an M-theory/type IIA perspective}},
  \href{http://dx.doi.org/10.21468/SciPostPhys.6.5.052}{\emph{SciPost Phys.}
  {\bf 6} (2019) 052}, [\href{https://arxiv.org/abs/1812.10451}{{\tt
  1812.10451}}].

\bibitem{Jefferson:2018irk}
P.~Jefferson, S.~Katz, H.-C. Kim and C.~Vafa, \emph{{On Geometric
  Classification of 5d SCFTs}},
  \href{http://dx.doi.org/10.1007/JHEP04(2018)103}{\emph{JHEP} {\bf 04} (2018)
  103}, [\href{https://arxiv.org/abs/1801.04036}{{\tt 1801.04036}}].

\bibitem{Apruzzi:2018nre}
F.~Apruzzi, L.~Lin and C.~Mayrhofer, \emph{{Phases of 5d SCFTs from M-/F-theory
  on Non-Flat Fibrations}},
  \href{http://dx.doi.org/10.1007/JHEP05(2019)187}{\emph{JHEP} {\bf 05} (2019)
  187}, [\href{https://arxiv.org/abs/1811.12400}{{\tt 1811.12400}}].

\bibitem{Bhardwaj:2018yhy}
L.~Bhardwaj and P.~Jefferson, \emph{{Classifying 5d SCFTs via 6d SCFTs: Rank
  one}},  \href{https://arxiv.org/abs/1809.01650}{{\tt 1809.01650}}.

\bibitem{Bhardwaj:2018vuu}
L.~Bhardwaj and P.~Jefferson, \emph{{Classifying 5d SCFTs via 6d SCFTs:
  Arbitrary rank}},
  \href{http://dx.doi.org/10.1007/JHEP10(2019)282}{\emph{JHEP} {\bf 10} (2019)
  282}, [\href{https://arxiv.org/abs/1811.10616}{{\tt 1811.10616}}].

\bibitem{Apruzzi:2019vpe}
F.~Apruzzi, C.~Lawrie, L.~Lin, S.~Sch\"afer-Nameki and Y.-N. Wang, \emph{{5d
  Superconformal Field Theories and Graphs}},
  \href{http://dx.doi.org/10.1016/j.physletb.2019.135077}{\emph{Phys. Lett. B}
  {\bf 800} (2020) 135077}, [\href{https://arxiv.org/abs/1906.11820}{{\tt
  1906.11820}}].

\bibitem{Apruzzi:2019opn}
F.~Apruzzi, C.~Lawrie, L.~Lin, S.~Schafer-Nameki and Y.-N. Wang, \emph{{Fibers
  add Flavor, Part I: Classification of 5d SCFTs, Flavor Symmetries and BPS
  States}}, \href{http://dx.doi.org/10.1007/JHEP11(2019)068}{\emph{JHEP} {\bf
  11} (2019) 068}, [\href{https://arxiv.org/abs/1907.05404}{{\tt 1907.05404}}].

\bibitem{Apruzzi:2019enx}
F.~Apruzzi, C.~Lawrie, L.~Lin, S.~Schafer-Nameki and Y.-N. Wang, \emph{{Fibers
  add Flavor, Part II: 5d SCFTs, Gauge Theories, and Dualities}},
  \href{http://dx.doi.org/10.1007/JHEP03(2020)052}{\emph{JHEP} {\bf 03} (2020)
  052}, [\href{https://arxiv.org/abs/1909.09128}{{\tt 1909.09128}}].

\bibitem{Bhardwaj:2019jtr}
L.~Bhardwaj, \emph{{On the classification of $5d$ SCFTs}},
  \href{https://arxiv.org/abs/1909.09635}{{\tt 1909.09635}}.

\bibitem{Apruzzi:2019kgb}
F.~Apruzzi, S.~Schafer-Nameki and Y.-N. Wang, \emph{{5d SCFTs from Decoupling
  and Gluing}}, \href{http://dx.doi.org/10.1007/JHEP08(2020)153}{\emph{JHEP}
  {\bf 08} (2020) 153}, [\href{https://arxiv.org/abs/1912.04264}{{\tt
  1912.04264}}].

\bibitem{Bhardwaj:2019fzv}
L.~Bhardwaj, P.~Jefferson, H.-C. Kim, H.-C. Tarazi and C.~Vafa, \emph{{Twisted
  Circle Compactifications of 6d SCFTs}},
  \href{https://arxiv.org/abs/1909.11666}{{\tt 1909.11666}}.

\bibitem{Bhardwaj:2019xeg}
L.~Bhardwaj, \emph{{Do all 5d SCFTs descend from 6d SCFTs?}},
  \href{https://arxiv.org/abs/1912.00025}{{\tt 1912.00025}}.

\bibitem{Eckhard:2020jyr}
J.~Eckhard, S.~Schafer-Nameki and Y.-N. Wang, \emph{{Trifectas for $T_N$ in
  5d}},  \href{https://arxiv.org/abs/2004.15007}{{\tt 2004.15007}}.

\bibitem{Bhardwaj:2020kim}
L.~Bhardwaj, \emph{{More 5d KK theories}},
  \href{https://arxiv.org/abs/2005.01722}{{\tt 2005.01722}}.

\bibitem{Aharony:1997bh}
O.~Aharony, A.~Hanany and B.~Kol, \emph{{Webs of (p,q) five-branes,
  five-dimensional field theories and grid diagrams}},
  \href{http://dx.doi.org/10.1088/1126-6708/1998/01/002}{\emph{JHEP} {\bf 01}
  (1998) 002}, [\href{https://arxiv.org/abs/hep-th/9710116}{{\tt
  hep-th/9710116}}].

\bibitem{Cabrera:2018ann}
S.~Cabrera and A.~Hanany, \emph{{Quiver Subtractions}},
  \href{http://dx.doi.org/10.1007/JHEP09(2018)008}{\emph{JHEP} {\bf 09} (2018)
  008}, [\href{https://arxiv.org/abs/1803.11205}{{\tt 1803.11205}}].

\bibitem{Cabrera:2018jxt}
S.~Cabrera, A.~Hanany and F.~Yagi, \emph{{Tropical Geometry and Five
  Dimensional Higgs Branches at Infinite Coupling}},
  \href{http://dx.doi.org/10.1007/JHEP01(2019)068}{\emph{JHEP} {\bf 01} (2019)
  068}, [\href{https://arxiv.org/abs/1810.01379}{{\tt 1810.01379}}].

\bibitem{Cabrera:2019izd}
S.~Cabrera, A.~Hanany and M.~Sperling, \emph{{Magnetic quivers, Higgs branches,
  and 6d $N$=(1,0) theories}},
  \href{http://dx.doi.org/10.1007/JHEP06(2019)071}{\emph{JHEP} {\bf 06} (2019)
  071}, [\href{https://arxiv.org/abs/1904.12293}{{\tt 1904.12293}}].

\bibitem{Bourget:2019aer}
A.~Bourget, S.~Cabrera, J.~F. Grimminger, A.~Hanany, M.~Sperling, A.~Zajac
  et~al., \emph{{The Higgs mechanism --- Hasse diagrams for symplectic
  singularities}}, \href{http://dx.doi.org/10.1007/JHEP01(2020)157}{\emph{JHEP}
  {\bf 01} (2020) 157}, [\href{https://arxiv.org/abs/1908.04245}{{\tt
  1908.04245}}].

\bibitem{Bourget:2019rtl}
A.~Bourget, S.~Cabrera, J.~F. Grimminger, A.~Hanany and Z.~Zhong, \emph{{Brane
  Webs and Magnetic Quivers for SQCD}},
  \href{http://dx.doi.org/10.1007/JHEP03(2020)176}{\emph{JHEP} {\bf 03} (2020)
  176}, [\href{https://arxiv.org/abs/1909.00667}{{\tt 1909.00667}}].

\bibitem{Cabrera:2019dob}
S.~Cabrera, A.~Hanany and M.~Sperling, \emph{{Magnetic quivers, Higgs branches,
  and 6d $ \mathcal{N} $ = (1, 0) theories --- orthogonal and symplectic gauge
  groups}}, \href{http://dx.doi.org/10.1007/JHEP02(2020)184}{\emph{JHEP} {\bf
  02} (2020) 184}, [\href{https://arxiv.org/abs/1912.02773}{{\tt 1912.02773}}].

\bibitem{Grimminger:2020dmg}
J.~F. Grimminger and A.~Hanany, \emph{{Hasse Diagrams for $\mathbf{3d}$
  $\mathbf{\mathcal{N}=4}$ Quiver Gauge Theories -- Inversion and the full
  Moduli Space}},  \href{https://arxiv.org/abs/2004.01675}{{\tt 2004.01675}}.

\bibitem{Bourget:2020asf}
A.~Bourget, J.~F. Grimminger, A.~Hanany, M.~Sperling, G.~Zafrir and Z.~Zhong,
  \emph{{Magnetic quivers for rank 1 theories}},
  \href{http://dx.doi.org/10.1007/JHEP09(2020)189}{\emph{JHEP} {\bf 09} (2020)
  189}, [\href{https://arxiv.org/abs/2006.16994}{{\tt 2006.16994}}].

\bibitem{Bourget:2020gzi}
A.~Bourget, J.~F. Grimminger, A.~Hanany, M.~Sperling and Z.~Zhong,
  \emph{{Magnetic Quivers from Brane Webs with O5 Planes}},
  \href{http://dx.doi.org/10.1007/JHEP07(2020)204}{\emph{JHEP} {\bf 07} (2020)
  204}, [\href{https://arxiv.org/abs/2004.04082}{{\tt 2004.04082}}].

\bibitem{yau2005classification}
S.~S.-T. Yau and Y.~Yu, \emph{Classification of 3-dimensional isolated rational
  hypersurface singularities with c*-action}, {\emph{The Rocky Mountain Journal
  of Mathematics} {\bf 35} (2005) 1795--1809}.

\bibitem{Davenport:2016ggc}
I.~C. Davenport and I.~V. Melnikov, \emph{{Landau-Ginzburg skeletons}},
  \href{http://dx.doi.org/10.1007/JHEP05(2017)050}{\emph{JHEP} {\bf 05} (2017)
  050}, [\href{https://arxiv.org/abs/1608.04259}{{\tt 1608.04259}}].

\bibitem{arnold2012singularities}
E.~Arnold, S.~Gusein-Zade and A.~Varchenko, \emph{Singularities of
  Differentiable Maps, Volume 2: Monodromy and Asymptotics of Integrals}.
\newblock Modern Birkh{\"a}user Classics. Birkh{\"a}user Boston, 2012.

\bibitem{Cecotti:2010fi}
S.~Cecotti, A.~Neitzke and C.~Vafa, \emph{{R-Twisting and 4d/2d
  Correspondences}},  \href{https://arxiv.org/abs/1006.3435}{{\tt 1006.3435}}.

\bibitem{Xie:2015rpa}
D.~Xie and S.-T. Yau, \emph{{4d N=2 SCFT and singularity theory Part I:
  Classification}},  \href{https://arxiv.org/abs/1510.01324}{{\tt 1510.01324}}.

\bibitem{Chen:2016bzh}
B.~Chen, D.~Xie, S.-T. Yau, S.~S.~T. Yau and H.~Zuo, \emph{{4D $\mathcal{N} =
  2$ SCFT and singularity theory. Part II: complete intersection}},
  \href{http://dx.doi.org/10.4310/ATMP.2017.v21.n1.a2}{\emph{Adv. Theor. Math.
  Phys.} {\bf 21} (2017) 121--145},
  [\href{https://arxiv.org/abs/1604.07843}{{\tt 1604.07843}}].

\bibitem{Wang:2016yha}
Y.~Wang, D.~Xie, S.~S. Yau and S.-T. Yau, \emph{{$4d$ $\mathcal{N} = 2$ SCFT
  from complete intersection singularity}},
  \href{http://dx.doi.org/10.4310/ATMP.2017.v21.n3.a6}{\emph{Adv. Theor. Math.
  Phys.} {\bf 21} (2017) 801--855},
  [\href{https://arxiv.org/abs/1606.06306}{{\tt 1606.06306}}].

\bibitem{Chen:2017wkw}
B.~Chen, D.~Xie, S.~S. Yau, S.-T. Yau and H.~Zuo, \emph{{4d $\mathcal{N}=2$
  SCFT and singularity theory Part III: Rigid singularity}},
  \href{http://dx.doi.org/10.4310/ATMP.2018.v22.n8.a2}{\emph{Adv. Theor. Math.
  Phys.} {\bf 22} (2018) 1885--1905},
  [\href{https://arxiv.org/abs/1712.00464}{{\tt 1712.00464}}].

\bibitem{Xie:2017pfl}
D.~Xie and S.-T. Yau, \emph{{Three dimensional canonical singularity and five
  dimensional $ \mathcal{N} $ = 1 SCFT}},
  \href{http://dx.doi.org/10.1007/JHEP06(2017)134}{\emph{JHEP} {\bf 06} (2017)
  134}, [\href{https://arxiv.org/abs/1704.00799}{{\tt 1704.00799}}].

\bibitem{Intriligator:1996ex}
K.~A. Intriligator and N.~Seiberg, \emph{{Mirror symmetry in three-dimensional
  gauge theories}},
  \href{http://dx.doi.org/10.1016/0370-2693(96)01088-X}{\emph{Phys. Lett. B}
  {\bf 387} (1996) 513--519}, [\href{https://arxiv.org/abs/hep-th/9607207}{{\tt
  hep-th/9607207}}].

\bibitem{Cremonesi:2013lqa}
S.~Cremonesi, A.~Hanany and A.~Zaffaroni, \emph{{Monopole operators and Hilbert
  series of Coulomb branches of $3d$ $\mathcal{N} = 4$ gauge theories}},
  \href{http://dx.doi.org/10.1007/JHEP01(2014)005}{\emph{JHEP} {\bf 01} (2014)
  005}, [\href{https://arxiv.org/abs/1309.2657}{{\tt 1309.2657}}].

\bibitem{Nakajima:2015txa}
H.~Nakajima, \emph{{Towards a mathematical definition of Coulomb branches of
  $3$-dimensional $\mathcal{N}=4$ gauge theories, I}},
  \href{http://dx.doi.org/10.4310/ATMP.2016.v20.n3.a4}{\emph{Adv. Theor. Math.
  Phys.} {\bf 20} (2016) 595--669},
  [\href{https://arxiv.org/abs/1503.03676}{{\tt 1503.03676}}].

\bibitem{Bullimore:2015lsa}
M.~Bullimore, T.~Dimofte and D.~Gaiotto, \emph{{The Coulomb Branch of 3d
  ${\mathcal{N}= 4}$ Theories}},
  \href{http://dx.doi.org/10.1007/s00220-017-2903-0}{\emph{Commun. Math. Phys.}
  {\bf 354} (2017) 671--751}, [\href{https://arxiv.org/abs/1503.04817}{{\tt
  1503.04817}}].

\bibitem{Braverman:2016wma}
A.~Braverman, M.~Finkelberg and H.~Nakajima, \emph{{Towards a mathematical
  definition of Coulomb branches of $3$-dimensional $\mathcal{N} = 4$ gauge
  theories, II}},
  \href{http://dx.doi.org/10.4310/ATMP.2018.v22.n5.a1}{\emph{Adv. Theor. Math.
  Phys.} {\bf 22} (2018) 1071--1147},
  [\href{https://arxiv.org/abs/1601.03586}{{\tt 1601.03586}}].

\bibitem{Argyres:2015gha}
P.~C. Argyres, M.~Lotito, Y.~L\"{u} and M.~Martone, \emph{{Geometric
  constraints on the space of $ \mathcal{N} $ = 2 SCFTs. Part II: construction
  of special K\"{a}hler geometries and RG flows}},
  \href{http://dx.doi.org/10.1007/JHEP02(2018)002}{\emph{JHEP} {\bf 02} (2018)
  002}, [\href{https://arxiv.org/abs/1601.00011}{{\tt 1601.00011}}].

\bibitem{Martone:2020nsy}
M.~Martone, \emph{{Towards the classification of rank-$r$ $\mathcal{N}=2$
  SCFTs. Part I: twisted partition function and central charge formulae}},
  \href{https://arxiv.org/abs/2006.16255}{{\tt 2006.16255}}.

\bibitem{Argyres:2020nrr}
P.~Argyres and M.~Martone, \emph{{Construction and classification of Coulomb
  branch geometries}},  \href{https://arxiv.org/abs/2003.04954}{{\tt
  2003.04954}}.

\bibitem{Argyres:2020wmq}
P.~C. Argyres and M.~Martone, \emph{{Towards a classification of rank r$
  \mathcal{N} $ = 2 SCFTs. Part II. Special Kahler stratification of the
  Coulomb branch}},
  \href{http://dx.doi.org/10.1007/JHEP12(2020)022}{\emph{JHEP} {\bf 12} (2020)
  022}, [\href{https://arxiv.org/abs/2007.00012}{{\tt 2007.00012}}].

\bibitem{Argyres:2012fu}
P.~C. Argyres, K.~Maruyoshi and Y.~Tachikawa, \emph{{Quantum Higgs branches of
  isolated N=2 superconformal field theories}},
  \href{http://dx.doi.org/10.1007/JHEP10(2012)054}{\emph{JHEP} {\bf 10} (2012)
  054}, [\href{https://arxiv.org/abs/1206.4700}{{\tt 1206.4700}}].

\bibitem{DelZotto:2014kka}
M.~Del~Zotto and A.~Hanany, \emph{{Complete Graphs, Hilbert Series, and the
  Higgs branch of the 4d $\mathcal{N} =$ 2 $(A_n,A_m)$ SCFTs}},
  \href{http://dx.doi.org/10.1016/j.nuclphysb.2015.03.017}{\emph{Nucl. Phys. B}
  {\bf 894} (2015) 439--455}, [\href{https://arxiv.org/abs/1403.6523}{{\tt
  1403.6523}}].

\bibitem{Buican:2015ina}
M.~Buican and T.~Nishinaka, \emph{{On the superconformal index of
  Argyres--Douglas theories}},
  \href{http://dx.doi.org/10.1088/1751-8113/49/1/015401}{\emph{J. Phys. A} {\bf
  49} (2016) 015401}, [\href{https://arxiv.org/abs/1505.05884}{{\tt
  1505.05884}}].

\bibitem{Song:2015wta}
J.~Song, \emph{{Superconformal indices of generalized Argyres-Douglas theories
  from 2d TQFT}}, \href{http://dx.doi.org/10.1007/JHEP02(2016)045}{\emph{JHEP}
  {\bf 02} (2016) 045}, [\href{https://arxiv.org/abs/1509.06730}{{\tt
  1509.06730}}].

\bibitem{Maruyoshi:2016tqk}
K.~Maruyoshi and J.~Song, \emph{{Enhancement of Supersymmetry via
  Renormalization Group Flow and the Superconformal Index}},
  \href{http://dx.doi.org/10.1103/PhysRevLett.118.151602}{\emph{Phys. Rev.
  Lett.} {\bf 118} (2017) 151602},
  [\href{https://arxiv.org/abs/1606.05632}{{\tt 1606.05632}}].

\bibitem{Maruyoshi:2016aim}
K.~Maruyoshi and J.~Song, \emph{{$ \mathcal{N}=1 $ deformations and RG flows of
  $ \mathcal{N}=2 $ SCFTs}},
  \href{http://dx.doi.org/10.1007/JHEP02(2017)075}{\emph{JHEP} {\bf 02} (2017)
  075}, [\href{https://arxiv.org/abs/1607.04281}{{\tt 1607.04281}}].

\bibitem{Agarwal:2016pjo}
P.~Agarwal, K.~Maruyoshi and J.~Song, \emph{{$ \mathcal{N} $ =1 Deformations
  and RG flows of $ \mathcal{N} $ =2 SCFTs, part II: non-principal
  deformations}}, \href{http://dx.doi.org/10.1007/JHEP12(2016)103}{\emph{JHEP}
  {\bf 12} (2016) 103}, [\href{https://arxiv.org/abs/1610.05311}{{\tt
  1610.05311}}].

\bibitem{Song:2017oew}
J.~Song, D.~Xie and W.~Yan, \emph{{Vertex operator algebras of Argyres-Douglas
  theories from M5-branes}},
  \href{http://dx.doi.org/10.1007/JHEP12(2017)123}{\emph{JHEP} {\bf 12} (2017)
  123}, [\href{https://arxiv.org/abs/1706.01607}{{\tt 1706.01607}}].

\bibitem{Agarwal:2017roi}
P.~Agarwal, A.~Sciarappa and J.~Song, \emph{{$ \mathcal{N} $ =1 Lagrangians for
  generalized Argyres-Douglas theories}},
  \href{http://dx.doi.org/10.1007/JHEP10(2017)211}{\emph{JHEP} {\bf 10} (2017)
  211}, [\href{https://arxiv.org/abs/1707.04751}{{\tt 1707.04751}}].

\bibitem{Benvenuti:2017bpg}
S.~Benvenuti and S.~Giacomelli, \emph{{Lagrangians for generalized
  Argyres-Douglas theories}},
  \href{http://dx.doi.org/10.1007/JHEP10(2017)106}{\emph{JHEP} {\bf 10} (2017)
  106}, [\href{https://arxiv.org/abs/1707.05113}{{\tt 1707.05113}}].

\bibitem{Agarwal:2018zqi}
P.~Agarwal, S.~Lee and J.~Song, \emph{{Vanishing OPE Coefficients in 4d $N=2$
  SCFTs}}, \href{http://dx.doi.org/10.1007/JHEP06(2019)102}{\emph{JHEP} {\bf
  06} (2019) 102}, [\href{https://arxiv.org/abs/1812.04743}{{\tt 1812.04743}}].

\bibitem{Beem:2013sza}
C.~Beem, M.~Lemos, P.~Liendo, W.~Peelaers, L.~Rastelli and B.~C. van Rees,
  \emph{{Infinite Chiral Symmetry in Four Dimensions}},
  \href{http://dx.doi.org/10.1007/s00220-014-2272-x}{\emph{Commun. Math. Phys.}
  {\bf 336} (2015) 1359--1433}, [\href{https://arxiv.org/abs/1312.5344}{{\tt
  1312.5344}}].

\bibitem{Beem:2017ooy}
C.~Beem and L.~Rastelli, \emph{{Vertex operator algebras, Higgs branches, and
  modular differential equations}},
  \href{http://dx.doi.org/10.1007/JHEP08(2018)114}{\emph{JHEP} {\bf 08} (2018)
  114}, [\href{https://arxiv.org/abs/1707.07679}{{\tt 1707.07679}}].

\bibitem{Beem:2019tfp}
C.~Beem, C.~Meneghelli and L.~Rastelli, \emph{{Free Field Realizations from the
  Higgs Branch}}, \href{http://dx.doi.org/10.1007/JHEP09(2019)058}{\emph{JHEP}
  {\bf 09} (2019) 058}, [\href{https://arxiv.org/abs/1903.07624}{{\tt
  1903.07624}}].

\bibitem{Beem:2019snk}
C.~Beem, C.~Meneghelli, W.~Peelaers and L.~Rastelli, \emph{{VOAs and rank-two
  instanton SCFTs}},
  \href{http://dx.doi.org/10.1007/s00220-020-03746-9}{\emph{Commun. Math.
  Phys.} {\bf 377} (2020) 2553--2578},
  [\href{https://arxiv.org/abs/1907.08629}{{\tt 1907.08629}}].

\bibitem{Gukov:1999ya}
S.~Gukov, C.~Vafa and E.~Witten, \emph{{CFT's from Calabi-Yau four folds}},
  \href{http://dx.doi.org/10.1016/S0550-3213(00)00373-4}{\emph{Nucl. Phys. B}
  {\bf 584} (2000) 69--108}, [\href{https://arxiv.org/abs/hep-th/9906070}{{\tt
  hep-th/9906070}}].

\bibitem{Nekrasov:1996cz}
N.~Nekrasov, \emph{{Five dimensional gauge theories and relativistic integrable
  systems}}, \href{http://dx.doi.org/10.1016/S0550-3213(98)00436-2}{\emph{Nucl.
  Phys. B} {\bf 531} (1998) 323--344},
  [\href{https://arxiv.org/abs/hep-th/9609219}{{\tt hep-th/9609219}}].

\bibitem{Kapustin:1999ha}
A.~Kapustin and M.~J. Strassler, \emph{{On mirror symmetry in three-dimensional
  Abelian gauge theories}},
  \href{http://dx.doi.org/10.1088/1126-6708/1999/04/021}{\emph{JHEP} {\bf 04}
  (1999) 021}, [\href{https://arxiv.org/abs/hep-th/9902033}{{\tt
  hep-th/9902033}}].

\bibitem{Witten:2003ya}
E.~Witten, \emph{{SL(2,Z) action on three-dimensional conformal field theories
  with Abelian symmetry}},  \href{https://arxiv.org/abs/hep-th/0307041}{{\tt
  hep-th/0307041}}.

\bibitem{Ferlito:2017xdq}
G.~Ferlito, A.~Hanany, N.~Mekareeya and G.~Zafrir, \emph{{3d Coulomb branch and
  5d Higgs branch at infinite coupling}},
  \href{http://dx.doi.org/10.1007/JHEP07(2018)061}{\emph{JHEP} {\bf 07} (2018)
  061}, [\href{https://arxiv.org/abs/1712.06604}{{\tt 1712.06604}}].

\bibitem{Hanany:2018uhm}
A.~Hanany and N.~Mekareeya, \emph{{The small E$_{8}$ instanton and the Kraft
  Procesi transition}},
  \href{http://dx.doi.org/10.1007/JHEP07(2018)098}{\emph{JHEP} {\bf 07} (2018)
  098}, [\href{https://arxiv.org/abs/1801.01129}{{\tt 1801.01129}}].

\bibitem{vanBeest:2020kou}
M.~van Beest, A.~Bourget, J.~Eckhard and S.~Schafer-Nameki, \emph{{(Symplectic)
  Leaves and (5d Higgs) Branches in the Poly(go)nesian Tropical Rain Forest}},
  \href{http://dx.doi.org/10.1007/JHEP11(2020)124}{\emph{JHEP} {\bf 11} (2020)
  124}, [\href{https://arxiv.org/abs/2008.05577}{{\tt 2008.05577}}].

\bibitem{Garcia-Etxebarria:2019cnb}
I.~n. Garc\'{i}a~Etxebarria, B.~Heidenreich and D.~Regalado, \emph{{IIB flux
  non-commutativity and the global structure of field theories}},
  \href{http://dx.doi.org/10.1007/JHEP10(2019)169}{\emph{JHEP} {\bf 10} (2019)
  169}, [\href{https://arxiv.org/abs/1908.08027}{{\tt 1908.08027}}].

\bibitem{Morrison:2020ool}
D.~R. Morrison, S.~Schafer-Nameki and B.~Willett, \emph{{Higher-Form Symmetries
  in 5d}}, \href{http://dx.doi.org/10.1007/JHEP09(2020)024}{\emph{JHEP} {\bf
  09} (2020) 024}, [\href{https://arxiv.org/abs/2005.12296}{{\tt 2005.12296}}].

\bibitem{Albertini:2020mdx}
F.~Albertini, M.~Del~Zotto, I.~Garc\'{i}a~Etxebarria and S.~S. Hosseini,
  \emph{{Higher Form Symmetries and M-theory}},
  \href{https://arxiv.org/abs/2005.12831}{{\tt 2005.12831}}.

\bibitem{DelZotto:2020esg}
M.~Del~Zotto, I.~n. Garc\'{i}a~Etxebarria and S.~S. Hosseini, \emph{{Higher
  Form Symmetries of Argyres-Douglas Theories}},
  \href{https://arxiv.org/abs/2007.15603}{{\tt 2007.15603}}.

\bibitem{Argyres:1995jj}
P.~C. Argyres and M.~R. Douglas, \emph{{New phenomena in SU(3) supersymmetric
  gauge theory}},
  \href{http://dx.doi.org/10.1016/0550-3213(95)00281-V}{\emph{Nucl. Phys. B}
  {\bf 448} (1995) 93--126}, [\href{https://arxiv.org/abs/hep-th/9505062}{{\tt
  hep-th/9505062}}].

\bibitem{Argyres:1995xn}
P.~C. Argyres, M.~Plesser, N.~Seiberg and E.~Witten, \emph{{New N=2
  superconformal field theories in four-dimensions}},
  \href{http://dx.doi.org/10.1016/0550-3213(95)00671-0}{\emph{Nucl. Phys. B}
  {\bf 461} (1996) 71--84}, [\href{https://arxiv.org/abs/hep-th/9511154}{{\tt
  hep-th/9511154}}].

\bibitem{Minahan:1996fg}
J.~A. Minahan and D.~Nemeschansky, \emph{{An N=2 superconformal fixed point
  with E(6) global symmetry}},
  \href{http://dx.doi.org/10.1016/S0550-3213(96)00552-4}{\emph{Nucl. Phys. B}
  {\bf 482} (1996) 142--152}, [\href{https://arxiv.org/abs/hep-th/9608047}{{\tt
  hep-th/9608047}}].

\bibitem{Xie:2012hs}
D.~Xie, \emph{{General Argyres-Douglas Theory}},
  \href{http://dx.doi.org/10.1007/JHEP01(2013)100}{\emph{JHEP} {\bf 01} (2013)
  100}, [\href{https://arxiv.org/abs/1204.2270}{{\tt 1204.2270}}].

\bibitem{Arras:2016evy}
P.~Arras, A.~Grassi and T.~Weigand, \emph{{Terminal Singularities, Milnor
  Numbers, and Matter in F-theory}},
  \href{http://dx.doi.org/10.1016/j.geomphys.2017.09.001}{\emph{J. Geom. Phys.}
  {\bf 123} (2018) 71--97}, [\href{https://arxiv.org/abs/1612.05646}{{\tt
  1612.05646}}].

\bibitem{Grassi:2018rva}
A.~Grassi and T.~Weigand, \emph{{On topological invariants of algebraic
  threefolds with ($\mathbb Q$-factorial) singularities}},
  \href{https://arxiv.org/abs/1804.02424}{{\tt 1804.02424}}.

\bibitem{Argyres:2016yzz}
P.~C. Argyres and M.~Martone, \emph{{4d $ \mathcal{N} $ =2 theories with
  disconnected gauge groups}},
  \href{http://dx.doi.org/10.1007/JHEP03(2017)145}{\emph{JHEP} {\bf 03} (2017)
  145}, [\href{https://arxiv.org/abs/1611.08602}{{\tt 1611.08602}}].

\bibitem{Aharony:2016kai}
O.~Aharony and Y.~Tachikawa, \emph{{S-folds and 4d N=3 superconformal field
  theories}}, \href{http://dx.doi.org/10.1007/JHEP06(2016)044}{\emph{JHEP} {\bf
  06} (2016) 044}, [\href{https://arxiv.org/abs/1602.08638}{{\tt 1602.08638}}].

\bibitem{Tachikawa:2017gyf}
Y.~Tachikawa, \emph{{On gauging finite subgroups}},
  \href{http://dx.doi.org/10.21468/SciPostPhys.8.1.015}{\emph{SciPost Phys.}
  {\bf 8} (2020) 015}, [\href{https://arxiv.org/abs/1712.09542}{{\tt
  1712.09542}}].

\bibitem{Apruzzi:2020pmv}
F.~Apruzzi, S.~Giacomelli and S.~Sch\"{a}fer-Nameki, \emph{{4d $\mathcal{N}=2$
  S-folds}}, \href{http://dx.doi.org/10.1103/PhysRevD.101.106008}{\emph{Phys.
  Rev. D} {\bf 101} (2020) 106008},
  [\href{https://arxiv.org/abs/2001.00533}{{\tt 2001.00533}}].

\bibitem{Hanany:2018vph}
A.~Hanany and G.~Zafrir, \emph{{Discrete Gauging in Six Dimensions}},
  \href{http://dx.doi.org/10.1007/JHEP07(2018)168}{\emph{JHEP} {\bf 07} (2018)
  168}, [\href{https://arxiv.org/abs/1804.08857}{{\tt 1804.08857}}].

\bibitem{Hanany:2018cgo}
A.~Hanany and M.~Sperling, \emph{{Discrete quotients of 3-dimensional $
  \mathcal{N}=4 $ Coulomb branches via the cycle index}},
  \href{http://dx.doi.org/10.1007/JHEP08(2018)157}{\emph{JHEP} {\bf 08} (2018)
  157}, [\href{https://arxiv.org/abs/1807.02784}{{\tt 1807.02784}}].

\bibitem{Hanany:2018dvd}
A.~Hanany and A.~Zajac, \emph{{Discrete Gauging in Coulomb branches of Three
  Dimensional $\mathcal{N}=4$ Supersymmetric Gauge Theories}},
  \href{http://dx.doi.org/10.1007/JHEP08(2018)158}{\emph{JHEP} {\bf 08} (2018)
  158}, [\href{https://arxiv.org/abs/1807.03221}{{\tt 1807.03221}}].

\bibitem{Closset:2020afy}
C.~Closset, S.~Giacomelli, S.~Schafer-Nameki and Y.-N. Wang, \emph{{5d and 4d
  SCFTs: Canonical Singularities, Trinions and S-Dualities}},
  \href{http://dx.doi.org/10.1007/JHEP05(2021)274}{\emph{JHEP} {\bf 05} (2021)
  274}, [\href{https://arxiv.org/abs/2012.12827}{{\tt 2012.12827}}].

\bibitem{Cordova:2016xhm}
C.~Cordova, T.~T. Dumitrescu and K.~Intriligator, \emph{{Deformations of
  Superconformal Theories}},
  \href{http://dx.doi.org/10.1007/JHEP11(2016)135}{\emph{JHEP} {\bf 11} (2016)
  135}, [\href{https://arxiv.org/abs/1602.01217}{{\tt 1602.01217}}].

\bibitem{Chang:2018xmx}
C.-M. Chang, \emph{{5d and 6d SCFTs Have No Weak Coupling Limit}},
  \href{http://dx.doi.org/10.1007/JHEP09(2019)016}{\emph{JHEP} {\bf 09} (2019)
  016}, [\href{https://arxiv.org/abs/1810.04169}{{\tt 1810.04169}}].

\bibitem{Cadavid:1995bk}
A.~C. Cadavid, A.~Ceresole, R.~D'Auria and S.~Ferrara,
  \emph{{Eleven-dimensional supergravity compactified on Calabi-Yau
  threefolds}},
  \href{http://dx.doi.org/10.1016/0370-2693(95)00891-N}{\emph{Phys. Lett.} {\bf
  B357} (1995) 76--80}, [\href{https://arxiv.org/abs/hep-th/9506144}{{\tt
  hep-th/9506144}}].

\bibitem{caibar1999minimal}
M.~Caibar, \emph{Minimal models of canonical singularities and their
  cohomology}.
\newblock PhD thesis, Ph. D. thesis, University of Warwick, 1999.

\bibitem{Caibarb3}
M.~Caibar, \emph{Minimal models of canonical 3-fold singularities and their
  betti numbers}, {\emph{International Mathematics Research Notices} {\bf 2005}
  (2005) 1563--1581}.

\bibitem{Ooguri:1996me}
H.~Ooguri and C.~Vafa, \emph{{Summing up D instantons}},
  \href{http://dx.doi.org/10.1103/PhysRevLett.77.3296}{\emph{Phys. Rev. Lett.}
  {\bf 77} (1996) 3296--3298},
  [\href{https://arxiv.org/abs/hep-th/9608079}{{\tt hep-th/9608079}}].

\bibitem{Bourget:2020xdz}
A.~Bourget, J.~F. Grimminger, A.~Hanany, R.~Kalveks, M.~Sperling and Z.~Zhong,
  \emph{{Magnetic Lattices for Orthosymplectic Quivers}},
  \href{http://dx.doi.org/10.1007/JHEP12(2020)092}{\emph{JHEP} {\bf 12} (2020)
  092}, [\href{https://arxiv.org/abs/2007.04667}{{\tt 2007.04667}}].

\bibitem{Katz:1997eq}
S.~Katz, P.~Mayr and C.~Vafa, \emph{{Mirror symmetry and exact solution of 4-D
  N=2 gauge theories: 1.}},
  \href{http://dx.doi.org/10.4310/ATMP.1997.v1.n1.a2}{\emph{Adv. Theor. Math.
  Phys.} {\bf 1} (1998) 53--114},
  [\href{https://arxiv.org/abs/hep-th/9706110}{{\tt hep-th/9706110}}].

\bibitem{DelZotto:2015rca}
M.~Del~Zotto, C.~Vafa and D.~Xie, \emph{{Geometric engineering, mirror symmetry
  and $ 6{\mathrm{d}}_{\left(1,0\right)}\to
  4{\mathrm{d}}_{\left(\mathcal{N}=2\right)} $}},
  \href{http://dx.doi.org/10.1007/JHEP11(2015)123}{\emph{JHEP} {\bf 11} (2015)
  123}, [\href{https://arxiv.org/abs/1504.08348}{{\tt 1504.08348}}].

\bibitem{Wang:2015mra}
Y.~Wang and D.~Xie, \emph{{Classification of Argyres-Douglas theories from M5
  branes}}, \href{http://dx.doi.org/10.1103/PhysRevD.94.065012}{\emph{Phys.
  Rev. D} {\bf 94} (2016) 065012},
  [\href{https://arxiv.org/abs/1509.00847}{{\tt 1509.00847}}].

\bibitem{Shapere:2008zf}
A.~D. Shapere and Y.~Tachikawa, \emph{{Central charges of N=2 superconformal
  field theories in four dimensions}},
  \href{http://dx.doi.org/10.1088/1126-6708/2008/09/109}{\emph{JHEP} {\bf 09}
  (2008) 109}, [\href{https://arxiv.org/abs/0804.1957}{{\tt 0804.1957}}].

\bibitem{Seiberg:1994rs}
N.~Seiberg and E.~Witten, \emph{{Electric - magnetic duality, monopole
  condensation, and confinement in N=2 supersymmetric Yang-Mills theory}},
  \href{http://dx.doi.org/10.1016/0550-3213(94)90124-4}{\emph{Nucl. Phys. B}
  {\bf 426} (1994) 19--52}, [\href{https://arxiv.org/abs/hep-th/9407087}{{\tt
  hep-th/9407087}}].

\bibitem{Lindstrom:1999pz}
U.~Lindstrom, M.~Rocek and R.~von Unge, \emph{{HyperKahler quotients and
  algebraic curves}},
  \href{http://dx.doi.org/10.1088/1126-6708/2000/01/022}{\emph{JHEP} {\bf 01}
  (2000) 022}, [\href{https://arxiv.org/abs/hep-th/9908082}{{\tt
  hep-th/9908082}}].

\bibitem{deWit:2001brd}
B.~de~Wit, M.~Rocek and S.~Vandoren, \emph{{Hypermultiplets, hyperKahler cones
  and quaternion Kahler geometry}},
  \href{http://dx.doi.org/10.1088/1126-6708/2001/02/039}{\emph{JHEP} {\bf 02}
  (2001) 039}, [\href{https://arxiv.org/abs/hep-th/0101161}{{\tt
  hep-th/0101161}}].

\bibitem{Alexandrov:2010np}
S.~Alexandrov, D.~Persson and B.~Pioline, \emph{{On the topology of the
  hypermultiplet moduli space in type II/CY string vacua}},
  \href{http://dx.doi.org/10.1103/PhysRevD.83.026001}{\emph{Phys. Rev. D} {\bf
  83} (2011) 026001}, [\href{https://arxiv.org/abs/1009.3026}{{\tt
  1009.3026}}].

\bibitem{Shimizu:2017kzs}
H.~Shimizu, Y.~Tachikawa and G.~Zafrir, \emph{{Anomaly matching on the Higgs
  branch}}, \href{http://dx.doi.org/10.1007/JHEP12(2017)127}{\emph{JHEP} {\bf
  12} (2017) 127}, [\href{https://arxiv.org/abs/1703.01013}{{\tt 1703.01013}}].

\bibitem{Chang:2019uag}
C.-M. Chang, M.~Fluder, Y.-H. Lin and Y.~Wang, \emph{{Proving the 6d Cardy
  Formula and Matching Global Gravitational Anomalies}},
  \href{https://arxiv.org/abs/1910.10151}{{\tt 1910.10151}}.

\bibitem{Minor1968}
J.~Milnor, \emph{Singular Points of Complex Hypersurfaces. (AM-61)}.
\newblock Princeton University Press, 1968.

\bibitem{Candelas:1989js}
P.~Candelas and X.~C. de~la Ossa, \emph{{Comments on Conifolds}},
  \href{http://dx.doi.org/10.1016/0550-3213(90)90577-Z}{\emph{Nucl. Phys. B}
  {\bf 342} (1990) 246--268}.

\bibitem{Sparks:2010sn}
J.~Sparks, \emph{{Sasaki-Einstein Manifolds}},
  \href{http://dx.doi.org/10.4310/SDG.2011.v16.n1.a6}{\emph{Surveys Diff.
  Geom.} {\bf 16} (2011) 265--324},
  [\href{https://arxiv.org/abs/1004.2461}{{\tt 1004.2461}}].

\bibitem{RANDELL1975347}
R.~C. Randell, \emph{The homology of generalized brieskorn manifolds},
  \href{http://dx.doi.org/https://doi.org/10.1016/0040-9383(75)90019-1}{\emph{Topology}
  {\bf 14} (1975) 347 -- 355}.

\bibitem{10.1007/BFb0070047}
P.~Orlik, \emph{On the homology of weighted homogeneous manifolds},  in
  \emph{Proceedings of the Second Conference on Compact Transformation Groups}
  (H.~T. Ku, L.~N. Mann, J.~L. Sicks and J.~C. Su, eds.), (Berlin, Heidelberg),
  pp.~260--269, Springer Berlin Heidelberg, 1972.

\bibitem{Seiberg:1996nz}
N.~Seiberg and E.~Witten, \emph{{Gauge dynamics and compactification to
  three-dimensions}},  in \emph{{Conference on the Mathematical Beauty of
  Physics (In Memory of C. Itzykson)}}, pp.~333--366, 6, 1996.
\newblock \href{https://arxiv.org/abs/hep-th/9607163}{{\tt hep-th/9607163}}.

\bibitem{Gaiotto:2008cd}
D.~Gaiotto, G.~W. Moore and A.~Neitzke, \emph{{Four-dimensional wall-crossing
  via three-dimensional field theory}},
  \href{http://dx.doi.org/10.1007/s00220-010-1071-2}{\emph{Commun. Math. Phys.}
  {\bf 299} (2010) 163--224}, [\href{https://arxiv.org/abs/0807.4723}{{\tt
  0807.4723}}].

\bibitem{Buican:2015hsa}
M.~Buican and T.~Nishinaka, \emph{{Argyres--Douglas theories, S$^1$ reductions,
  and topological symmetries}},
  \href{http://dx.doi.org/10.1088/1751-8113/49/4/045401}{\emph{J. Phys. A} {\bf
  49} (2016) 045401}, [\href{https://arxiv.org/abs/1505.06205}{{\tt
  1505.06205}}].

\bibitem{Benini:2009gi}
F.~Benini, S.~Benvenuti and Y.~Tachikawa, \emph{{Webs of five-branes and N=2
  superconformal field theories}},
  \href{http://dx.doi.org/10.1088/1126-6708/2009/09/052}{\emph{JHEP} {\bf 09}
  (2009) 052}, [\href{https://arxiv.org/abs/0906.0359}{{\tt 0906.0359}}].

\bibitem{Benini:2010uu}
F.~Benini, Y.~Tachikawa and D.~Xie, \emph{{Mirrors of 3d Sicilian theories}},
  \href{http://dx.doi.org/10.1007/JHEP09(2010)063}{\emph{JHEP} {\bf 09} (2010)
  063}, [\href{https://arxiv.org/abs/1007.0992}{{\tt 1007.0992}}].

\bibitem{Dey:2020hfe}
A.~Dey, \emph{{Three Dimensional Mirror Symmetry beyond $ADE$ quivers and
  Argyres-Douglas theories}},  \href{https://arxiv.org/abs/2004.09738}{{\tt
  2004.09738}}.

\bibitem{Gaiotto:2014kfa}
D.~Gaiotto, A.~Kapustin, N.~Seiberg and B.~Willett, \emph{{Generalized Global
  Symmetries}}, \href{http://dx.doi.org/10.1007/JHEP02(2015)172}{\emph{JHEP}
  {\bf 02} (2015) 172}, [\href{https://arxiv.org/abs/1412.5148}{{\tt
  1412.5148}}].

\bibitem{Aharony:2013hda}
O.~Aharony, N.~Seiberg and Y.~Tachikawa, \emph{{Reading between the lines of
  four-dimensional gauge theories}},
  \href{http://dx.doi.org/10.1007/JHEP08(2013)115}{\emph{JHEP} {\bf 08} (2013)
  115}, [\href{https://arxiv.org/abs/1305.0318}{{\tt 1305.0318}}].

\bibitem{Freed:2006yc}
D.~S. Freed, G.~W. Moore and G.~Segal, \emph{{Heisenberg Groups and
  Noncommutative Fluxes}},
  \href{http://dx.doi.org/10.1016/j.aop.2006.07.014}{\emph{Annals Phys.} {\bf
  322} (2007) 236--285}, [\href{https://arxiv.org/abs/hep-th/0605200}{{\tt
  hep-th/0605200}}].

\bibitem{Gaiotto:2015una}
D.~Gaiotto and H.-C. Kim, \emph{{Duality walls and defects in 5d $
  \mathcal{N}=1 $ theories}},
  \href{http://dx.doi.org/10.1007/JHEP01(2017)019}{\emph{JHEP} {\bf 01} (2017)
  019}, [\href{https://arxiv.org/abs/1506.03871}{{\tt 1506.03871}}].

\bibitem{vanCoevering:2009zz}
C.~van Coevering, \emph{{Examples of asymptotically conical Ricci-flat Kahler
  manifolds}}, \href{http://dx.doi.org/10.1007/s00209-009-0631-7}{\emph{Math.
  Z.} {\bf 267} (2011) 465--496}.

\bibitem{Bhardwaj:2013qia}
L.~Bhardwaj and Y.~Tachikawa, \emph{{Classification of 4d N=2 gauge theories}},
  \href{http://dx.doi.org/10.1007/JHEP12(2013)100}{\emph{JHEP} {\bf 12} (2013)
  100}, [\href{https://arxiv.org/abs/1309.5160}{{\tt 1309.5160}}].

\bibitem{Douglas:1996sw}
M.~R. Douglas and G.~W. Moore, \emph{{D-branes, quivers, and ALE instantons}},
  \href{https://arxiv.org/abs/hep-th/9603167}{{\tt hep-th/9603167}}.

\bibitem{Lawrie:2012gg}
C.~Lawrie and S.~Schafer-Nameki, \emph{{The Tate Form on Steroids: Resolution
  and Higher Codimension Fibers}},
  \href{http://dx.doi.org/10.1007/JHEP04(2013)061}{\emph{JHEP} {\bf 04} (2013)
  061}, [\href{https://arxiv.org/abs/1212.2949}{{\tt 1212.2949}}].

\bibitem{Saxena:2019wuy}
V.~Saxena, \emph{{Rank-two 5d SCFTs from M-theory at isolated toric
  singularities: a systematic study}},
  \href{https://arxiv.org/abs/1911.09574}{{\tt 1911.09574}}.

\bibitem{Cremonesi:2014xha}
S.~Cremonesi, G.~Ferlito, A.~Hanany and N.~Mekareeya, \emph{{Coulomb Branch and
  The Moduli Space of Instantons}},
  \href{http://dx.doi.org/10.1007/JHEP12(2014)103}{\emph{JHEP} {\bf 12} (2014)
  103}, [\href{https://arxiv.org/abs/1408.6835}{{\tt 1408.6835}}].

\bibitem{Hayashi:2018lyv}
H.~Hayashi, S.-S. Kim, K.~Lee and F.~Yagi, \emph{{Dualities and 5-brane webs
  for 5d rank 2 SCFTs}},
  \href{http://dx.doi.org/10.1007/JHEP12(2018)016}{\emph{JHEP} {\bf 12} (2018)
  016}, [\href{https://arxiv.org/abs/1806.10569}{{\tt 1806.10569}}].

\bibitem{Hayashi:2019jvx}
H.~Hayashi, S.-S. Kim, K.~Lee and F.~Yagi, \emph{{Complete prepotential for 5d
  $ \mathcal{N} $ = 1 superconformal field theories}},
  \href{http://dx.doi.org/10.1007/JHEP02(2020)074}{\emph{JHEP} {\bf 02} (2020)
  074}, [\href{https://arxiv.org/abs/1912.10301}{{\tt 1912.10301}}].

\bibitem{Hanany:2017ooe}
A.~Hanany and R.~Kalveks, \emph{{Quiver Theories and Formulae for Nilpotent
  Orbits of Exceptional Algebras}},
  \href{http://dx.doi.org/10.1007/JHEP11(2017)126}{\emph{JHEP} {\bf 11} (2017)
  126}, [\href{https://arxiv.org/abs/1709.05818}{{\tt 1709.05818}}].

\bibitem{Arnold1975}
V.~I. Arnold, \emph{{CRITICAL} {POINTS} {OF} {SMOOTH} {FUNCTIONS} {AND} {THEIR}
  {NORMAL} {FORMS}},
  \href{http://dx.doi.org/10.1070/rm1975v030n05abeh001521}{\emph{Russian
  Mathematical Surveys} {\bf 30} (oct, 1975) 1--75}.

\bibitem{Cecotti:2011gu}
S.~Cecotti and M.~Del~Zotto, \emph{{On Arnold's 14 `exceptional' N=2
  superconformal gauge theories}},
  \href{http://dx.doi.org/10.1007/JHEP10(2011)099}{\emph{JHEP} {\bf 10} (2011)
  099}, [\href{https://arxiv.org/abs/1107.5747}{{\tt 1107.5747}}].

\bibitem{Benvenuti:2018bav}
S.~Benvenuti, \emph{{A tale of exceptional $3d$ dualities}},
  \href{http://dx.doi.org/10.1007/JHEP03(2019)125}{\emph{JHEP} {\bf 03} (2019)
  125}, [\href{https://arxiv.org/abs/1809.03925}{{\tt 1809.03925}}].

\bibitem{Dedushenko:2019mnd}
M.~Dedushenko and Y.~Wang, \emph{{4d/2d $\rightarrow $ 3d/1d: A song of
  protected operator algebras}},  \href{https://arxiv.org/abs/1912.01006}{{\tt
  1912.01006}}.

\bibitem{Nanopoulos:2010bv}
D.~Nanopoulos and D.~Xie, \emph{{More Three Dimensional Mirror Pairs}},
  \href{http://dx.doi.org/10.1007/JHEP05(2011)071}{\emph{JHEP} {\bf 05} (2011)
  071}, [\href{https://arxiv.org/abs/1011.1911}{{\tt 1011.1911}}].

\bibitem{Gaiotto:2009we}
D.~Gaiotto, \emph{{N=2 dualities}},
  \href{http://dx.doi.org/10.1007/JHEP08(2012)034}{\emph{JHEP} {\bf 08} (2012)
  034}, [\href{https://arxiv.org/abs/0904.2715}{{\tt 0904.2715}}].

\bibitem{Cecotti:2013lda}
S.~Cecotti, M.~Del~Zotto and S.~Giacomelli, \emph{{More on the N=2
  superconformal systems of type $D_p(G)$}},
  \href{http://dx.doi.org/10.1007/JHEP04(2013)153}{\emph{JHEP} {\bf 04} (2013)
  153}, [\href{https://arxiv.org/abs/1303.3149}{{\tt 1303.3149}}].

\bibitem{Beratto:2020wmn}
E.~Beratto, S.~Giacomelli, N.~Mekareeya and M.~Sacchi, \emph{{3d mirrors of the
  circle reduction of twisted $A_{2N}$ theories of class $\mathsf{S}$}},
  \href{https://arxiv.org/abs/2007.05019}{{\tt 2007.05019}}.

\bibitem{Buican:2014hfa}
M.~Buican, S.~Giacomelli, T.~Nishinaka and C.~Papageorgakis,
  \emph{{Argyres-Douglas Theories and S-Duality}},
  \href{http://dx.doi.org/10.1007/JHEP02(2015)185}{\emph{JHEP} {\bf 02} (2015)
  185}, [\href{https://arxiv.org/abs/1411.6026}{{\tt 1411.6026}}].

\bibitem{Gaiotto:2008ak}
D.~Gaiotto and E.~Witten, \emph{{S-Duality of Boundary Conditions In N=4 Super
  Yang-Mills Theory}},
  \href{http://dx.doi.org/10.4310/ATMP.2009.v13.n3.a5}{\emph{Adv. Theor. Math.
  Phys.} {\bf 13} (2009) 721--896},
  [\href{https://arxiv.org/abs/0807.3720}{{\tt 0807.3720}}].

\bibitem{Gaiotto:2012uq}
D.~Gaiotto and S.~S. Razamat, \emph{{Exceptional Indices}},
  \href{http://dx.doi.org/10.1007/JHEP05(2012)145}{\emph{JHEP} {\bf 05} (2012)
  145}, [\href{https://arxiv.org/abs/1203.5517}{{\tt 1203.5517}}].

\bibitem{Closset:2019juk}
C.~Closset and M.~Del~Zotto, \emph{{On 5d SCFTs and their BPS quivers. Part I:
  B-branes and brane tilings}},  \href{https://arxiv.org/abs/1912.13502}{{\tt
  1912.13502}}.

\bibitem{Klebanov:1998hh}
I.~R. Klebanov and E.~Witten, \emph{{Superconformal field theory on
  three-branes at a Calabi-Yau singularity}},
  \href{http://dx.doi.org/10.1016/S0550-3213(98)00654-3}{\emph{Nucl. Phys. B}
  {\bf 536} (1998) 199--218}, [\href{https://arxiv.org/abs/hep-th/9807080}{{\tt
  hep-th/9807080}}].

\bibitem{Fazzi:2019gvt}
M.~Fazzi and A.~Tomasiello, \emph{{Holography, Matrix Factorizations and
  K-stability}}, \href{http://dx.doi.org/10.1007/JHEP05(2020)119}{\emph{JHEP}
  {\bf 05} (2020) 119}, [\href{https://arxiv.org/abs/1906.08272}{{\tt
  1906.08272}}].

\bibitem{Hanany:2016gbz}
A.~Hanany and R.~Kalveks, \emph{{Quiver Theories for Moduli Spaces of Classical
  Group Nilpotent Orbits}},
  \href{http://dx.doi.org/10.1007/JHEP06(2016)130}{\emph{JHEP} {\bf 06} (2016)
  130}, [\href{https://arxiv.org/abs/1601.04020}{{\tt 1601.04020}}].

\bibitem{Argyres:2016xmc}
P.~Argyres, M.~Lotito, Y.~Lu and M.~Martone, \emph{{Geometric constraints on
  the space of $ \mathcal{N}$ = 2 SCFTs. Part III: enhanced Coulomb branches
  and central charges}},
  \href{http://dx.doi.org/10.1007/JHEP02(2018)003}{\emph{JHEP} {\bf 02} (2018)
  003}, [\href{https://arxiv.org/abs/1609.04404}{{\tt 1609.04404}}].

\bibitem{1994alg.geom..3004A}
K.~{Altmann}, \emph{{The versal Deformation of an isolated toric Gorenstein
  Singularity}}, {\emph{arXiv e-prints} (Mar., 1994) alg--geom/9403004},
  [\href{https://arxiv.org/abs/alg-geom/9403004}{{\tt alg-geom/9403004}}].

\bibitem{Gauntlett:2004hh}
J.~P. Gauntlett, D.~Martelli, J.~F. Sparks and D.~Waldram, \emph{{A New
  infinite class of Sasaki-Einstein manifolds}},
  \href{http://dx.doi.org/10.4310/ATMP.2004.v8.n6.a3}{\emph{Adv. Theor. Math.
  Phys.} {\bf 8} (2004) 987--1000},
  [\href{https://arxiv.org/abs/hep-th/0403038}{{\tt hep-th/0403038}}].

\bibitem{Borokhov:2002cg}
V.~Borokhov, A.~Kapustin and X.-k. Wu, \emph{{Monopole operators and mirror
  symmetry in three-dimensions}},
  \href{http://dx.doi.org/10.1088/1126-6708/2002/12/044}{\emph{JHEP} {\bf 12}
  (2002) 044}, [\href{https://arxiv.org/abs/hep-th/0207074}{{\tt
  hep-th/0207074}}].

\bibitem{Cremonesi:2017jrk}
S.~Cremonesi, \emph{{3d supersymmetric gauge theories and Hilbert series}},
  {\emph{Proc. Symp. Pure Math.} {\bf 98} (2018) 21--48},
  [\href{https://arxiv.org/abs/1701.00641}{{\tt 1701.00641}}].

\bibitem{Benvenuti:2006qr}
S.~Benvenuti, B.~Feng, A.~Hanany and Y.-H. He, \emph{{Counting BPS Operators in
  Gauge Theories: Quivers, Syzygies and Plethystics}},
  \href{http://dx.doi.org/10.1088/1126-6708/2007/11/050}{\emph{JHEP} {\bf 11}
  (2007) 050}, [\href{https://arxiv.org/abs/hep-th/0608050}{{\tt
  hep-th/0608050}}].

\bibitem{Closset:2016arn}
C.~Closset and H.~Kim, \emph{{Comments on twisted indices in 3d supersymmetric
  gauge theories}},
  \href{http://dx.doi.org/10.1007/JHEP08(2016)059}{\emph{JHEP} {\bf 08} (2016)
  059}, [\href{https://arxiv.org/abs/1605.06531}{{\tt 1605.06531}}].

\bibitem{Razamat:2014pta}
S.~S. Razamat and B.~Willett, \emph{{Down the rabbit hole with theories of
  class $ \mathcal{S} $}},
  \href{http://dx.doi.org/10.1007/JHEP10(2014)099}{\emph{JHEP} {\bf 10} (2014)
  099}, [\href{https://arxiv.org/abs/1403.6107}{{\tt 1403.6107}}].

\bibitem{vanBeest:2020civ}
M.~Van~Beest, A.~Bourget, J.~Eckhard and S.~Sch\"afer-Nameki, \emph{{(5d
  RG-flow) Trees in the Tropical Rain Forest}},
  \href{http://dx.doi.org/10.1007/JHEP03(2021)241}{\emph{JHEP} {\bf 03} (2021)
  241}, [\href{https://arxiv.org/abs/2011.07033}{{\tt 2011.07033}}].

\end{thebibliography}\endgroup
\bibliographystyle{JHEP}
\end{document}